\documentclass[preprint,aps,tighten]{revtex4}

\input epsf

\def\Journal#1#2#3#4{{#1} {\bf #2}, #3 (#4)}

\def\NPB{{\em Nucl. Phys.} B}
\def\NPA{{\em Nucl. Phys.} A}
\def\PLB{{\em Phys. Lett.}  B}
\def\PRL{\em Phys. Rev. Lett.}
\def\PRD{{\em Phys. Rev.} D}
\def\PRC{{\em Phys. Rev.} C}
\def\ZPC{{\em Z. Phys.} C}
\def\CPC{\em Comput. Phys. Commun.}
\def\JCP{\em J. Comp. Phys.}
\def\PR{\em Phys. Rep.}
\def\RMP{\em Rev. Mod. Phys.}
\def\JPG{\em J. Phys. G}

\begin{document}

\title{Thermalization of gluons in ultrarelativistic heavy ion collisions
by including three-body interactions in a parton cascade}

\author{Zhe Xu$^{1,2}$ \footnote{E-mail: Zhe.Xu@theo.physik.uni-giessen.de}
and Carsten Greiner $^{2}$
\footnote{E-mail: carsten.greiner@th.physik.uni-frankfurt.de}}
\affiliation{$^1$Institut f\"ur Theoretische Physik, 
Justus-Liebig-Universit\"at Giessen, D-35392 Giessen, Germany  \vspace*{2mm}
\\
$^2$Institut f\"ur Theoretische Physik, Johann Wolfgang Goethe 
Universit\"at Frankfurt, D-60054 Frankfurt am Main, Germany}

\date{July 2004}

\begin{abstract}
We develop a new 3+1 dimensional Monte Carlo cascade solving the
kinetic on-shell Boltzmann equations for partons including the
inelastic $gg \leftrightarrow ggg$ pQCD processes. The back
reaction channel is treated -- for the first time -- fully
consistently within this scheme. An extended stochastic method
is used to solve the collision integral. The frame
dependence and convergency are studied for a fixed tube with
thermal initial conditions. The detailed numerical analysis 
shows that the stochastic method is fully covariant and that
convergency is achieved more efficiently than within a standard
geometrical formulation of the collision term, especially for
high gluon interaction rates. The cascade is then applied to
simulate parton evolution and to investigate thermalization of
gluons for a central Au+Au collision at RHIC energy. For this
study the initial conditions are assumed to be generated by
independent minijets with $p_T > p_0=2$ GeV. With that choice
it is demonstrated that overall kinetic equilibration is driven
mainly by the inelastic processes and is achieved on a scale of
$1$ fm/c. The further evolution of the expanding gluonic matter
in the central region then shows almost an ideal hydrodynamical
behavior. In addition, full chemical equilibration of the gluons
follows on a longer timescale of about $3$ fm/c. 
\end{abstract}


\maketitle

\section{Introduction}
\label{sec:intro}
The main subject of the heavy ion experiments at the Relativistic
Heavy Ion Collider (RHIC) at BNL and at the Large Hadron Collider
(LHC) at CERN is to create a new state of matter, the Quark Gluon
Plasma (QGP), which is expected to be a transient thermal system
of interacting quarks and gluons. Due to the confinement free
quarks and gluons cannot be detected. The search for QGP has to
be carried out by analyzing certain proposed hadronic and
electromagnetic signatures \cite{McL86,Sh80,MR82,KMR86,MS86,KM81,KLS91,G90}.
However, the
possible signatures of the QGP may also come in part from the late
time dynamics of a hadron gas formed after the phase transition
\cite{MSSG89,GCG98,GL01,GGJ88,VPKH88,GGBCM99,GGX03}. 
Therefore one needs detailed informations
about the creation of the QGP, its lifetime and the hadronization
in order to draw reliable conclusions.

Recent measurements \cite{RHIC} at RHIC of the elliptic flow
parameter $v_2$ for semi-central collisions suggest that - in
comparison to fits based on simple ideal hydrodynamical models
\cite{KHHH01} - the evolving system builds up a
sufficiently early pressure and potentially also achieves (local)
equilibrium. On the other hand, the system in the reaction is at
least initially far from any (quasi-)equilibrium configuration. 
To address the crucial question of thermalization of gluons and
quarks, a number of theoretical analyses have been worked out
either using the relaxation time approximation 
\cite{B84,HW96,W96,W96s}
or performing full $3+1$ dimensional Monte Carlo cascade
simulations based on the solution of the Boltzmann equations for
quarks and gluons \cite{GM92,Zh98,MG00,BMGMN00,BMS03}. The first parton
cascade, VNI, inspired by pQCD including binary elastic
scatterings ($2\leftrightarrow 2$) and gluon radiation and
fusion ($1\leftrightarrow 2$) was developed by Geiger and
M\"uller \cite{GM92}. In the simulation for a central Au+Au
collision at RHIC energy \cite{G92} they concluded that a
thermalized QGP will be formed at $\tau \approx 1.8$ fm/c. 
However, the onset of potential hydrodynamical behavior during
the parton evolution was not demonstrated in their analyses. In
addition, the treatment of the propagation of off-shell partons
in their approach is not clear from a physical point of view.
Recently, Molnar and Gyulassy studied the buildup of the elliptic
flow at RHIC \cite{MG02} applying an on-shell parton cascade, MPC
\cite{MG00} (an improved version of ZPC \cite{Zh98}), in which up
to now only elastic gluon interactions are included. In their
analysis the early pressure can be achieved only if an unrealistic,
much higher cross section is being employed. Furthermore, it is
known that the elastic (and forward directed) $gg\leftrightarrow gg$
collisions cannot drive the system to kinetic equilibrium, as
pointed out in Ref. \cite{SS01}. This would suggest that the collective
flow phenomena observed at RHIC cannot be described via pQCD.
On the other hand, the possible importance of the inelastic
interactions on overall thermalization was raised in the so-called
``bottom up thermalization'' picture \cite{BMSS01}. It is
intuitively clear that gluon multiplication should not only lead
to chemical equilibration \cite{BDMTW93}, but also should lead to
a faster kinetic equilibration \cite{XS94,BV01}. This represents
one (but not all) important motivation for developing a consistent
algorithm to handle inelastic processes like $gg\leftrightarrow ggg$.

In solving the transport equations, in most of the cascade models 
cross sections are interpreted geometrically to model the collision
processes. It turns out that in dense matter when the interaction
length $\sqrt{\sigma/\pi}$ is not much smaller than the mean free
path of particles, causality violation \cite{K84,KBHMP95} will
arise in these cascade models and will lead to numerical artifacts 
\cite{ZHGP98,CH02}. One way to reduce these artifacts is to apply the
common test particle method (or ``particle subdivisions'')
\cite{W82,WMGPS89}, in which the interaction length of the test
particles is reduced by $\sqrt{N_{test}}$, while the mean free path
is unchanged. $N_{test}$ denotes the number of the test particles
per real particle. However, the limitation of these transport models
is obvious: Inelastic collision processes with more than two
incoming particles cannot be straightforwardly implemented since
it is in general difficult to determine, for instance, a $3\to 2$
process geometrically. Therefore, until now, the role of the
inelastic processes in the formation of the QGP has not been studied
fully quantitatively.

An alternative collision algorithm suggested in \cite{DB91,L93,C02}
dealt with the transition rate instead of the geometrical
interpretation of cross section and determined proceeding 
collision processes in a stochastic manner by sampling possible
transitions in a certain volume and time interval. This collision
algorithm opens up the possibility to include the inelastic
collision processes into transport simulations solving the
Boltzmann equations
\begin{equation}
\label{boltzmann}
\left ( \frac{\partial}{\partial t} + \frac{{\mathbf p}_1}{E_1}
\frac{\partial}{\partial {\mathbf r}} \right )\, 
f_1({\mathbf r}, {\mathbf p}_1, t) = {\cal C}_{22} + {\cal C}_{23} + \cdots\,,
\end{equation}
where ${\cal C}_{22}$ and ${\cal C}_{23}$ denote the collision
term of $2 \leftrightarrow 2$ and $2\leftrightarrow 3$ processes.
In this paper we will present a newly developed on-shell parton
cascade using this sort of stochastic collision algorithm. Also
the oftenly employed scheme based on the geometrical interpretation
of cross section is discussed and compared with the stochastic
algorithm. In particular, we concentrate on the study of the
(unphysical) frame dependence. The new transport scheme will then
be applied to simulate the parton evolution for a central
ultrarelativistic heavy ion collision at highest RHIC energy. The
emphasis is put on the investigation of gluon thermalization and
their collective dynamics. For this investigation the initial
conditions are assumed to be generated by independent minijets
\cite{KLL87,EKL89}. Other initial conditions, like the much
discussed ``color glass condensate'' \cite{MV94}, can also
be implemented, but we leave this for a future work. For the
present study we consider quarks and gluons as classical
Boltzmann particles throughout the paper. The Pauli blocking
and gluon enhancement can, in principle, be implemented 
and will also be discussed elsewhere.

The paper is organized as follows. In Sec. \ref{sec:twobody}
we consider two-body collision processes and contrast the
geometrical with the stochastic collision algorithm. The dynamical
evolution of a system within a fixed box is carried out to study
global kinetic equilibration.  In addition, such calculations are
mandatory to debug the operation of the code and to look for
the limitation of the algorithms. The implementation of the
inelastic collision processes is described in Sec. \ref{sec:multi}.
There, we carry out box calculations to study global kinetic
and chemical equilibration. In Sec. \ref{sec:qgp} we study
thermalization of a parton system in a box with initial conditions 
sampled according to the production of minijets as expected in
a central heavy ion collision at RHIC. The Lorentz or frame
(nondependence) and the convergency of results extracted from
cascade simulations are investigated in Sec. \ref{sec:frame}.
In Sec. \ref{sec:rhic} we then finally present first results
of cascade simulations for a central Au+Au collision at RHIC
energy. (The readers, who are solely interested in the operation
and results of the full $3+1$ dimensional cascade, may directly
pass to Sec. \ref{sec:rhic}.) We summarize in
Sec. \ref{sec:summary} and give an outlook for future works.
In Appendices \ref{app_colltime} and \ref{app_optim} more details
of the geometrical collision algorithm are given. We list the
pQCD partonic scattering cross sections in Appendix \ref{ppcs}
for two-body processes and in Appendix \ref{ggcs} for
$gg\leftrightarrow ggg$ processes. In Appendix \ref{sampl} the
numerical recipes for Monte Carlo samplings are presented. 

\section{Two-body collision processes}
\label{sec:twobody}
We consider a system consisting of classical, ultrarelativistic
particles which are interacting via two-body collisions. The main
emphasis is put on the numerical realization of such collision
sequences in a relativistic transport simulation, which is
theoretically based on the solution of the Boltzmann equations
(\ref{boltzmann}) with the following collision term given by 
\begin{eqnarray}
\label{cterm1}
{\cal C}_{22} &=& \frac{1}{2E_1} \int \frac{d^3p_2}{(2\pi)^3 2E_2} \,
\frac{1}{\nu} \int \frac{d^3p'_1}{(2\pi)^3 2E'_1} 
\frac{d^3p'_2}{(2\pi)^3 2E'_2} f'_1 f'_2 | {\cal M}_{1'2'\to 12} |^2
(2\pi)^4 \delta^{(4)} (p'_1+p'_2-p_1-p_2) \nonumber \\
&& -\frac{1}{2E_1} \int \frac{d^3p_2}{(2\pi)^3 2E_2} \,
\frac{1}{\nu} \int \frac{d^3p'_1}{(2\pi)^3 2E'_1} 
\frac{d^3p'_2}{(2\pi)^3 2E'_2} f_1 f_2 | {\cal M}_{12\to 1'2'} |^2
(2\pi)^4 \delta^{(4)} (p_1+p_2-p'_1-p'_2) \,. \nonumber \\
\end{eqnarray}
$\nu$ will be set to $2$ when considering the double counting
if $1'$ and $2'$ are identical particles. Otherwise $\nu$ is
set to $1$.

Since no mean field is considered throughout the present study,
the evolution of particles is intuitively straightforward:
Particles move along straight line between two collision events.
After a particular collision the momenta of colliding particles
are changed statistically according to the differential cross
section. The determination of the collision sequence is, however,
not unique and depends on the particular numerical implementation.
We present in this section two numerical methods dealing with the
realization of binary collisions. Comparisons between these two
methods will be made in detail when investigating
kinetic equilibration in a fixed box. We also study any potential
(but unphysical) frame dependence of transport simulations within
both schemes and how to minimize possible deficiencies. These
results will be presented later in Sec. \ref{sec:frame}.

\subsection{The geometrical method}
\label{sec:meth1}
In the first method a collision happens when two incoming
particles approach as close to each other that their closest
distance is smaller than  $\sqrt{\sigma_{22}/\pi}$, where
$\sigma_{22}$ denotes the total cross section for the
colliding particles. In other words, the collision probability
is either $1$ or $0$, depending on how close the collision
partners come together. Since the total cross section is
interpreted geometrically, we label this procedure the
``geometrical method''. In this picture of the closest
approach,which is already employed in parton cascade models
like ZPC \cite{Zh98}, MPC \cite{MG00} and PCPC \cite{BMGMN00},
collisions do happen one by one as time proceeds. The next
collision event can be determined by comparing the individual
times marking the occurrence of the various and possible
collisions.

Unlike the total cross section the closest distance is,
however, not invariant under Lorentz transformation. This
leads to the situation that a particle pair collides in one
frame, but might not in another frame, which is unphysical.
One faces here a violation of covariance, which is a historic
problem in microscopic simulation within relativistic transport
models. In the present scheme we define the closest distance
in the center of mass frame of the individual particle pair
\cite{CH02}
and thus make it to be a Lorentz invariant quantity by hand.
In spite of this definition the covariance of the Boltzmann
equation is still not fulfilled, because the time ordering
of collisions might be changed under Lorentz transformation
\cite{K84,KBHMP95}. Still, for a sufficiently dilute
system the geometrical method works rather robust. We will
continue discussing this  problem of covariance violation
later in this section and also in Sec. \ref{sec:frame}.
Besides the problem just mentioned, the {\it ordering time} 
of one particular collision itself which orders the
occurrence of all collisions in a particular frame, called
lab frame, is not well defined. Since we determine the
closest distance of two incoming particles in their center
of mass frame, it is reasonable to define the collision
points for the two particles also in this frame at the 
closest distance and at the same time. Consequently both
particles, if they do collide, change their momenta at the
same time in their center of mass frame, but generally
at different times in the lab frame. (We now denote these
individual two times by ``collision times''.) One can now
define the ordering time at some stage between these two
collision times. There is, however, no unambiguous
prescription. In general, different choices for the ordering
time will lead to different collision sequences. This, as
numerically verified, does not strongly affect the
behaviors of physical (ensemble averaged) quantities shown
below. In our simulation we choose the smaller one of the
two collision times as the ordering time.
In ZPC \cite{Zh98} and MPC \cite{MG00} the ordering time was
taken as the average of the two collision times.

\begin{figure}[h]
\centerline{\epsfysize=10cm \epsfbox{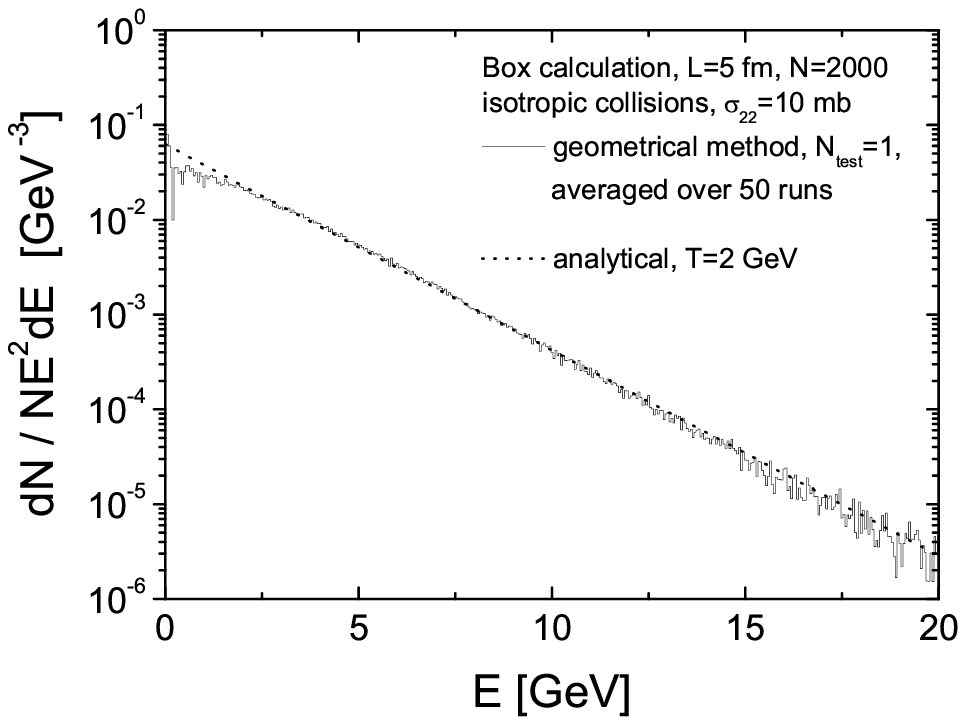}}
\caption{Energy distribution at final time ($t=5$ fm/c) of a system
consisting of $N=2000$ massless particles in a fixed box.
The initial energy distribution is set to be a delta-function at $6$ GeV.
The size of the box is $5$ fm $\times$ $5$ fm $\times$ $5$ fm.
We here apply the geometrical collision algorithm. The collisions
are taken as isotropic and the total cross section is fixed to be a constant
$\sigma_{22}=10$ mb. The dotted line denotes the analytical result of
temperature $T=2$ GeV. The numerical distribution is
obtained from an ensemble of $50$ independent realizations.}
\label{box1}
\end{figure}

In order to demonstrate the correct operation of the
numerical realization of the geometrical method, we will
choose a situation when the outcome is known analytically.
For this purpose we carry out ``box calculations'', in which
a particle ensemble with a nonequilibrium initial condition
is enclosed in a fixed box and will evolve dynamically until
an appropriate final time. The collisions of particles 
against the walls of the box are simply done via mechanical
reflections. For sufficiently long times, the system should
get kinetically equilibrated at the end. For a classical,
ultrarelativistic ideal gas the energy distribution has the
Boltzmann form
\begin{equation}
\label{diste}
\frac{dN}{N E^2 dE} = \frac{1}{2 T^3} e^{-E/T},
\end{equation}
which guides as an analytical reference for the numerical
results. The temperature $T$ can be obtained from the simple
relation between energy and particle density
\begin{equation}
\label{gas}
\epsilon = 3 n T \,,
\end{equation}
where $\epsilon$ and $n$ are solely given by the initial
conditions. Initially, particles are now distributed
homogeneously within the box and their momentum distribution
is chosen highly anisotropic via
\begin{equation}
\label{init}
\frac{dN}{N dp_T dp_z} = \delta(p_T-6 \mbox{ GeV})\, \delta(p_z) .
\end{equation}
In Fig. \ref{box1} the final energy distribution from such box
calculations for a system of $N=2000$ massless particles is
depicted. The size of the box is set to be
$5$ fm $\times$ $5$ fm $\times$ $5$ fm. We consider isotropic
collisions and take a constant total cross section of
$\sigma_{22}=10$ mb. The final time is set to be $5$ fm/c. (As
one will shortly realize, this chosen time is sufficient long
for the system to become equilibrated.) To improve statistics
we have collected particles from $50$ independent realizations.
The dotted line depicted in Fig. \ref{box1} denotes the analytical
distribution (\ref{diste}) with temperature $T=2$ GeV. We see a
nice agreement between the numerical result and the analytical
distribution except a slight, but characteristic deviation at
low energies. We will come back to explain this discrepancy
immediately. 

\begin{figure}[h]
\centerline{\epsfysize=10cm \epsfbox{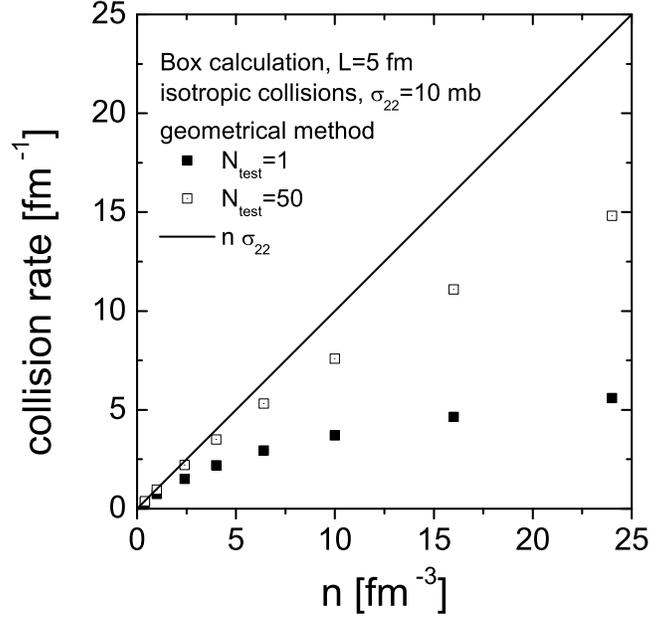}}
\caption{Collision rates for given particle densities. The size of the box
is $5$ fm $\times$ $5$ fm $\times$ $5$ fm. We apply here the geometrical
collision algorithm. The collisions are isotropic and the total cross
section is fixed to a constant $\sigma_{22}=10$ mb. The particle system
is taken initially as thermal with a temperature of $T=1$ GeV. The solid
line shows the expected relationship between collision rate and particle
density: $R = n \sigma_{22}$. The solid squares show the calculated
collision rates without test particles ($N_{test}=1$) and the open squares
show the results with $50$ test particles per real particle ($N_{test}=50$).
}
\label{box2}
\end{figure}

Such a successful passing of the previous test is necessary
for every collision algorithm, but it is still not a sufficient
argument to guarantee whether the presented algorithm is
operating correctly. One has to ask any numerical algorithm for
its limitation of correctly describing the underlying problem.
To be specific when considering the collision integral
(\ref{cterm1}), it is not obvious whether the geometrical
interpretation of the total cross section is a reasonable choice
to account for the Boltzmann process. In fact such a description
has some shortcomings concerning causality violations which have
been pointed out for example in Ref. \cite{KBHMP95}. Especially for
the algorithm presented above we have to face the fact that the
collision times of colliding particles are different in the lab
frame. This will lead to a noticeable reduction of the collision
rate compared to one given by the collision integral: Assume
that the difference of the collision times is $\Delta t_c$.
Consequently the particle with larger collision time should not
collide again during this interval $\Delta t_c$, otherwise
causality would be violated. As pointed out in Appendix
\ref{app_colltime}, for a system in equilibrium the ensemble
averaged time delay $<\Delta t_c>$ depends only on the total
cross section and increases with the increasing total cross
section. This will lead to an artificial increase of the mean free path
and thus to a decrease of the collision rate. In other words,
the collision rate decreases when noncausal collisions are
forbidden. This problem has also been pointed out in Ref. \cite{ZHGP98,CH02}.
We can demonstrate this effect employing box
calculations, in which we consider an initially kinetic
equilibrated gas distributed homogenously within the box.
The size of the box is taken to be the same as in
Fig. \ref{box1}. We employ isotropic collisions with a constant
cross section of $\sigma_{22}=10$ mb. In Fig. \ref{box2}
collision rates are depicted as solid squares for several
particle densities. The collision rate is obtained here as the
time average of the collision number. While the box size is
fixed, we vary the particle number to get different densities.
The solid line shows the expected relationship between the
collision rate and particle density in equilibrium $R = n \sigma_{22}$.
We see a clear decrease of the collision rate when the expected
mean free path $1/n \sigma_{22}$ is not much larger than the
interaction length $\sqrt{\sigma_{22}/\pi}$. Such a numerical
artifact would strongly slow down the kinetic thermalization
of an initially highly nonequilibrium state, as, for instance,
in case of ultrarelativistic heavy ion collisions. As also
clearly seen from Fig. \ref{box2}, the collision rate tends to
saturate at high density. The reason for this is that the
collision rate has an upper limit which is exactly the inverse
of the average collision time difference $<\Delta t_c>/2$
depending only on the total cross section as mentioned before.
One can compute $<\Delta t_c>/2$ analytically. The detailed
calculation is given in Appendix \ref{app_colltime}. It turns
out that $<\Delta t_c>/2=0.12$ fm/c for $\sigma_{22}=10$ mb.
This indicates that the saturation value of the collision rate
would be $8.3$ $\mbox{fm}^{-1}$ at high density.

\begin{figure}[h]
\centerline{\epsfysize=10cm \epsfbox{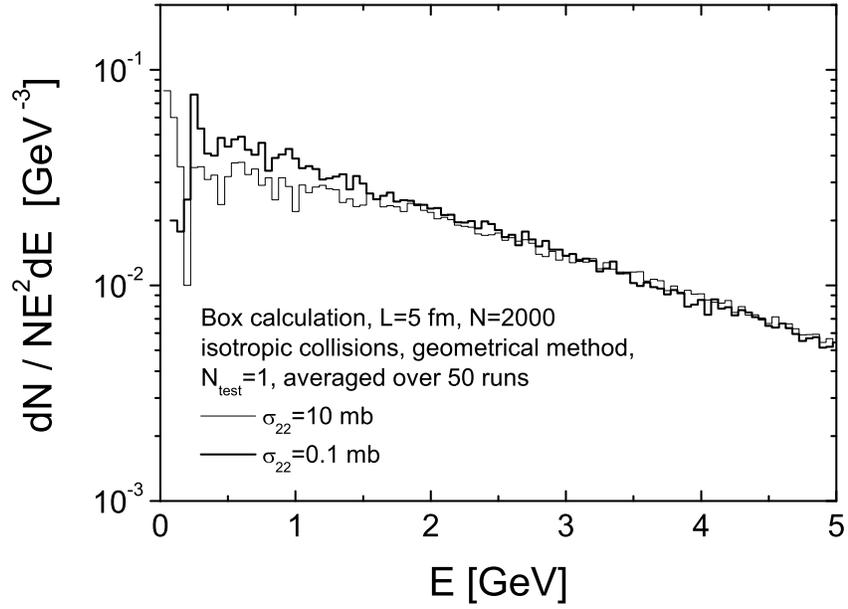}}
\caption{Energy distributions from box calculations. The thin histogram
shows the same distribution as in Fig. \ref{box1}. The thick histogram shows
the result with a smaller total cross section of $\sigma_{22}=0.1$ mb.
}
\label{box3}
\end{figure}

We now return to the slight discrepancy at low energy as
noticed in Fig. \ref{box1} and consider this as a consequence
of the same effect of the relativistic time spread of collisions
pointed out above, since in
this particular situation the particle density is so high that the
mean free path is one order of magnitude smaller than
the interaction length. To
confirm this suspicion, we carry out similar calculations as in
Fig. \ref{box1}, but with a tiny cross section of $\sigma_{22}=0.1$ mb.
The energy distribution, depicted as thick histogram, is shown in
Fig. \ref{box3} compared with the distribution (thin histogram)
obtained by using $\sigma_{22}=10$ mb. One does not see the
artificial distortion in the spectrum at low
energies any more when the cross section and hence the relativistic
time spread is small. As a conclusion, the relativistic time spread
effect not only decreases the collision rate, but also slightly
distorts the system out of equilibrium.

To suppress this  numerical artifact and hence to conserve Lorentz
covariance we employ the widely used test particle, or ``subdivision'',
technique \cite{W82,WMGPS89} based on the scaling
\begin{equation}
\label{testpartcl}
n \to n\, N_{test} \quad \mbox{and} \quad \sigma \to \sigma/N_{test},
\end{equation}
where $N_{test}$ is the number of test particles belonging to one
real particle. While the mean free path is unchanged by the scaling,
the interaction length is reduced by a factor of $\sqrt{N_{test}}$.
This consequently reduces the relativistic time spread which vanishes
in the limit $N_{test} \to \infty$. The open squares in Fig. \ref{box2}
denote the results by using $N_{test}=50$. The tendency of convergency
towards the ideal limit is visible.

\begin{figure}[h]
\centerline{\epsfysize=10cm \epsfbox{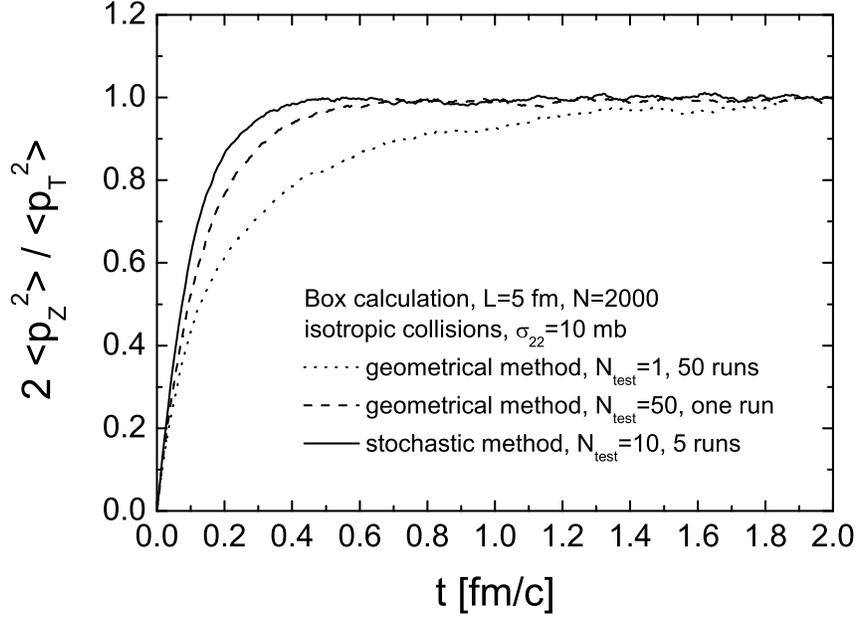}}
\caption{Time evolution of the momentum anisotropy from box calculations. 
The initial condition and parameters are set to be the same as
in Fig. \ref{box1}. The dotted (dashed) curve shows the results obtained by 
employing the geometrical method without test particles (with $50$ test
particles per real particle). The solid curve shows the result obtained
by employing the stochastic method with ten test particles per real
particle.
}
\label{box4}
\end{figure}

In Fig. \ref{box4} we show the time evolution of the momentum
anisotropy defined as the fraction of the average transverse momentum
squared over the average longitudinal momentum squared. The initial
conditions and parameters are set to be the same as in Fig. \ref{box1}.
The dotted line depicts the result without applying the test particle
method ($N_{test}=1$) and the dashed line shows the result with
$N_{test}=50$. The results confirm our reasoning that the relativistic
effect of spreading of the two collision times for a colliding particle
pair increases the relaxation time for achieving kinetic equilibrium.

\subsection{The stochastic method}
\label{sec:meth2}
In the last section we have determined the collision probability of
two incoming particles by means of the geometrical interpretation of
the total cross section. Instead, one can also derive the collision
probability directly from the collision term of the Boltzmann equation
\cite{DB91,L93,C02}. When assuming two particles in a
spatial volume element $\Delta^3 x$ with momenta in the range
(${\bf p}_1, {\bf p}_1+\Delta^3 p_1$) and 
(${\bf p}_2, {\bf p}_2+\Delta^3 p_2$), the collision rate per unit
phase space for such particle pair can be read off from Eq. (\ref{cterm1})
\begin{eqnarray}
\label{collrate22}
\frac{\Delta N_{coll}^{2\to 2}}{\Delta t \frac{1}{(2\pi)^3} \Delta^3 x
\Delta^3 p_1} &=& \frac{1}{2E_1}
\frac{\Delta^3 p_2}{(2\pi)^3 2E_2} f_1 f_2 \nonumber \\
&& \times \frac{1}{\nu} \int \frac{d^3 p^{'}_1}{(2\pi)^3 2E^{'}_1}
\frac{d^3 p^{'}_2}{(2\pi)^3 2E^{'}_2} | {\cal M}_{12\to 1'2'} |^2 (2\pi)^4
\delta^{(4)} (p_1+p_2-p^{'}_1-p^{'}_2).
\end{eqnarray}
Expressing distribution functions as
\begin{equation}
\label{distf}
f_i=\frac{\Delta N_i}{\frac{1}{(2\pi)^3} \Delta^3 x \Delta^3 p_i},
\quad i=1,2,
\end{equation}
and employing the usual definition of cross section \cite{GLW80}
for massless particles
\begin{equation}
\label{cs22}
\sigma_{22}= \frac{1}{2s} \frac{1}{\nu} \int 
\frac{d^3 p^{'}_1}{(2\pi)^3 2E^{'}_1} \frac{d^3 p^{'}_2}{(2\pi)^3 2E^{'}_2}
| {\cal M}_{12\to 1'2'} |^2 (2\pi)^4 \delta^{(4)} (p_1+p_2-p^{'}_1-p^{'}_2)
\,,
\end{equation}
one obtains the absolute collision probability in a unit box $\Delta^3 x$
and unit time $\Delta t$
\begin{equation}
\label{p22}
P_{22} = \frac{\Delta N_{coll}^{2\to 2}}{\Delta N_1 \Delta N_2} = 
v_{rel} \sigma_{22} \frac{\Delta t}{\Delta^3 x}\,.
\end{equation}
$v_{rel}=s/2E_1E_2$ denotes the relative velocity,
where $s$ is the invariant mass of the particle pair. Unlike in
the geometrical method where the collision probability is either
$0$ or $1$, $P_{22}$ now can be any number between $0$ and $1$.
(Notice that, in practice, one should choose suitable $\Delta^3 x$
and $\Delta t$ to make $P_{22}$ to be consistently less than $1$.)
Whether the collision will happen or not is sampled stochastically
as follows: We compare $P_{22}$ with a random number between $0$
and $1$. If the random number is less than $P_{22}$, the collision
will occur. Otherwise there is no collision between the two
particles within the present time step. We call this collision
algorithm the ``stochastic method''. Since in the limit
$\Delta t \to 0$ and $\Delta^{3}x \to 0$ the numerical solutions
using the stochastic method converge to the exact solutions of the
Boltzmann equation \cite{B89}, we divide in practice the space into
sufficient small spatial cells. For a true situation $\Delta t $
and $\Delta^{3}x$ have to be taken smaller than the typical scales
of spatial and temporal inhomogeneities of the particle densities.
Only particles from the same cell can collide with each other. If
a particle pair collides, the collision time will be sampled
uniformly within the interval ($t, t+\Delta t$). The collision
times for both colliding particles are here the same. The particle
system propagates now from one time step to the next. This is
different compared to the transport simulation scheme utilizing
the geometrical method.

In general we also might employ, in addition, the test particle
technique in order to reduce statistical fluctuations of the
collision events in cells. Accordingly the collision probability
is changed to
\begin{equation}
\label{p22t}
P^{'}_{22} = v_{rel} \frac{\sigma_{22}}{N_{test}}
\frac{\Delta t}{\Delta^3 x}
\end{equation}
by the scaling $\sigma \to \sigma/N_{test}$.

In the following we discuss the Lorentz invariance of the
stochastic algorithm in the limit $\Delta^3 x \to 0$,
$\Delta t \to 0$ and $N_{test} \to \infty$.
Since $\Delta t \Delta^3 x$, $\Delta^3 p/\Delta E$, the
distribution function $f$ and the total cross section are Lorentz
scalars, it is easy to realize from Eq. (\ref{collrate22}) that the
collision number $\Delta N_{coll}^{2\to 2}$ is a scalar under
Lorentz transformations. Furthermore this is also true for
$\Delta N_i$, the particle number counted within a phase space
interval at time $t$. Hence, the collision rate 
$\Delta N_{coll}^{2\to 2}/ \Delta N_i \Delta \tau$ as well as
the collision probability $P_{22}$ are scalars under Lorentz
transformations. 
Therefore, in the limit $\Delta^3 x \to 0$, $\Delta t \to 0$
and $N_{test} \to \infty$ the stochastic method yields per se
a Lorentz covariant algorithm. However, in practice, a non-zero
subvolume $\Delta^3 x$ and a non-zero timestep $\Delta t$
disturb full Lorentz invariance explicitely. Any potential, but
unphysical frame dependence will be discussed later in
Sec. \ref{sec:frame}.

\begin{figure}[h]
\centerline{\epsfysize=10cm \epsfbox{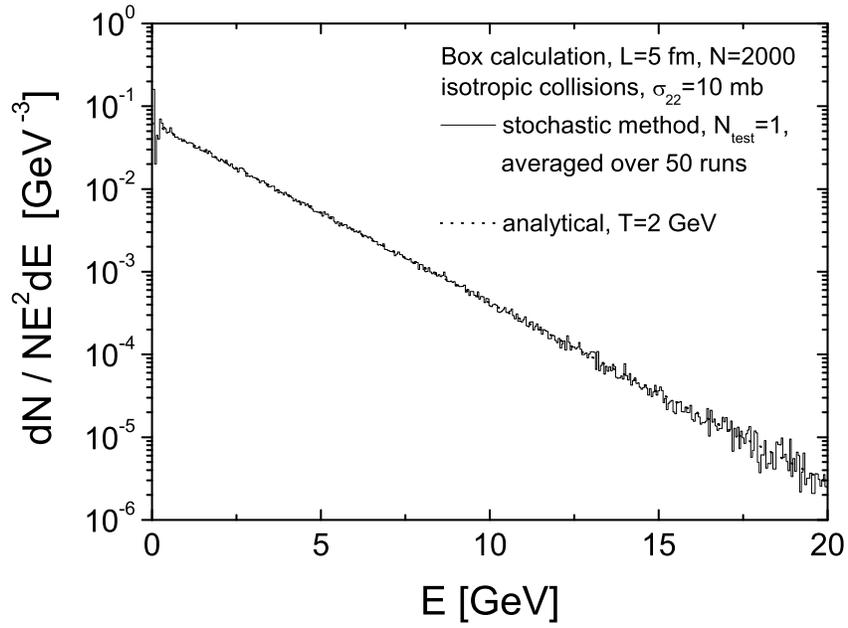}}
\caption{Energy distribution from box calculations. The initial conditions 
and parameters are set to be the same as in Fig. \ref{box1}. We apply here
the stochastic collision algorithm. The box is divided into equal cells. The
length of a cell is $1$ fm.
}
\label{box5}
\end{figure}

To test and demonstrate the stochastic method we again pursue
box calculations. The initial conditions are the same as in
Fig. \ref{box1}. The size of the box is set as before to be
$5$ fm $\times$ $5$ fm $\times$ $5$ fm. Since we consider a
spatially homogeneous initial situation of particles and this
configuration will not change very much during particle
propagation, we choose a straightforward static cell
configuration and divide the box into equal cells. The cell
length is set to be $1$ fm in the calculations. We consider
isotropic collisions and use a constant total cross section
of $\sigma_{22}=10$ mb. Figure \ref{box5} shows the final energy
distribution obtained by an average over $50$ independent
runs (with $N_{test}=1$). One clearly recognizes that the
stochastic collision algorithm also passes this basic test.
The agreement between the numerical and analytical
distribution is perfect and we do not see any distortion in
the spectrum in contrast to the situation experienced in
Fig. \ref{box1}.

\begin{figure}[h]
\centerline{\epsfysize=10cm \epsfbox{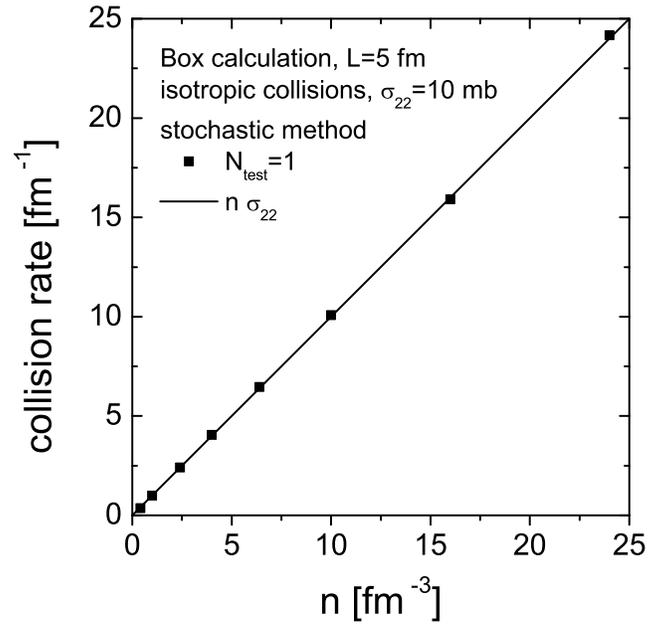}}
\caption{Collision rates for given particle densities. The initial conditions 
and parameters are set to be the same as in Fig. \ref{box2}. We apply here
the stochastic collision algorithm. The cell configuration is the same as in
Fig. \ref{box5}.
}
\label{box6}
\end{figure}

Since the stochastic method is based directly on the formal
collision rate, thus the numerical realized collision rate
should be met in transport simulations if the sampled
statistics in each cell is sufficiently high. We extract the
collision rates from box calculations employing the stochastic
method and show the results in Fig. \ref{box6} as solid squares.
The box size and cell configuration are set to be the same as
in Fig. \ref{box5}. The system is taken at thermal equilibrium
for the initial condition. One nicely recognizes that the
squares lay on the expected line. (We do mention here that
the box size is fixed and we vary the particle number to
simulate different particle densities. For instance, a
density of $1$ $\mbox{fm}^{-3}$ corresponds to a total particle
number of $125$, which means on average one particle per cell.
For still lower densities not investigated, one would have to
work in addition with a suitable amount of test particles.)
 
\begin{figure}[h]
\centerline{\epsfysize=10cm \epsfbox{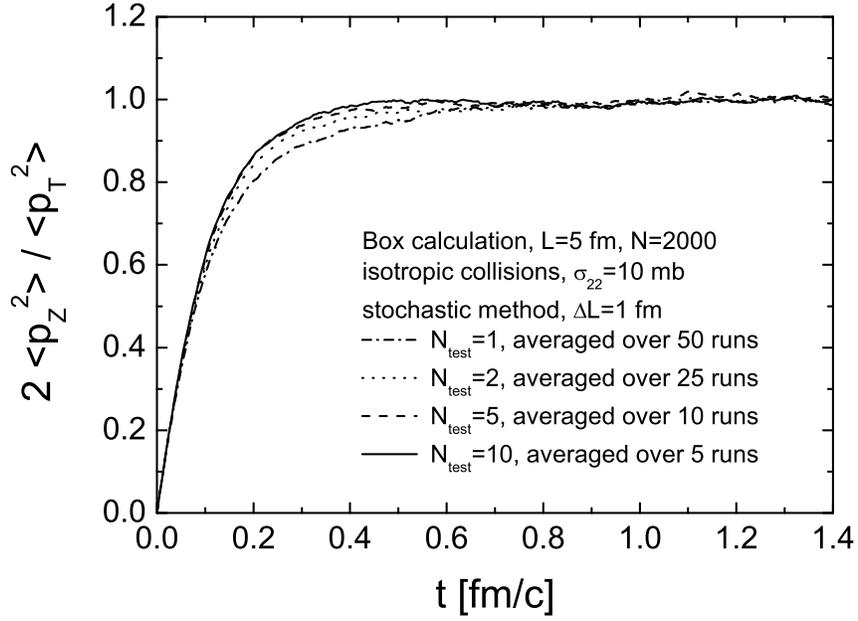}}
\caption{Time evolution of momentum anisotropy from box calculations. 
The initial conditions and parameters are set to be the same as
in Fig. \ref{box1} (or Fig. \ref{box5}). The stochastic method is used here.
The cell configuration is the same as in Fig. \ref{box5}.
The curves show the results with different number of test particles.
}
\label{box7}
\end{figure}

For a system which is initially out of equilibrium the lack of
statistics in cells will affect the dynamical evolution of the
system, since now all cells are correlated during the relaxation
time. To study the effect we repeat the same simulations
performed for Fig. \ref{box5} starting with that particular
nonequilibrium initial condition (\ref{init}) and calculate the time
evolution of momentum anisotropy. We use here the test particle
method to control statistical fluctuations. Figure \ref{box7} shows
the time evolution of the anisotropy for different test particle
numbers $N_{test}$. We see that the lack of statistics in cells
leads to a slight slowdown in the momentum relaxation. This
effect is reduced by using larger values for $N_{test}$, which
in turn results in lower statistical fluctuations.

Let us summarize with some comparisons between the two
simulation methods of treating collisions as presented in this
section. In the simulation employing the stochastic method, 
the collision rate is correctly realized if the statistics
in the individual cell is sufficiently high. In contrast,
the collision rate will be numerically suppressed in the
simulation using the geometrical method, when the mean free
path is not much larger than the interaction length among
test particles.  In simulations with both algorithms the test
particle technique has to be applied in addition in order to solve
the Boltzmann equation with sufficient accuracy.
For dense and strongly interacting system, convergence of the
numerical results with increasing test particle number turns
out to be more efficient in simulations employing the stochastic
method than in simulations employing the geometrical method,
as shown in Fig. \ref{box4}.
In transport simulations applying the stochastic
method we have to face the difficulty of dynamically
configurating the space into small cells, which is not
necessary in the geometrical method. Furthermore, the time step
has to be chosen much smaller than the cell volume to avoid a
strong change of the density distribution in cells. This, of
course, reduces the computing efficiency. In general one should
choose such a collision algorithm, so that numerical expense is
small. However, the stochasic method offers an advanced
technique when dealing with inelastic collision processes, which
is the subject of the next section, whereas it might be rather
impossible to get a unique and consistent geometrical picture
for multiparticle transition processes like $2\leftrightarrow 3$
for instance. A further comparison between the two algorithms
will be discussed in Sec. \ref{sec:frame} concerning any
potential, but unphysical Lorentz frame dependence of the
algorithms.

\section{Particle multiplication and annihilation processes}
\label{sec:multi}
In this section we will now immediately extend the
stochastic method to the more complicated particle
multiplication and annihilation processes involving more than
two particles. These processes are essential to drive the
system towards chemical equilibrium and also do contribute to
kinetic equilibration. The simplest processes are
$2\leftrightarrow 3$. In physical terms such processes will be
specified then later in the paper as gluon Bremsstrahlung and 
its back reaction.
We note that the stochastic method has already
been employed for $2\leftrightarrow 3$ processes in deuteron
production $pnN\leftrightarrow dN$ \cite{DB91} and antibaryon
production via, e.g., $\rho + \rho + \omega \leftrightarrow \bar{B} + B$
\cite{C02} with much simpler and factorized matrix elements.
The true complication in the following is to incorporate the
true Bremsstrahlung matrix element.
Now we will discuss their numerical
implementations. The implementation of higher order processes
is straightforward within the extended stochastic algorithm.

The collision term corresponding the $2\leftrightarrow 3$
processes of identical particles is given by the expression
\begin{eqnarray}
\label{cterm2}
{\cal C}_{23} &=& \frac{1}{2E_1} \frac{1}{2!} 
\int \frac{d^3p_2}{(2\pi)^3 2E_2} \frac{d^3p_3}{(2\pi)^3 2E_3} \, \frac{1}{2!}
\int \frac{d^3p'_1}{(2\pi)^3 2E'_1} \frac{d^3p'_2}{(2\pi)^3 2E'_2} \times
\nonumber \\
&& \times f'_1 \, f'_2 \, | {\cal M}_{1'2'\to 123} |^2 \,
(2\pi)^4 \, \delta^{(4)} (p'_1+p'_2-p_1-p_2-p_3) \nonumber \\
&& +\frac{1}{2E_1} \int \frac{d^3p_2}{(2\pi)^3 2E_2} \, \frac{1}{3!}
\int \frac{d^3p'_1}{(2\pi)^3 2E'_1} \frac{d^3p'_2}{(2\pi)^3 2E'_2} 
\frac{d^3p'_3}{(2\pi)^3 2E'_3} \times \nonumber \\
&& \times f'_1 \, f'_2 \, f'_3 \, | {\cal M}_{1'2'3'\to 12} |^2 \,
(2\pi)^4 \, \delta^{(4)} (p'_1+p'_2+p'_3-p_1-p_2) \nonumber \\
&& -\frac{1}{2E_1} \frac{1}{2!} 
\int \frac{d^3p_2}{(2\pi)^3 2E_2} \frac{d^3p_3}{(2\pi)^3 2E_3} \, 
\frac{1}{2!} \int \frac{d^3p'_1}{(2\pi)^3 2E'_1} 
\frac{d^3p'_2}{(2\pi)^3 2E'_2} \times \nonumber \\
&& \times f_1 \, f_2 \, f_3 \, | {\cal M}_{123\to 1'2'} |^2 \,
(2\pi)^4 \, \delta^{(4)} (p_1+p_2+p_3-p'_1-p'_2) \nonumber \\
&& -\frac{1}{2E_1} \int \frac{d^3p_2}{(2\pi)^3 2E_2} \, \frac{1}{3!}
\int \frac{d^3p'_1}{(2\pi)^3 2E'_1} \frac{d^3p'_2}{(2\pi)^3 2E'_2} 
\frac{d^3p'_3}{(2\pi)^3 2E'_3} \times \nonumber \\
&& \times f_1 \, f_2 \, | {\cal M}_{12\to 1'2'3'} |^2 \,
(2\pi)^4 \, \delta^{(4)} (p_1+p_2-p'_1-p'_2-p'_3) \,.
\end{eqnarray}
The collision probability $P_{23}$ for a particle multiplication
process can be derived analogously to Eq. (\ref{p22}) as
\begin{equation}
\label{p23}
P_{23} = v_{rel} \frac{\sigma_{23}}{N_{test}} \frac{\Delta t}{\Delta^3 x},
\end{equation}
where the total cross section $\sigma_{23}$ is defined as
\begin{equation}
\label{cs23}
\sigma_{23} = \frac{1}{2s} \frac{1}{3!} \int 
\frac{d^3 p^{'}_1}{(2\pi)^3 2 E^{'}_1} \frac{d^3 p^{'}_2}{(2\pi)^3 2 E^{'}_2}
\frac{d^3 p^{'}_3}{(2\pi)^3 2 E^{'}_3} | {\cal M}_{12 \to 1'2'3'} |^2 (2\pi)^4
\delta^{(4)}(p_1+p_2-p^{'}_1-p^{'}_2-p^{'}_3) .
\end{equation}
One can also extend the geometrical method to the multiplication
processes. But it is in general impossible to obtain a unified scheme
for the annihilation processes in a consistent geometrical picture.
In contrast, the extension to $3\to2$ processes via the stochastic
method is straightforward. We write the collision rate stemming from
Eq. (\ref{cterm2}) per unit phase space in a form like Eq. (\ref{collrate22})
\begin{eqnarray}
\label{collrate32}
\frac{\Delta N_{coll}^{3\to 2} / N_{test}}{\Delta t \frac{1}{(2\pi)^3}
\Delta^3 x \Delta^3 p_1} && = \frac{1}{2E_1}
\frac{\Delta^3 p_2}{(2\pi)^3 2E_2} \frac{\Delta^3 p_3}{(2\pi)^3 2E_3}
\frac{f_1}{N_{test}} \frac{f_2}{N_{test}} \frac{f_3}{N_{test}} \nonumber \\
&& \times \frac{1}{2!} \int \frac{d^3 p^{'}_1}{(2\pi)^3 2E^{'}_1}
\frac{d^3 p^{'}_2}{(2\pi)^3 2E^{'}_2} | {\cal M}_{123\to 1'2'} |^2 (2\pi)^4
\delta^{(4)} (p_1+p_2+p_3-p^{'}_1-p^{'}_2) \,, \nonumber \\
\end{eqnarray}
where $f_i, i=1,2,3,$ denote now the phase space density of the
test particles. Inserting Eq. (\ref{distf}) into Eq. (\ref{collrate32})
gives the collision probability of a $3 \to 2$ process
\begin{equation}
\label{p32}
P_{32} = \frac{\Delta N_{coll}^{3\to 2}}{\Delta N_1 \Delta N_2 \Delta N_3}
= \frac{1}{8E_1 E_2 E_3} \frac{I_{32}}{N_{test}^2}
\frac{\Delta t}{(\Delta^3 x)^2}
\end{equation}
for given momenta of the incoming particles in a particular space
cell. $I_{32}$ is defined as the integral
$\frac{1}{2!} \int d^3p'_1 d^3p'_2 \cdots$ in Eq. (\ref{collrate32})
over the final states.

Danielewicz and Bertsch \cite{DB91} obtained a similar expression
for $P_{32}$
\begin{equation}
\label{p32db}
P_{32} = v_{12} \frac{\sigma_{12}}{N_{test}} \frac{{\cal V}_3}{N_{test}}
\frac{\Delta t}{(\Delta^3 x)^2} ,
\end{equation}
when investigating the production of deuterons in a nonrelativistic
transport model of low energy heavy ion reactions, where they
approximately factorized the matrix element into a term describing
a two-body collision and a term mimicking particle fusion.
$\sigma_{12}$ is the total cross section for the two-body collision
and ${\cal V}_3$ can be interpreted as a volume: Once three
particles are within this volume, a $3\to 2$ transition may be
considered to occur. The volume scales with
${\cal V}_3 \to {\cal V}_3/N_{test}$
when employing test particles. Therefore it is intuitively clear
why the quantity $I_{32}$ in Eq. (\ref{p32}) scales with $1/N^2_{test}$.
In contrast to Eq. (\ref{p32db}), expression (\ref{p32}) is a more
general one formulated in a unified manner, and is correct for any
given matrix elements without any approximations.

As an example, when considering isotropic $2\leftrightarrow 3$
collisions for identical particles, integrals over momentum space
for $\sigma_{23}$ and $I_{32}$ can be easily calculated
analytically and one obtains
\begin{equation}
I_{32}=192\pi^2 \sigma_{23} \, .
\end{equation}

\begin{figure}[h]
\centerline{\epsfysize=10cm \epsfbox{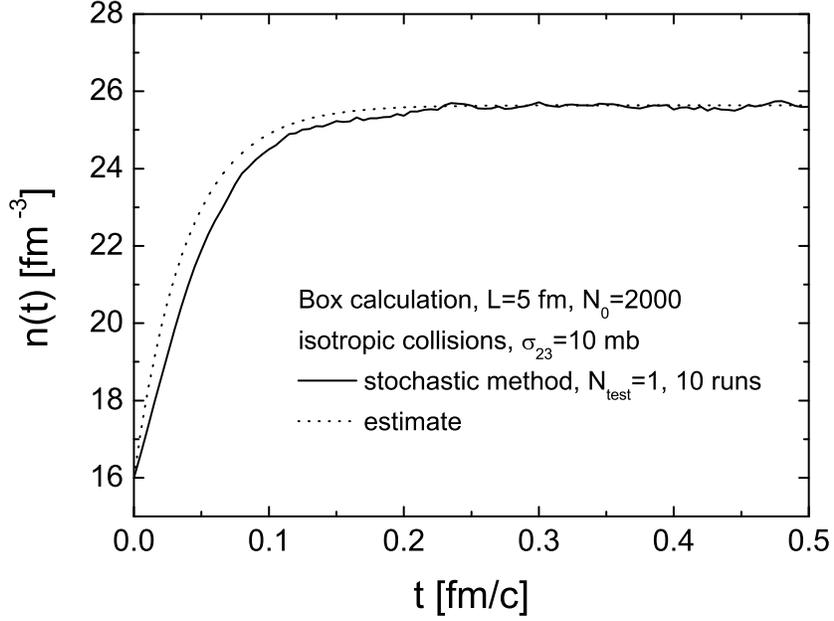}}
\caption{Time evolution of the particle density from box calculations.
The initial conditions and parameters are set to be the same as in
Fig. \ref{box1} (or Fig. \ref{box5}). We consider isotropic inelastic 
collisions ($2\leftrightarrow 3$) with a constant cross section of 
$\sigma_{23}=10$ mb and employ the stochastic collision algorithm. 
The cell configuration 
is the same as in Fig. \ref{box5}. The dotted line denotes the estimate
using a simple time relaxation approximation.
}
\label{box9}
\end{figure}

\begin{figure}[h]
\centerline{\epsfysize=10cm \epsfbox{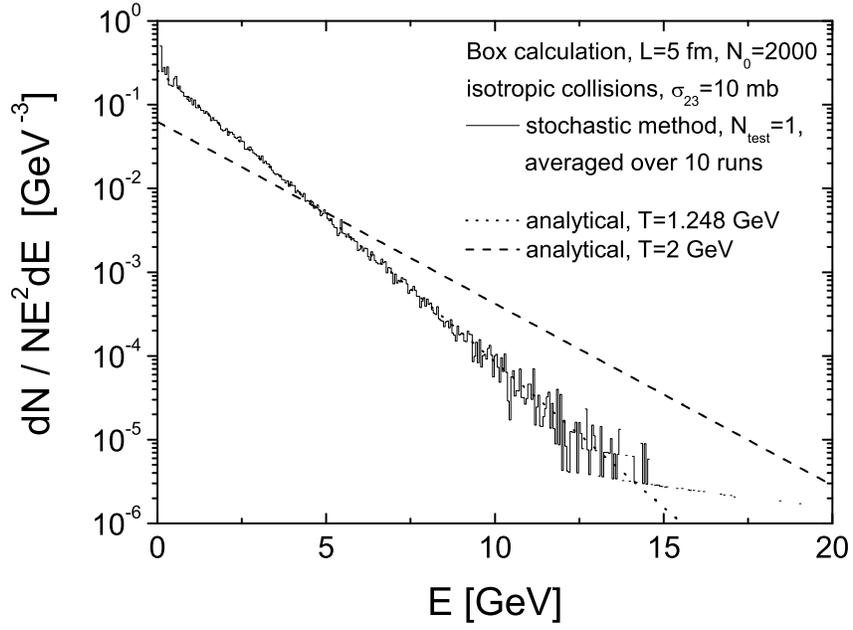}}
\caption{Energy distribution from the same calculations as in Fig. \ref{box9}.
The histogram shows the numerical result. The dotted line shows the analytical
expectation and the dashed line shows the analytical distribution
(the same as in Fig. \ref{box5}) if the particle number would be conserved.
}
\label{box8}
\end{figure}

\begin{figure}[h]
\centerline{\epsfysize=10cm \epsfbox{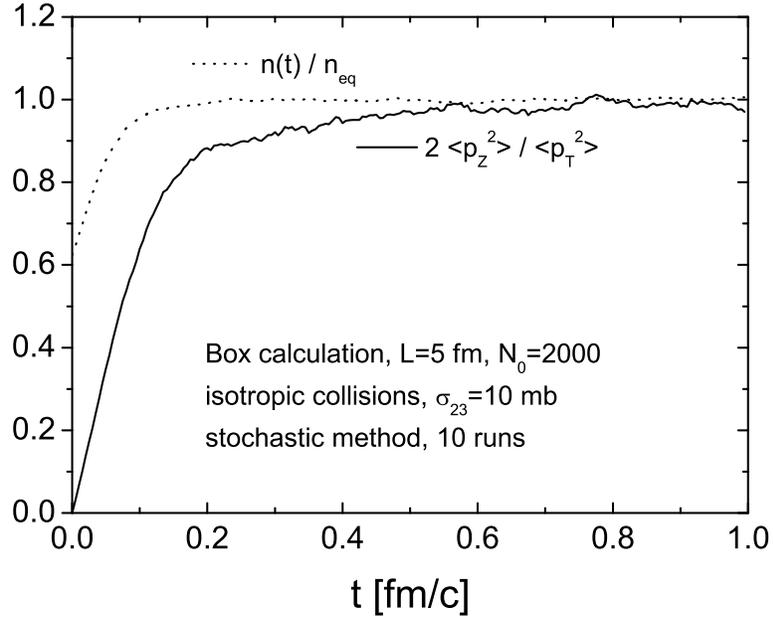}}
\caption{Time evolution of the fugacity $[n(t)]/n_{eq}$ versus 
the momentum anisotropy from the same calculation as in Fig. \ref{box9}.
}
\label{box10}
\end{figure}

Applying the probabilities (\ref{p23}) and (\ref{p32}) we are
now able to study kinetic and chemical equilibration in a box.
We assume a system consisting of identical particles and
consider only isotropic $2\leftrightarrow 3$ collisions.
$\sigma_{23}$ is set to be $10$ mb. As in the box calculations
refering to Fig. \ref{box1}, initially the system is chosen to
be strongly out of equilibrium according to Eq. (\ref{init}).
The particles are distributed homogeneously in the box. The box
has a volume of $5$ fm $\times$ $5$ fm $\times$ $5$ fm and is
divided into equal cells. The cell length is $1$ fm. Initially
the system contains $N_0=2000$ massless particles. Newly
produced particles will be positioned randomly within the
individual cells where the transitions occur. Before we come
to the results, let us determine the final particle density
and temperature to be expected when the system becomes
thermally equilibrated. For an ultrarelativistic (one component)
Maxwell-Boltzmann gas the following relations
\begin{equation}
\label{thermal}
\epsilon = 3 n_{eq} T \quad \mbox{and} \quad n_{eq} = \frac{T^3}{\pi^2}
\end{equation}
hold in equilibrium. One can solve $T$ and $n_{eq}$ for an
energy density given by the initial condition. In our case,
according to Eq. (\ref{init}), we obtain $T=1.248$ GeV and
$n_{eq}= 25.64$ $\mbox{fm}^{-3}$ which is larger than the initial
particle density $n(t_0)=16$ $\mbox{fm}^{-3}$. Figure \ref{box9}
depicts the time evolution of the particle density obtained
from the box calculation. The results are obtained by averaging
ten independent runs. We see that the particle density
increases smoothly towards its final value which agrees fully
with the analytical expectation. The dotted curve presents
an estimate made by using the following relaxation approximation 
\begin{equation}
\label{relax}
n(t) = n_{eq} + (n(t_0)-n_{eq}) \ e^{-\frac{t-t_0}{\theta}} ,
\end{equation}
where $\theta$ stands for the relaxation time. In general,
for any complex equilibration, this quantity will be time 
dependent. For the estimate the relaxation time is taken by
a simple fixed value at equilibrium $\theta=1/n_{eq} \sigma_{23}$
which slightly overestimates the relaxation, as also seen
in Fig. \ref{box9}. In Fig. \ref{box8} the final energy
distribution is depicted by the histogram. The dotted line
denotes the analytical distribution with the expected
temperature $T=1.248$ GeV. The numerical result agrees
again perfectly with the analytical distribution. The fact
that the final particle density and the final temperature
obtained from the inverse slope of the energy spectrum are
identical to the two analytical values demonstrates that
detailed balance between the multiplication and
annihilation processes is fully realized in our simulations.
In Fig. \ref{box10} we compare the time evolutions of the
normalized particle density (the fugacity) and the momentum
anisotropy. It turns out that for the given initial
conditions the kinetic equilibration is slightly slower
compared to the chemical equilibration.
We notice that the quantity
$2<p_z^2>/<p_T^2>=2 \int{d^3p \, p_z^2 \, f} / \int{d^3p
\, p_T^2 \, f}$
is more sensitive to fluctuations than
$n=\int{d^3p/(2\pi)^3 \, f}$, which is the reason
why in Fig. \ref{box10} the curve of the fugacity is smoother than
that of the momentum anisotropy.

\section{Quark Gluon Plasma in box}
\label{sec:qgp}
A quark gluon plasma (QGP) is suggested as a kineticly and
chemically equilibrated system of deconfined quarks and
gluons. Such state of matter is presumed to have been formed
after the big bang and also expected to exist temporarily 
during the course of an ultrarelativistic heavy ion collision
in the laboratory. The main goal of the heavy ion collision
experiments at RHIC and of the future experiments at LHC is
to find evidence of such a new state of matter, the existence
of quark gluon plasma. From the theoretical point of view it
is also very interesting to address the possibility of the
formation of QGP under different theoretical assumptions of
the initial conditions, and to investigate the further
evolution of the quark gluon system in space and time. A
cascade type transport simulation solving relativistic
Boltzmann equations for quarks and gluons with Monte Carlo
technique is just well suited for such a study. Whereas the
current parton cascade models, MPC \cite{MG00},
PCPC \cite{BMGMN00} and VNI/BMS \cite{BMS03}, have not
included the $2\leftrightarrow 3$ processes, we can apply
the extended stochastic collision algorithm presented in the
last section to build up a parton cascade describing the
space-time evolution of interacting quarks and gluons
including $gg\leftrightarrow ggg$ within the framework of
perturbative QCD. As a first application, we restrict
ourselves in this section to investigate the formation of a
quark gluon plasma in a fixed box. The convenience is that a
thermalized parton system should be formed in any case after
some time. Although this situation cannot be given in
reality, one can still address the way of equilibration for
different particle species. Furthermore, box calculations
offer an essential test for the numerical realization of
detailed balance of $gg\leftrightarrow ggg$ and
$gg\leftrightarrow q\bar q$ processes. A realistic space-time
approach for the simulation of parton evolution during the
early stage after an ultrarelativistic heavy ion collision
will be presented in Sec. \ref{sec:rhic}.

The parton interactions include all two-body processes:
($1$) $gg \leftrightarrow gg$, ($2$) $gg \leftrightarrow q {\bar q}$
($3$) $gq \leftrightarrow gq$, ($4$) $qq \leftrightarrow qq$,
($5$) $qq' \leftrightarrow qq'$,
($6$) $q {\bar q} \leftrightarrow q {\bar q}$,
($7$) $q {\bar q} \leftrightarrow q' {\bar q'}$,
and three-body processes ($8$) $gg \leftrightarrow ggg$.
The matrix elements squared in leading order of the perturbative
QCD are taken from Refs. \cite{ORG78,GB82}. We regularize
the infrared divergences by using the Debye screening mass
\cite{W96s} $m_D^2$ for gluons
\begin{equation}
\label{md2}
m_D^2 = 16\pi \alpha_s \int \frac{d^3p}{(2\pi)^3} \frac{1}{p} 
( N_c f_g + n_f f_q)
\end{equation}
and the quark medium mass $m_q^2$ for quarks
\begin{equation}
\label{mq2}
m_q^2 = 4\pi \alpha_s \frac{N_c^2-1}{2N_c} \int \frac{d^3p}{(2\pi)^3}
\frac{1}{p} ( f_g + f_q) ,
\end{equation}
where $N_c=3$ for SU(3) of QCD and $n_f$ is the number
of quark flavor. All formulas for the differential cross
sections are listed in Appendices \ref{ppcs} and \ref{ggcs}. Here
we write down only the differential cross sections (or the matrix
element squared) of the dominant processes for achieving kinetic
and chemical equilibration \cite{W96,BDMTW93}:
\begin{eqnarray}
&& \frac{d\sigma^{gg\to gg}}{dq_{\perp}^2} =
\frac{9\pi\alpha_s^2}{(q_{\perp}^2+m_D^2)^2}\,, \\
&& \frac{d\sigma^{gg\to q\bar q}}{dq_{\perp}^2} =
\frac{\pi\alpha_s^2}{3s(q_{\perp}^2+m_q^2)} \,, \\
\label{m23}
&& | {\cal M}_{gg \to ggg} |^2 = \left ( \frac{9 g^4}{2} 
\frac{s^2}{({\bf q}_{\perp}^2+m_D^2)^2} \right ) 
\left ( \frac{12 g^2 {\bf q}_{\perp}^2}
{{\bf k}_{\perp}^2 [({\bf k}_{\perp}-{\bf q}_{\perp})^2+m_D^2]} \right ) \, ,
\end{eqnarray}
where $g^2=4\pi\alpha_s$. The matrix element (\ref{m23})
describing the $gg \leftrightarrow ggg$ transitions is
factorized into a part for elastic scattering and a part
for gluon radiation (or gluon fusion). ${\bf q}_{\perp}$
and ${\bf k}_{\perp}$ denote, respectively, the perpendicular
component of the momentum transfer and that of the momentum of
the radiated gluon in the c.m. frame. In a dense medium the
radiation of soft gluons is assumed to be suppressed due to
the Landau-Pomeranchuk effect: The emission of a
soft gluon should be completed before it scatters again. This
leads to a lower cutoff of $k_{\perp}$ via a step function
$\Theta (k_{\perp} \Lambda_g - \cosh y )$, where $y$ is the
rapidity of the radiated gluon in the c.m. frame and
$\Lambda_g$ denotes the gluon mean free path which is the
inverse of the gluon collision rate $\Lambda_g=1/R_g$. $R_g$
is the sum of the rate of the following transitions: $gg \to gg$,
$gg \to q{\bar q}$, $gq \to gq$, $gg \to ggg$, and $ggg \to gg$.

\begin{figure}[h]
\centerline{\epsfysize=12cm \epsfbox{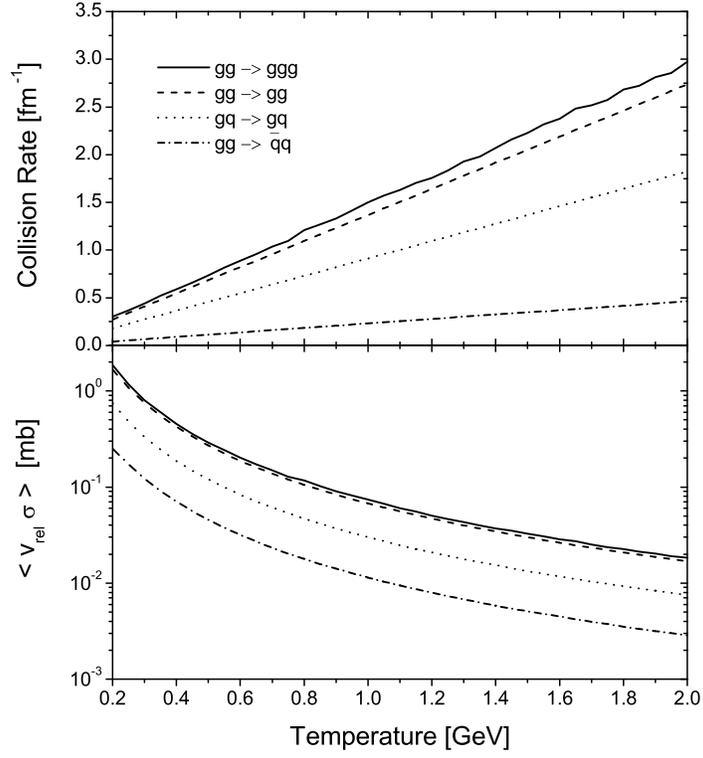}}
\caption{Gluon collision rates and thermally averaged $<v_{rel}\sigma>$
as function of temperature. The solid, dashed, dotted and dash-dotted
line show the temperature dependence for $gg\to ggg$, $gg\to gg$,
$gq\to gq$ and $gg\to q{\bar q}$
transitions respectively. We consider here two quark flavors and employ
a constant coupling $\alpha_s=0.3$ (for the cross sections and the 
screening masses).
}
\label{ratevcs}
\end{figure}

The collision rate is an important quantity governing the time
scale of kinetic and chemical equilibration. In Fig. \ref{ratevcs}
we depict the thermally averged cross section $<v_{rel} \sigma>$
and the gluon collision rates as function of temperature for
$gg \to gg$, $gg\to q{\bar q}$, $gq \to gq$, and $gg \to ggg$ transitions.
$<v_{rel} \sigma>$ are calculated numerically, for which we take
the screening masses obtained at equilibrium ($f_g=f_q=e^{-E/T}$)
\begin{equation}
m_D^2=(3+n_f)\frac{8}{\pi}\alpha_s T^2 \quad \mbox{and} \quad
m_q^2=\frac{16}{3\pi}\alpha_s T^2 \, .
\end{equation}
In the calculations we consider two quark flavors ($n_f=2$)
and employ a constant coupling $\alpha_s=0.3$. The corresponding
collision rates are obtained by $R=n_g <v_{rel}\sigma>$, where
$n_g=\nu_g T^3/\pi^2$ is the gluon density in thermal
equilibrium. $\nu_g=2\times 8$ denotes the degeneracy of gluons.
Because of our simple minded inclusion of the Landau-Pomeranchuk
effect, the cross section $\sigma_{gg\to ggg}$ depends on the sum
of the rates
$R_g=R_{gg\to gg}+R_{gg\to q{\bar q}}+R_{gg\to ggg}+R_{ggg\to gg}$,
in which, however, $R_{gg\to ggg}$ and
$R_{ggg\to gg}(=R_{gg\to ggg} \mbox{ in equilibrium})$ depend
again on $\sigma_{gg\to ggg}$. This problem is solved by a
selfconsistent, iterative computation. Inspecting Fig. \ref{ratevcs}
we see that the collision rates are proportional to the temperature,
which indicates that the $<v_{rel}\sigma>$ are inversely
proportional to $T^2$. This behavior stems from the fact that the
cross section $\sigma_{gg\to gg}$ and $\sigma_{gq\to gq}$ depend
mainly on $1/m^2_D$ and the cross section $\sigma_{gg\to ggg}$ and
$\sigma_{gg\to q{\bar q}}$ mainly on $1/s$. Furthermore we realize
that the collision rate of the three-body processes is in the same
order as the rate of two-body gluon collisions.

We now come to some numerical details when simulating the parton
equilibration in a fixed box. As shown in Appendix \ref{ggcs}, the
computations of $\sigma_{23}$ and $I_{32}$ over momentum space are
reduced to a four- (\ref{csgg3}) and a two-dimensional (\ref{i32})
integration respectively. Even then, the computations are still
time-consuming when $\sigma_{23}$ and $I_{32}$ have to be
calculated for every gluon doublet and triplet in cells, since the
number of integrations is proportional to $n^2$ and $n^3$
respectively ($n$ being the total gluon number in an individual
cell). In order to reduce the computing time, one first thinks
of tabulating $\sigma_{23}$ as well as $I_{32}$. In simulations
we then make interpolations using these tabulated data sets. This
gives a convenient way for obtaining $\sigma_{23}$ because the
underlying integral depends on only two parameters, $m_D^2/s$ and
$\Lambda_g \sqrt s$, as mentioned in Appendix \ref{ggcs}. The
same data sets have been used for calculating $\sigma_{23}$ in
thermal equilibrium as shown in Fig. \ref{ratevcs}. In contrast
to the case for $\sigma_{23}$, $I_{32}$ depends on five parameters
(see Appendix \ref{ggcs}).
A tabulation of $I_{32}$ is thus crude due to the limitation of
the storage, which leads to large errors by interpolations.
Therefore we decide to calculate $I_{32}$ in simulations using
the Monte Carlo algorithm VEGAS \cite{PFTV} with low computing
expense (two iterations and $100$ function calls). Furthermore,
instead of evaluating probabilities of all possible collisions,
we follow the scheme of Refs. \cite{DB91,L93} and choose
randomly ${\cal N}$ out of the possible doublets or triplets,
since in our case the transition probabilities of any channel are
in fact very small within one time step. In order to achieve the
correct collision rate, we have to accordingly amplify the
corresponding collision probabilities to be
\begin{equation}
\label{prob_mod}
P_{22} \to P_{22} \frac{n(n-1)/2}{{\cal N}_{22}} \,, \quad
P_{23} \to P_{23} \frac{n(n-1)/2}{{\cal N}_{23}} \,, \quad
P_{32} \to P_{32} \frac{n(n-1)(n-2)/6}{{\cal N}_{32}} \,.
\end{equation}
The choices of ${\cal N}_{22}$, ${\cal N}_{23}$ and ${\cal N}_{32}$
are arbitrary. In the following simulations we set
${\cal N}_{22}={\cal N}_{23}={\cal N}_{32}=n$. 

The initial condition for the box calculations is taken by
sampling multiple minijet production in heavy ion collisions at
RHIC energy $\sqrt{s}=200$ GeV. Minijets denote on-shell partons
with transverse momentum being greater than $p_0$, where $p_0$
is a parameter separating the hard, perturbative, from the soft,
nonperturbative, nucleon interactions. In calculations we set
$p_0$ to be $2$ GeV. It had been proposed a long time ago in
Ref. \cite{EKL89} that at RHIC energy the produced minijets take half
of the transverse energy. The momentum spectrum of the
minijets has a power-law behavior and thus the initial condition
of the minijets is strongly out of equilibrium. In the following
studies we are interested in the way of how thermalization of
different parton species proceeds and also interested in
the timescales of kinetic and chemical equilibration.

We assume that a nucleus-nucleus collision can be simply modeled
as a sequence of binary nucleon-nucleon collisions. Then the
initial momentum distribution of the produced partons is obtained
according to the differential jet cross section in nucleon-nucleon
collisions \cite{WG91}
\begin{equation}
\label{csjet}
\frac{d\sigma_{jet}}{dp_T^2dy_1dy_2} = K \sum_{a,b}
x_1f_a(x_1,p_T^2)x_2f_b(x_2,p_T^2) \frac{d\sigma_{ab}}{d\hat t} \, ,
\end{equation} 
where $p_T$ is the transverse momentum and $y_1$ and $y_2$ are
the momentum rapidities of the produced partons. $x_1$ and $x_2$
are the Feynman variables denoting the longitudinal momentum
fractions carried by the partons respectively. $d\sigma_{ab}$
stands for the leading order perturbative parton-parton cross
sections. The phenomenological factor $K$, set to be $2$,
accounts for higher-order corrections. We employ the
Gl\"uck-Reya-Vogt parametrization \cite{GRV95} for the parton
structure functions $f_a(x,p_T^2)$. For the box calculations
we consider gluons stremming from a central rapidity region
$y\in [-0.5:0.5]$ as the only initial parton species, since at
the central rapidity region the partons with small $x$ dominate
and these are almost gluons. The initial number of gluons is
assumed to be $500$.

The primary minijets produced in a real high energy heavy ion
collision are distributed within a thin disc due to the
Lorentz contraction. Instead of such a space-time
configuration, we assume a homogeneous spatial distribution
of partons in the box for simplicity. This allows us to still
use a static cell configuration. Moreover,
all particles are assumed to be formed at t=0 fm/c.
We will discuss the
space-time distribution of the primary minijets later in
Sec. \ref{sec:rhic} when considering the parton evolution
in a real heavy ion collision. The size of the box is set to
be $3$ fm $\times$ $3$ fm $\times$ $3$ fm and the box is
divided into equal cells. The length of a cell is set to be
$1$ fm. These settings are tuned as that there will be enough
gluons (about $15$) in each cell during the whole evolution.
(For quarks strong statistic fluctuation occurs at the
beginning of the evolution due to the initial lack of quarks.)

We employ a constant coupling of $\alpha_s=0.3$ in the rest
of this section for evaluating the screening masses and the
cross sections. The screening masses $m_D^2$ and $m_q^2$ are
calculated dynamically according to Eqs. (\ref{md2}) and (\ref{mq2}).
The integrations are computed as
\begin{equation}
\int \frac{d^3p}{(2\pi)^3\, p}\, f \to \frac{1}{V}\sum_i\frac{1}{p_i} \,,
\end{equation}
where the sum runs over all particles in a volume $V$, which
should be, in general, small in order to maintain the local
homogeneity. Since the initial position of partons is distributed
homogeneously, we extend the sum over all particles in the fixed box.

The gluon collision rate, which will be employed for evaluating
$\sigma_{23}$ and $I_{32}$, can be obtained from the calculated
collision probabilities, since the sum of the probabilities of
all possible collisions gives the average total collision number
within the current time step. We then have
\begin{eqnarray}
&& R_{gg\to f} = \frac{\sum_i P_i^{gg\to f}}{\frac{1}{2} N_g \Delta t}\,,
\qquad f=gg, q{\bar q}, ggg,
\\
&& R_{ggg\to gg} = \frac{\sum_i P_i^{ggg\to gg}}{\frac{1}{2} N_g \Delta t}
\\
\mbox{and}
&& R_{gq\to gq} = \frac{\sum_i P_i^{gq\to gq}}{N_g \Delta t} \,,
\end{eqnarray}
where the sums run over possible particle doublets or triplets in
the individual cells and also over all cells. $N_g$ denotes the
total gluon number in the box.
On the other hand, the $P_i^{gg\to ggg}$ and $P_i^{ggg \to gg}$
depend again on $\sigma_{23}$ and $I_{32}$ respectively. Therefore,
a correct calculation for $\sigma_{23}$ and $I_{32}$ as well as
$P_i^{gg\to ggg}$ and $P_i^{ggg \to gg}$ should be a selfconsistent,
iterative computation. However, since such computations are too
time consuming, we employ the gluon collision rate, obtained at
the last time step, to calculate $\sigma_{23}$ and $I_{32}$
within the current time step.

When the parton system becomes fully equilibrated at the
later evolution, the final values of gluon and quark number
should be given by
\begin{eqnarray}
\label{ng}
&& N^{eq}_g = \nu_g \frac{T^3}{\pi^2} V \, ,\\
\label{nq}
&& N^{eq}_q = 2 \nu_q \frac{T^3}{\pi^2} V \,,
\end{eqnarray}
where $\nu_g=2\times 8$ and $\nu_q=2\times 3 \times n_f$ are
the degeneracy factors of a gluon and quark respectively.
The factor $2$ in Eq. (\ref{nq}) indicates the sum of quark
and antiquark. Employing the relation
\begin{equation}
E = 3 (N^{eq}_g + N^{eq}_q) T \,,
\end{equation}
which holds in thermal equilibrium, we obtain the final temperature
\begin{equation}
\label{tmp}
T = \left ( \frac{E}{V} \frac{\pi^2}{3(\nu_g+2\nu_q)} \right )^{\frac{1}{4}}\,.
\end{equation}
The total energy $E$ can be determined by the specified initial
momentum distribution of minijets, Eq. (\ref{csjet}). Considering
only up and down quarks ($n_f=2$) we get a final temperature of
about $430$ MeV and thus $m_D^2 \approx 0.7 \mbox{GeV}^2$ and
$m_q^2 \approx 0.1 \mbox{GeV}^2$ for $\alpha_s=0.3$.

\begin{figure}[h]
\centerline{\epsfysize=10cm \epsfbox{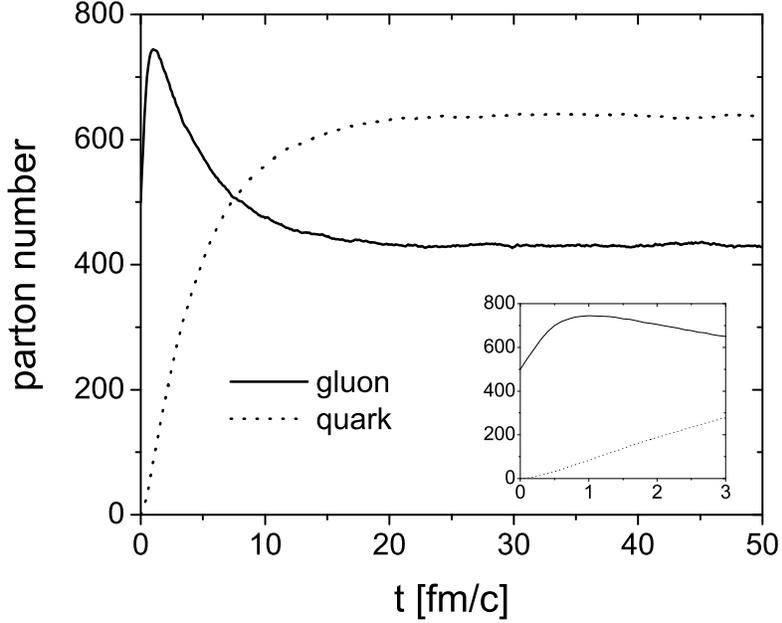}}
\caption{Time evolution of the gluon and quark number in box calculations.
We consider here gluons and quarks with two flavors as parton species.
Collision processes are the elementary two-body parton-parton scatterings
and three-body processes $gg\leftrightarrow ggg$ in leading order of
perturbative QCD. The coupling is assumed to be a constant of $\alpha_s=0.3$.
The initial momentum distribution of particles is taken from the minijets
production in central rapidity interval $y\in (-0.5:0.5)$ in a nucleon-nucleon
collision at RHIC energy $\sqrt s=200$ GeV. The initial particles are
gluons and distributed homogenously in the box. The size of the box is
$3$ fm $\times$ $3$ fm $\times$ $3$ fm and the box is divided into equal
cells. The length of a cell is $1$ fm. The initial gluon number is set
to be $500$. The results are obtained from an average over $60$ runs.
}
\label{box-nm}
\end{figure}

Figure \ref{box-nm} shows time evolutions of the gluon and quark
number. Sixty independent realizations are collected to obtain
sufficient statistics. We see that the time evolution of the
gluon number has two stages. At first the gluon number
increases rapidly to a maximum and then relaxes towards its
equilibrium value on a slower scale. The quark number starts
from zero because of the initial absence of quark species and
increases smoothly towards its equilibrium value. The gluon
and quark number do reach their final values simultaneously.
These behaviors of $N_g(t)$ and $N_q(t)$ reveal the well-known
scenario of two-stage chemical equilibration: The gluon system
equilibrates at first as if no quarks were there and then
cools down gradually by producing quark-antiquark pairs until
the quarks reach the equilibrium. Such two-stage equilibration
could also happen in a real high energy heavy ion collision
\cite{S92}.

Next we compare the equilibrium values of gluon and quark
number of Fig. \ref{box-nm} with the analytical values which
one would expect directly from the initial conditions. The
final temperature in one individual run can be obtained by
inserting the total amount of energy into expression
(\ref{tmp}). Averaged over $60$ runs we have $<T>=427.84$ MeV.
Inserting the averaged temperature into Eqs. (\ref{ng}) and
(\ref{nq}) gives $<N^{eq}_g>=428$ and $<N^{eq}_q>=643$. The
values extracted from Fig. \ref{box-nm} are $N_g=430$ and
$N_q=640$. We see that the agreements are pretty good, which
demonstrates that our new cascade algorithm is indeed very
successful in keeping the detailed balance even for the
considered complexity of employing pQCD motivated cross
sections. We also calculate the equilibrium number of gluons
when no quarks are considered ($n_f=0$). In the present
situation this is $\bar N^{eq}_g=852$, which is somewhat
greater than the maximum of gluon number read off from
Fig. \ref{box-nm}, since in the latter case gluons are
already lost due to the production of quark-antiquark pairs
starting at the beginning of the evolution.

\begin{figure}[h]
\centerline{\epsfysize=10cm \epsfbox{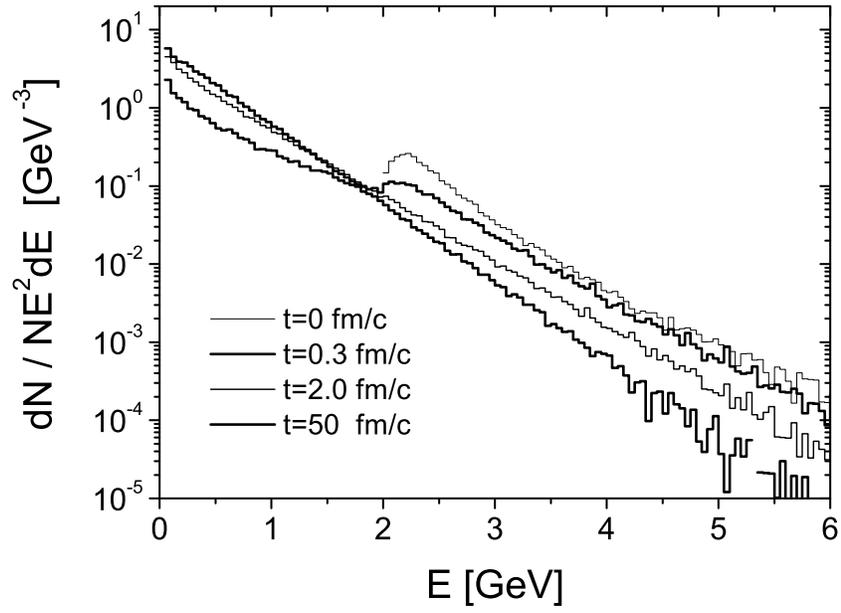}}
\caption{Energy distributions at different times from 
the same calculation as in Fig. \ref{box-nm}.
}
\label{box-distE}
\end{figure}

In Fig. \ref{box-distE} we depict the energy distributions of
the partons (gluons and quarks) at different times. The initial
($t=0$ fm/c) distribution possesses a cutoff at $E=p_0=2$ GeV and
is highly nonthermal. Immediately after the onset of interactions,
soft gluons with smaller energy do emerge by the process $gg\to ggg$
and thermalize very quickly.
We see that at $0.3$ fm/c the energy distribution for partons
with smaller energy than $2$ GeV is largely populated.
The hard particles with larger
energy are still out of equilibrium. There is still a hump at
$2$ GeV. This hump will vanish gradually and at $2$ fm/c the
total distribution becomes exponential. One can refer to this
stage as the onset of kinetic equilibration. The energy
distribution at a final time of $t=50$ fm/c is also depicted in
Fig. \ref{box-distE}. We have compared this spectrum to the analytical
form Eq. (\ref{diste}) with the averaged temperature $<T>=427.84$ MeV
obtained from the initial input. (The analytical distribution is
not shown in Fig. \ref{box-distE}.) The agreement is very good.

\begin{figure}[h]
\centerline{\epsfysize=10cm \epsfbox{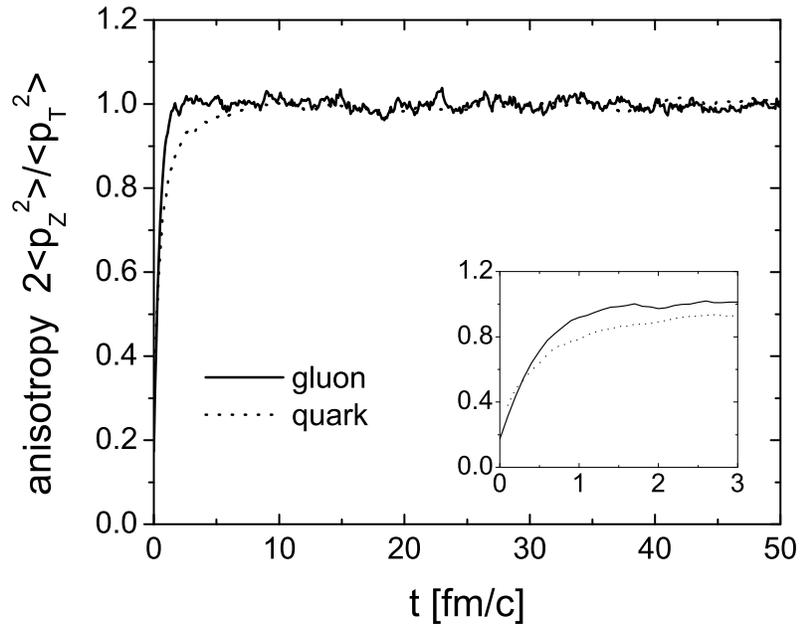}}
\caption{Time evolution of the momentum anisotropy for gluons and quarks from
the same calculation as in Fig. \ref{box-nm}.
}
\label{box-aniso}
\end{figure}

\begin{figure}[h]
\centerline{\epsfysize=10cm \epsfbox{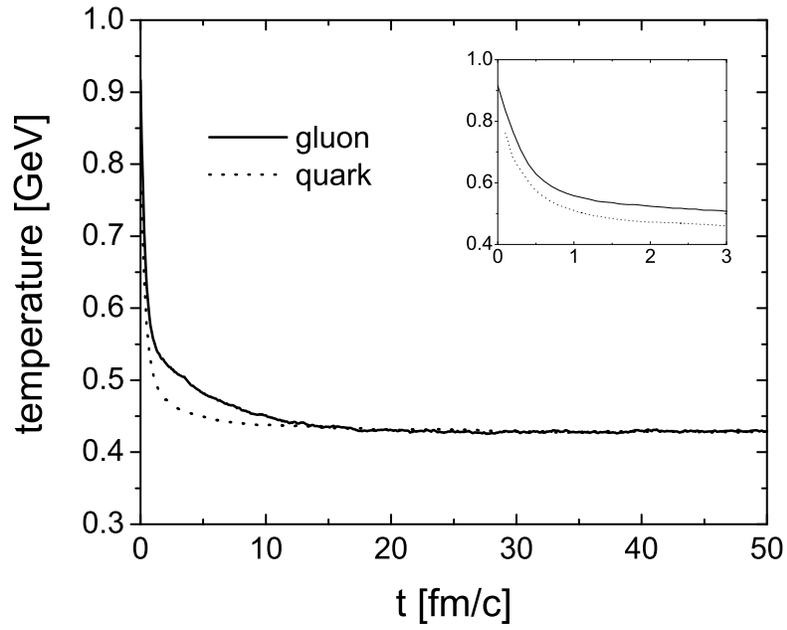}}
\caption{Time evolution of the temperature for gluons and quarks from
the same calculation as in Fig. \ref{box-nm}.
}
\label{box-temp}
\end{figure}

To study the kinetic equilibration in more detail, we calculate
the time evolutions of the momentum anisotropy 
\begin{equation}
\frac{2<p_z^2>_g}{<p_T^2>_g}(t) \,, \qquad
\frac{2<p_z^2>_q}{<p_T^2>_q}(t) 
\end{equation}
for gluons and quarks, which are shown in Fig. \ref{box-aniso}.
We see that the momentum of the gluons and quarks becomes
isotropic at almost same time of about $1-2$ fm/c which is
just the timescale when the energy spectrum gets exponential,
as shown in Fig. \ref{box-distE}. However, if one looks at the
time evolutions of the effective temperatures in Fig. \ref{box-temp},
which are defined as $T_g(t):=E_g(t)/3\,N_g(t)$ and
$T_q(t):=E_q(t)/3\,N_q(t)$, one notices that between $0$ fm/c
and $10$ fm/c the temperature of quarks is lower than the one
of gluons. The reason is that the quarks stem mainly
by the $gg\to q\bar q$ quark pair production and the cross
section $\sigma_{gg\to q{\bar q}}$ is inversely proportional
to $s$. Therefore, when the quark production is still more dominant
compared to the annihilation process, more quark-antiquark
pairs with smaller energies are produced than those with
larger energies, compared to the equilibrated Boltzmann distribution.
Correspondingly, there would be a slight
suppression in the energy spectrum of quarks at high energy
and in the energy spectrum of gluons at low energy during
the ongoing chemical equilibration. It takes time for
the gluon-quark mixture to obtain an identical temperature
via the gluon-quark interactions. This identical, final temperature
is extracted from Fig. \ref{box-temp}, $T_g=T_q=429$ MeV, and agrees
perfectly with the expectation of $<T>=427.84$ MeV.

\begin{figure}[h]
\centerline{\epsfysize=10cm \epsfbox{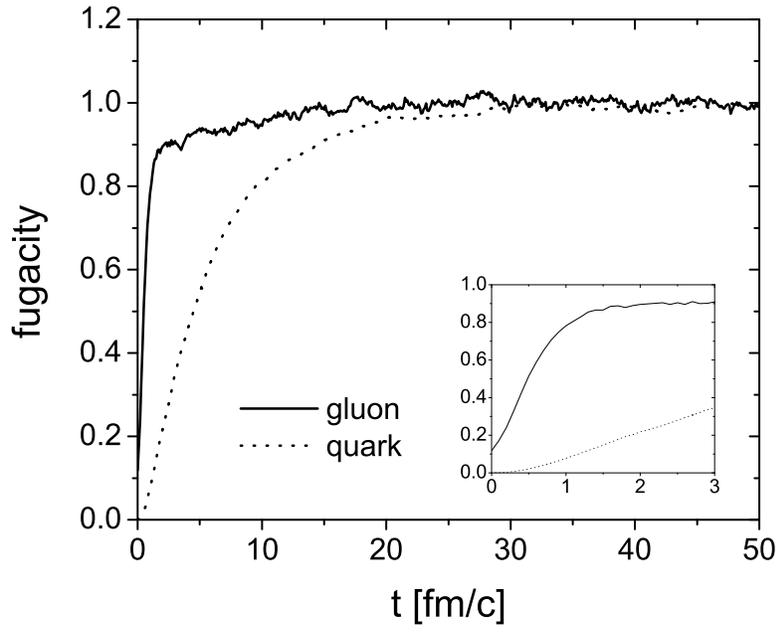}}
\caption{Time evolution of the fugacity for gluons and quarks from
the same calculation as in Fig. \ref{box-nm}.
}
\label{box-fuga}
\end{figure}

The parton fugacity is defined as follows
\begin{equation}
\label{fuga}
\lambda_g (t):= \frac{N_g(t)}{\tilde N^{eq}_g(t)} \quad \mbox{and} \quad
\lambda_q (t):= \frac{N_q(t)}{\tilde N^{eq}_q(t)} \, ,
\end{equation}
where
\begin{equation}
\tilde N^{eq}_g(t) := \nu_g \frac{T_g^3(t)}{\pi^2} V \quad \mbox{and}\quad
\tilde N^{eq}_q(t) := 2 \nu_q \frac{T_q^3(t)}{\pi^2} V \,. 
\end{equation}
In Fig. \ref{box-fuga} the time evolutions of the fugacity are
depicted for gluons (solid curve) and quarks (dotted curve).
We see that while the gluons approach the chemical equilibrium
at about $3$ fm/c, the quarks do equilibrate later at $20$ fm/c.
The two-stage chemical equilibration is clearly demonstrated
in Fig. \ref{box-fuga}.

\begin{figure}[h]
\centerline{\epsfysize=10cm \epsfbox{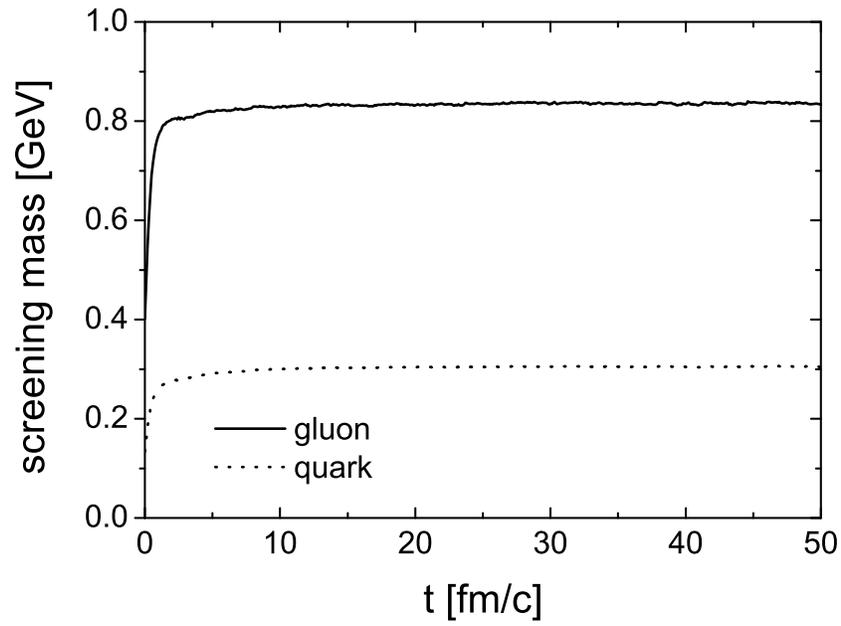}}
\caption{Time evolution of the screening mass for gluons and quarks from 
the same calculation as in Fig. \ref{box-nm}.
}
\label{box-smass}
\end{figure}

\begin{figure}[h]
\centerline{\epsfysize=10cm \epsfbox{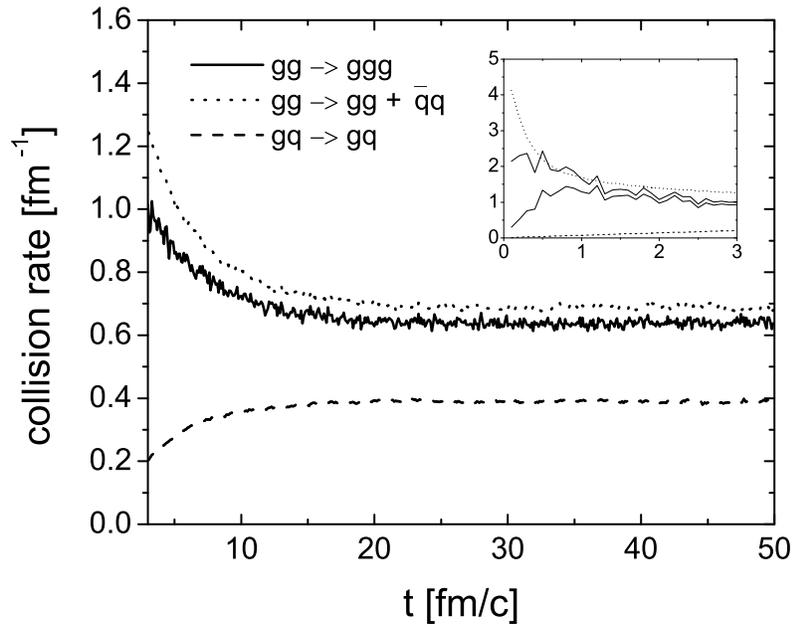}}
\caption{Time evolution of the collision rates for gluons in the different
channels. The results are obtained from the same calculation as in
Fig. \ref{box-nm}.
}
\label{box-collrate}
\end{figure}

We also depict the time evolutions of the screening masses in
Fig. \ref{box-smass} and of the gluon collision rates in
Fig. \ref{box-collrate}. The comparisons of the extracted
equilibrium values from the figures with the analytical
values give perfect agreements. In the small window of
Fig. \ref{box-collrate} the collision rate of
$gg \to ggg$(upper) and $ggg\to gg$(lower) are shown by
solid lines. We see that the two processes occur with the
same rates at about $2\sim 3$ fm/c, which is just the time
scale when the gluons become chemically equilibrated. The
identical time scale is also obtained from Fig. \ref{box-fuga}.
We did not depict the time evolution of the rate of $ggg\to gg$
process from $3$ fm/c to $50$ fm/c, since it is almost
identical with that of $gg\to ggg$ process.

From the present study of creating QGP in a box some
speculations are made when we consider parton evolution in
a real ultrarelativistic heavy ion collision.
(1) Two-stage equilibration is a good scenario describing
parton thermalization in high energy heavy ion collisions.
(2) The cross section $\sigma_{gg\to ggg}$ is in the same
order as $\sigma_{gg \to gg}$ and thus the
$gg\leftrightarrow ggg$ processes should play an important
role in chemical and as well as kinetic equilibration.
Analyses based on a full $3+1$ dimensional transport simulation
of the parton evolution after a high energy heavy ion collision
will be presented in Sec. \ref{sec:rhic}.

\section{Testing the frame independence}
\label{sec:frame}
The relativistic kinetic equation
\begin{equation}
\label{boltzmann2}
p^{\mu}\partial_{\mu}f=I_{coll}
\end{equation}
is a Lorentz covariant expression. Therefore the covariance of
its solution should not be affected by the choice of the frame,
in which the many-body dynamics is actually described. Frame
independence must also be fulfilled for any physical observables
which can be expressed as Lorentz scalars. However, the equation
(\ref{boltzmann2}) cannot be solved exactly in practice by 
applying a transport algorithm. Strictly speaking, the frame
independence is not fulfilled in any cascade-typ simulations. Our
aim in this section is to study potential frame dependence in our
description employing collision algorithms presented in Secs.
\ref{sec:twobody} and \ref{sec:multi}. We will also demonstrate
the increasing insensitivity of the particularly chosen frame and
the convergency of the numerical results when adding more and
more test particles into the dynamics.

As explained in Sec. \ref{sec:twobody}, the geometrical method
is based on the geometrical interpretation of the total cross
section and the time ordering of the collision events is
generally frame dependent when the mean free path of particles is
in the same order as the mean interaction length. In contrast,
in simulations employing the stochastic method, which deals with
the transition rate, a time ordering of the collision
sequence is not needed because collision events
will be sampled stochastically within a time step.
Still, one has to be aware that a nonzero subvolume of cells and
a nonzero timestep disturb the Lorentz invariance.
Zhang and Pang had studied already the
frame dependence of parton cascade results in Ref. \cite{ZHP97}
applying a parton cascade code with a similar geometrical
collision scheme as presented by us. They argued that results
from parton cascade simulations are not sensitive to the choice
of the frame when the collision criterion is formulated in the
center of mass frame of two incoming partons. We will demonstrate
the issues in detail in the following considerations and
calculations.

\subsection{One dimensional expansion in a tube}
\label{expansion}
For the purpose of studying the frame dependence we do not need
consider a special situation. However, as emphasized in the
Introduction, the here presented cascade model will be applied
to simulate the parton evolution in ultrarelativistic heavy ion
collisions. Therefore it makes sense to consider a
one dimensional expanding system as testing ground, since at the
initial stage of an ultrarelativistic heavy ion collision the
partonic system will undergo mainly a longitudinal expansion.
For convenience, particles of the test system are classical
Boltzmann particles instead of quarks and gluons. Furthermore,
in the present section we will employ isotropic collisions and
a constant cross section. In order to mimic a perfect
longitudinal expansion we embed all particles into a cylindrical
tube with infinite length. The reflections of particles against
the tube wall are operated in a same way as performed in the box
calculations.

Initially, particles are considered to be thermal in their local
spatial element. We use a Bjorken-type boost invariant initial
conditions \cite{B83}
\begin{equation}
\label{bjorken}
f({\mathbf x}, {\mathbf p}, \tau) =
e^{-\frac{p_{\perp} \cosh(y-\eta)}{T(\tau)}} \, ,
\end{equation}
where $\tau$ is the proper time $\tau = \sqrt{t^2-z^2}$ and $y$
and $\eta$ denote, respectively, momentum and space-time rapidity
\begin{equation}
\label{rapidity}
y=\frac{1}{2} \ln \frac{E+p_z}{E-p_z} \quad , \quad 
\eta =\frac{1}{2} \ln \frac{t+z}{t-z} \, .
\end{equation}
Due to the assumption of the boost invariance, quantities such
as particle density $n$, energy density $\epsilon$ and
temperature $T$ depend only on the proper time $\tau$. For an
ideal, longitudinal and boost-invariant hydrodynamical expansion
we obtain
\begin{eqnarray}
\label{ntau}
n(\tau) &=& n(\tau_0)\, \frac{\tau_0}{\tau} \,,\\
\label{etau}
\epsilon(\tau) &=& \epsilon(\tau_0)\, \left ( \frac{\tau_0}{\tau} 
\right )^{4/3} \,,\\
\label{ttau}
T(\tau) &=& T(\tau_0)\,  \left ( \frac{\tau_0}{\tau} \right )^{1/3} \,.
\end{eqnarray}
Besides the study of the frame dependence we also attempt to
address the possibility of buildup of an approximately ideal
hydrodynamical expansion in cascade simulations when the
collision rate is considered to be very high. The time
dependences (\ref{ntau}), (\ref{etau}), and (\ref{ttau}) then
serve as ideal references when comparing them with results
extracted from the numerical simulations.

To be able to apply the stochastic method, the tube needs to
be subdivided into sufficient small cells. A static cell
structure as configurated in the box calculations is not
suitable any more for an expanding system. However, since the
expansion is only one dimensional, we can still employ a static
configuration in the transverse plane. Instead of a lattice
structure, (which will also work,) we make use of the symmetry
in the given situation and consider a spider web like structure
in the transverse plane. Particularly we divide the polar angle
$\phi$ and the radial length squared $r^2$ equally within the
interval $[0,2\pi]$ and $[0,R^2]$, respectively, where $R$
denotes the radius of the cylindrical tube. This division gives
a same transverse area $\Delta F=\Delta\phi \Delta r^2/2$ for all
cells. Longitudinally we have to construct a comoving cell
configuration which adapts to the expanding system, since, as a
reminder, the spatial inhomogeneity of particles in the local
cells should be small within one time step. Using the thermal
distribution function (\ref{bjorken}) it can be simply realized
by means of the Cooper-Frye formula \cite{CF74} that the particle
number per unit space-time rapidity $dN/d\eta$ calculated at time
$t$ in a frame (and also at $\tau$ as well) is constant, i.e.,
time independent, when the system expands hydrodynamically. This
gives us the guideline to divide the tube longitudinally into
equal small $\eta$ bins. We mark the individual cells $[\eta_i,\eta_{i+1}]$
with the central value $\eta=(\eta_i+\eta_{i+1})/2$ and the size
$\Delta \eta_c=\eta_{i+1}-\eta_i$. Then the longitudinal length
of a particular cell reads
\begin{equation}
\label{etabin}
\Delta z(t) = t \, \left [
\tanh(\eta+\Delta \eta_c /2)-\tanh(\eta-\Delta \eta_c /2) \right ]
\end{equation}
and increases linearly in time. At time $t$, when going outwards
from the expansion center towards the front edges, the cells
becomes more and more narrow. Since the particle diffusion within
a time step should not destroy the homogeneity in the local cells
very much, the time step has to be chosen smaller than the
shortest longitudinal size among all cells. In simulations we set
the time step to be half of the shortest $\Delta z$ of the cell
located at the front edge
\begin{equation}
\label{dt}
\Delta t \,(t) = 0.5\, \Delta z_{min}(t) = 0.5\, t \, \left [ 
\tanh(\eta_m+\Delta \eta_c /2)-\tanh(\eta_m-\Delta \eta_c /2) \right ] \,,
\end{equation}
where $\eta_m$ denotes the outermost $\eta$ bin.

With Eq. (\ref{dt}) we obtain the collision probability for a two-body
process in the central cell ($\eta=0$)
\begin{equation}
P_{22}=v_{rel} \sigma_{22} \frac{\Delta t}{\Delta^3 x}
=v_{rel} \sigma_{22} \frac{0.5\,\left [ 
\tanh(\eta_m+\Delta \eta_c /2)-\tanh(\eta_m-\Delta \eta_c /2) \right ]}
{\Delta x_{\perp} \,2\,\tanh(\Delta \eta_c /2)} \,.
\end{equation}
For the parameters $\sigma_{22}=10$ mb, $\Delta x_{\perp}=2.5 \mbox{ fm}^2$,
$\eta_m=3.0$ and $\Delta \eta_c=0.2$, the collision probability
$P_{22}$ in the central region is expected to be a small value,
$P_{22} < 0.004$. In order to make an estimate of the collision
probability in the noncentral cells we go to their local comoving
frames for convenience, since the collision probability is
invariant under Lorentz transformations. The time in the local
frame of a $\eta$ bin is $\tau=t/\gamma$, where $\gamma=\cosh \eta$
denotes the Lorentz factor. Suppose that the system undergoes
one-dimensional hydrodynamical expansion, the collision rate
$R=n<v_{rel}\sigma_{22}>$ in the local frame of a moving noncentral
cell is higher than that in the central cell by factor $\gamma$,
since the particle density is just $\gamma$-times higher according
to Eq. (\ref{ntau}). [Note that the estimate becomes complicated when
the total cross section depends on $s$ instead of a constant, since
the distribution of $s$ is a function of the temperature and the
temperatures in the central and noncentral cell are different at
time $t$ according to Eq. (\ref{ttau}).] On the other hand the
transformed time step $\Delta \tau$ is $\gamma$-times smaller than 
$\Delta t$. Therefore the averaged collision number, which is a
Lorentz scalar, is the same in all cells within a time step
$\Delta t$. Furthermore, for the given cell configuration there are
on average the same number of particles in each cell. This leads
to the conclusion that for an approximate one dimensional
hydrodynamical expansion and choosing a constant cross section, the
mean collision probability of two incoming particles (for an
ensemble average) is the same wherever the collision will occur.
Due to the fact that the collision probability is small we employ
the method as explained in Sec. \ref{sec:qgp} to reduce the computing
time: We choose randomly $n$ collision pairs ($n$ being the
particle number in a cell) instead of $n(n-1)/2$ possible doublets.
The collision probability of each chosen pair is then amplified by
a factor of $(n-1)/2$.

For the numerical simulations we consider a tube with a radius of 
$R=5$ fm. All particles will be produced initially at $\tau_0=0.1$
fm/c and are distributed homogeneously within a space-time rapidity
region $\eta \in[-3:3]$. The initial temperature at $\tau_0$ is set
to be $T_0=2.6$ GeV and thus the initial particle density is
\begin{equation}
\frac{dN}{d\eta}(\tau_0)= \pi R^2 \, \frac{T_0^3}{\pi^2} \, \tau_0=1748 \,.
\end{equation}
We have chosen these parameters to achieve initially a dense system.
For the cell configuration we set 
\begin{equation}
\label{cell}
\Delta \phi = 2\pi /8 , \quad \Delta r^2 = R^2/4 \mbox{ fm}^2 \quad
\mbox{and} \quad \Delta \eta_c =0.2 \, .
\end{equation}
The transverse area of cells is thus about $2.5 \mbox{ fm}^2$ and
the particle number in one cell is around $11$.

The total cross section of the two-body collisions is set to be 
$\sigma_{22}=10$ mb if only $2\leftrightarrow 2$ processes are
included. We also carry out calculations including both
$2\leftrightarrow 2$ and $2\leftrightarrow 3$ processes. To be
able to make comparisons between simulations without and with
inelastic processes, we set the cross sections in the latter
case to be $\sigma_{22}=5$ mb and $\sigma_{23}=5/2$ mb, which
will lead to the same number of absolute transitions per unit
time in both cases. The angular distributions of the transitions
are considered to be isotropic.

\begin{figure}[h]
\centerline{\epsfysize=8cm \epsfbox{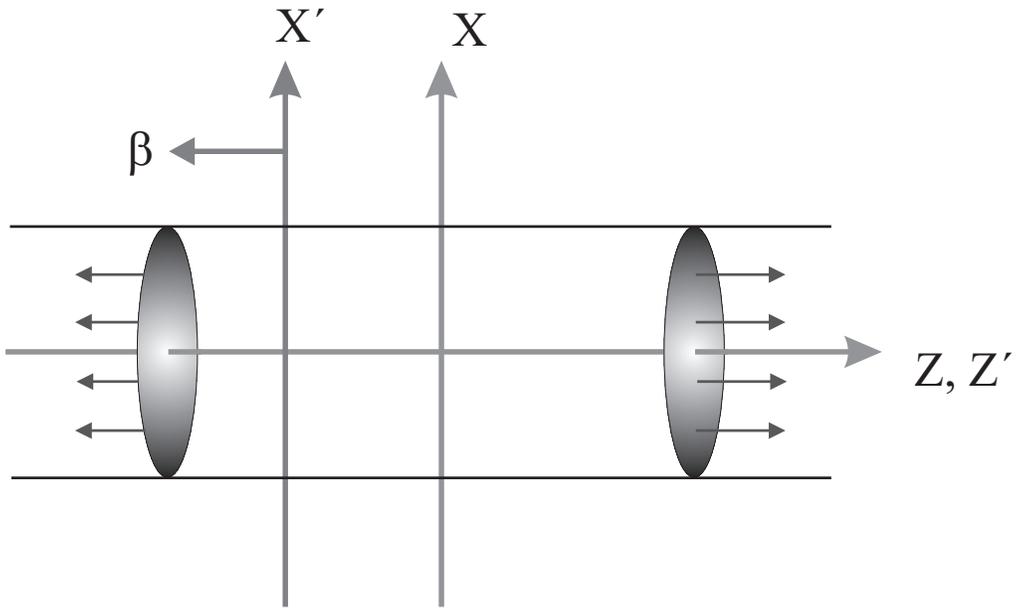}}
\caption{One dimensional expansion in a tube. The lab frame is labeled by
$X$, $Y$, and $Z$, the boosted frame by $X'$, $Y'$, and $Z'$ which is moving
with a velocity of $\beta$ relative to the lab frame.
}
\label{tube0}
\end{figure}

To study the frame dependence we will simulate the expansion in
a so-called lab frame, whose origin agrees with the center of the
expanding system and in a boosted reference frame, which is moving
relatively to the lab frame with velocity $\beta=-\tanh \eta_0$.
The situation is illustrated in Fig. \ref{tube0}. In the simulations
we set $\eta_0=2$. Since particles are initialized longitudinally
within a limited spatial region in rapidity, the pictures of the
expansion in the two frames will be quite different. The expansion
in the lab frame is symmetric, while in the boosted frame the right
part of the system expands faster than the left part at late times.
Therefore the expansion itself is frame dependent at late times due
to the limitation of the particle initialization. We will
concentrate on a so-called central region which is a cylinder
around $\eta=0$ in the lab frame and correspondingly around
$\eta_0$ in the boosted frame with a size of $\Delta \eta=1$. The
time evolutions of observables such as $n(\tau)$, $\epsilon(\tau)$,
$T(\tau)$, and others will be extracted in this central region in
the two frames and will be compared. We will present the results
in Sec. \ref{results}.

Particles are initialized in the lab frame. At first we sample
$\eta$ by its uniform distribution within $\eta \in [-3:3]$ at
the starting time $\tau_0$. We then obtain the time and
longitudinal position of the particle
\begin{equation}
t_0 = \tau_0 \cosh \eta , \quad z_0 = \tau_0 \sinh \eta \,.
\end{equation}
The transverse positions $x_0$ and $y_0$ are sampled uniformly
within the tube. Finally we determine the initial momentum
according to the thermal distribution (\ref{bjorken}) at $\tau_0$
for given $\eta$. The initial positions and momenta of particles
in the boosted frame are obtained by Lorentz transformations from
the lab frame.

\subsection{Improved cell configuration}
\label{improv}
\begin{figure}[h]
\centerline{\epsfysize=10cm \epsfbox{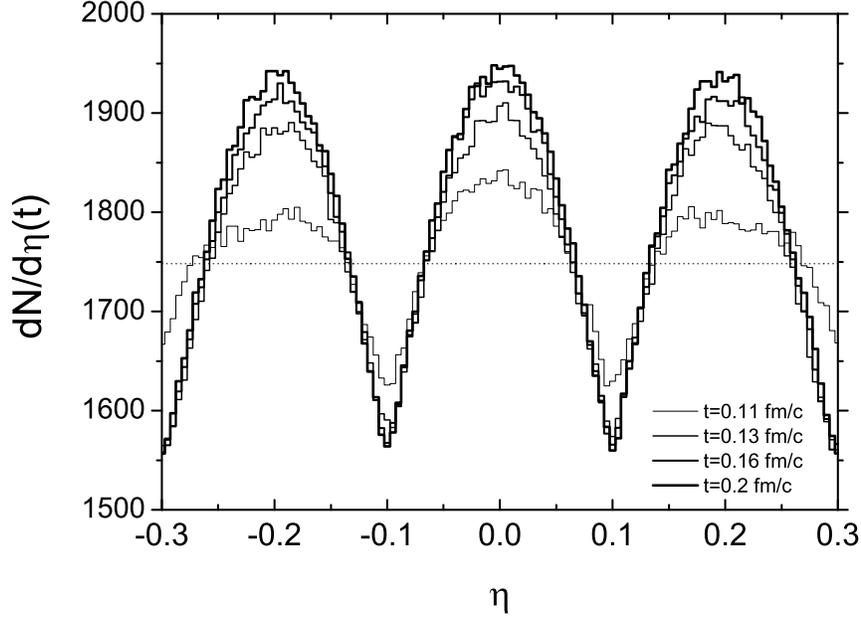}}
\caption{Space-time rapidity distributions at different times
($t=0.11$, $0.13$, $0.16$, and $0.2$ fm/c from histogram with smallest
amplitude to histogram with largest amplitude) from a simulation of
one dimensional expansion in a tube. We consider a thermal and
boost-invariant initial condition for evolving particles: Particles
are produced initially on a hyperbola of $\tau_0=0.1$ fm/c and are
distributed homogenously within a space-time rapidity interval
$\eta \in [-3:3]$, $dN/d\eta (\tau_0)=1748$, which is depicted by
the dotted straight line. The initial temperature is set to be
$T(\tau_0)=2.6$ GeV. The radius of the tube is $R=5$ fm.
We consider $2\leftrightarrow 2$ collisions with isotropic cross
section and a constant total cross section of $\sigma_{22}=10$ mb. The
stochastic method is used in the simulation. The $\eta$ bins of the
cell configuration are set to be $\Delta \eta_c =0.2$. No test 
particles ($N_{test}=1$) are used in the simulation.
The distributions are obtained by an average over $10^4$ independent
realizations.
}
\label{tube19}
\end{figure}

Before we concentrate on the further analysis, we have to make
sure that the cell configuration constructed in the last section
is really suitable for an expanding system simulated by employing
the stochastic method. To demonstrate this we perform a one 
dimensional expansion in the lab frame with the
parameters set in the last section and extract $dN/d\eta$
distribution at time $t$. One expects that the distribution will
be constant over a large region, since this was the basis
motivation for the cell construction. Figure \ref{tube19} shows the
$dN/d\eta$ distribution within an interval of $\eta \in [-0.3:0.3]$
at time $0.11$, $0.13$, $0.16$ and $0.2$ fm/c. The dotted line
depicts the initial value $dN/d\eta(\tau_0)=1748$. Astonishingly,
at first sight one notices a clear structure in the distribution
within the $\eta$ bins. (Remember that the size of the $\eta$ bins
is set to be $\Delta \eta_c=0.2$.) One can also realize that this
structure approaches a characteristic final shape at late times.
The meaning of the structure is that particles in a cell are
spatially centered. This has no physical reason, but comes from a
numerical artifact due to the finite size of the cell structure,
which can be understood as follows: We concentrate on the central
$\eta$ bin, $\eta \in [-0.1:0.1]$, and assume that the expanding
system is in local thermal equilibrium. Any change of the
$dN/d\eta$ distribution in the central $\eta$ bin is caused from
collisions among the particles and from the ongoing particle
diffusion. Even in the central $\eta$ bin the collective motion
is still outwards in spite of the small flow velocity. There are
more particles moving outwards than particles moving towards the
center. Suppose two extreme cases of collision occurring in the
central $\eta$ bin: In case 1 two particles are moving towards
the center and are approaching each other. In case 2 two
particles are moving outwards and back-to-back. Due to the
considered isotropic scattering the momentum distribution of the
particles after the collision is same in both cases. Since, on
average, the case 2 happens more frequently than the case 1, one
can draw the conclusion that collisions in an $\eta$ bin tend to
bring more particles back into the center than to push them
towards the outside when the collective flow in an $\eta$ bin is
indeed directed outwards. This is the reason for the artificial
structure of the $dN/d\eta$ distribution in the small $\eta$ bins.
On the other hand, since the distribution of $dN/d\eta$ is no more
constant, the particle diffusion from the center outwards is now
stronger than the diffusion towards the center. The diffusion is
thus counterbalancing the particle centralization and the
$dN/d\eta$ distribution will approach a final shape when the
balance between the diffusion and the centralization is fully
established.

\begin{figure}[h]
\centerline{\epsfysize=10cm \epsfbox{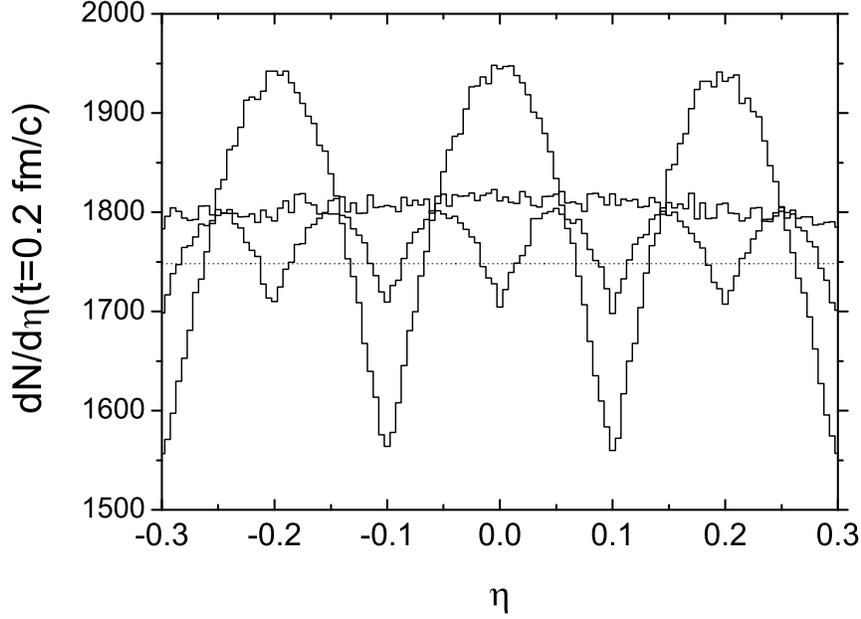}}
\caption{Space-time rapidity distributions at time $t=0.2$ fm/c in
tube calculations. The initial condition and collision cross
section are the same as in Fig. \ref{tube19}. The stochastic method
is employed in the simulations. The result showing structure
with larger(smaller) period is obtained from the simulation with
$\Delta \eta_c =0.2$($0.1$). In the simulation with
$\Delta \eta_c =0.1$ we use $2$ test particles per real particle in 
order to achieve the same statistics in each cell as that in the
simulation with $\Delta \eta_c =0.2$ and $N_{test}=1$. The histogram,
which is nearly constant, is obtained from the simulation with improved
moving cell configuration of $\Delta \eta_c =0.2$ and $N_{test}=1$.
The dotted line shows the initial distribution $dN/d\eta=1748$. All
the distributions are received by an average over $10^4$ independent
realizations.
}
\label{tube20}
\end{figure}

In Fig. \ref{tube20} we compare the $dN/d\eta$ distributions at
time $t=0.2$ fm/c from the simulations with $\Delta \eta_c =0.2$
and with a smaller size of $\Delta \eta_c =0.1$. In the 
simulation with $\Delta \eta_c =0.1$ we employ $2$ test particles
per real particle in order to obtain the same statistics as in
the case with $\Delta \eta_c =0.2$ and $N_{test}=1$. We see a
weaking in the structure of $dN/d\eta$, though the structure does
still exist. In the limit $\Delta \eta_c \to 0$, however,
the characteristic substructure in the $dN/d\eta$ distribution will
vanish, since the velocity of the intrinsic collective flow
in the $\eta$ bins goes to zero. Therefore decreasing the size of
the $\eta$ bins and using more test particles would be a natural
way to reduce this numerical artifact. However, the more test
particles, the longer will the computing time be. A more elaborate
way which does not need further test particles is to move the cell
configuration randomly from time to time. For instance, we move
the central $\eta$ bin $\eta \in [-0.1:0.1]$ to $[-0.1+\xi:0.1+\xi]$,
where $\xi$ is a random number distributed within
$[0:\Delta \eta_c=0.2]$. Although particles in each $\eta$ bin
will be still centered within each time step after collisions, but
because of the random shift of the cell configuration the associated
center of the bin for a particular particle is also moving, so that
there is no absolute center for the particle. Therefore, on average,
the effect of the centralization will be washed out. In
Fig. \ref{tube20} we depict the $dN/d\eta$ distribution from
simulations employing the improved moving cell configuration with
$\Delta \eta_c =0.2$. We see that the distribution is nearly constant
and does not show any unwanted substructure. In Fig. \ref{tube20} we
also notice a tiny enhancement of the $dN/d\eta$ distribution when
compared with the initial distribution $dN/d\eta =1748$. We will
come back to this further artifact in the next subsection.

\subsection{Results}
\label{results}

\subsubsection{$2\leftrightarrow 2$ processes without test particles}
\label{22woTP}
\begin{figure}[h]
\centerline{\epsfysize=14cm \epsfbox{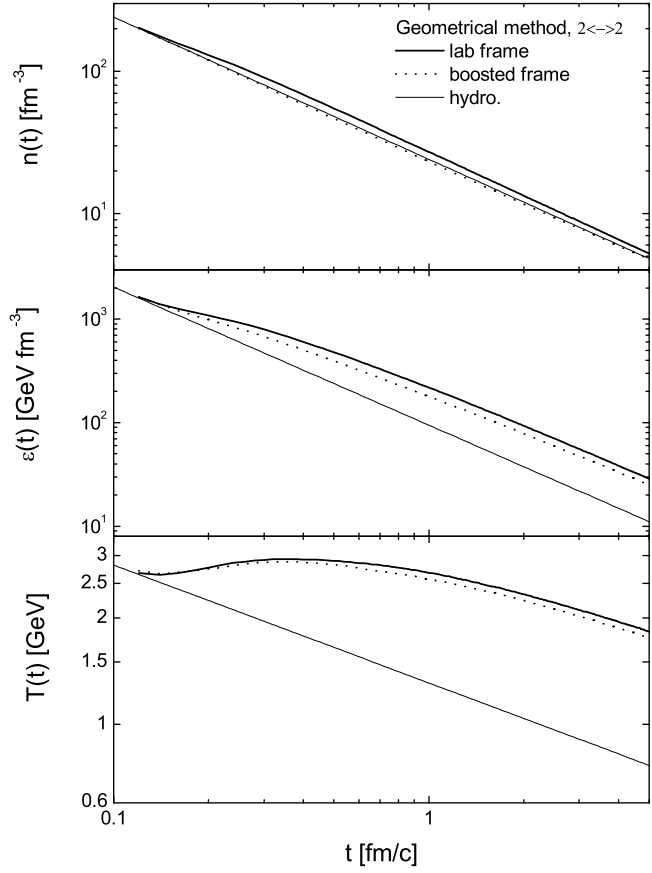}}
\caption{Time evolution of the particle density, energy density, and
temperature extracted in the central space-time rapidity region
$\eta \in [-0.5:0.5]$ from simulations of one dimensional expansion
in the lab and boosted frame of a tube. The geometrical method is
employed in the simulations. The initial condition and collision
cross section are the same as in Fig. \ref{tube19}. No test 
particles ($N_{test}=1$) are used. Only $2\leftrightarrow 2$ processes
are included. The results are obtained by an
anerage over $20$ independent realizations. The thin lines indicate
time evolutions of the quantities in the hydrodynamical limit.
}
\label{tube1}
\end{figure}

\begin{figure}[h]
\centerline{\epsfysize=14cm \epsfbox{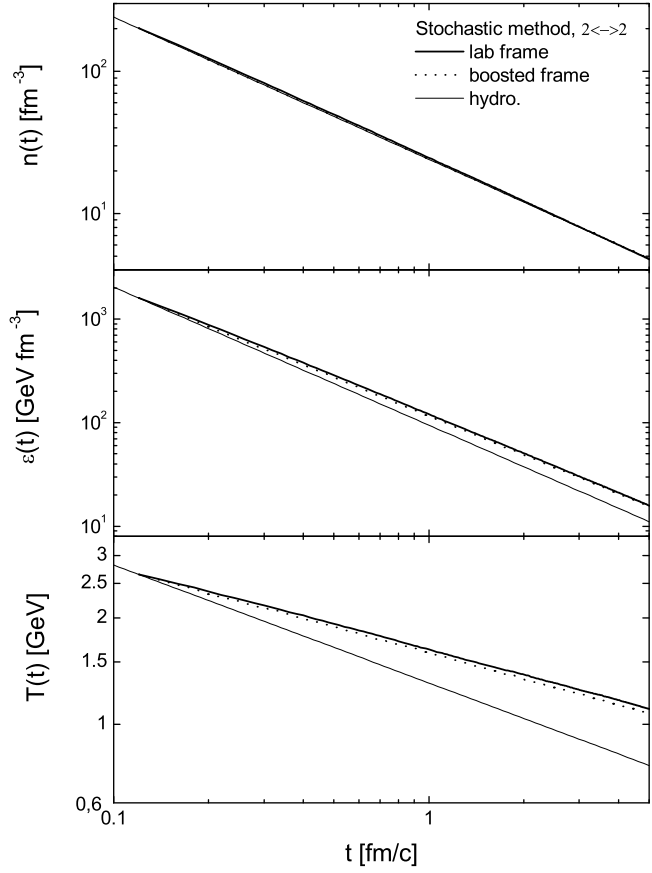}}
\caption{Time evolution of the particle density, energy density, and
temperature extracted in the central space-time rapidity region
$\eta \in [-0.5:0.5]$ from simulations employing the stochastic method
in the lab and boosted frame of a tube. The initial condition and collision
cross section are the same as in Fig. \ref{tube19}. No test 
particles ($N_{test}=1$) are used. Only $2\leftrightarrow 2$
processes are implemented. We apply the moving cell configuration with
$\Delta \eta_c=0.2$. The results are obtained by an average over $20$ 
independent realizations.
}
\label{tube2}
\end{figure}

At first we present the results from simulations without test
particles. Figures \ref{tube1} and \ref{tube2} show the time
evolution of the particle density, energy density and temperature
in the central space-time rapidity region in the two frames.
The results are extracted from the simulations employing the
geometrical and stochastic method respectively and are obtained
by averaging $20$ independent realizations. The
effective temperature is defined as $T=\epsilon/3n$ and corresponds
to the statistical temperature when the system is at local kinetic
equilibrium. Otherwise $T$ can be regarded as the mean energy per
particle. In the simulation with the stochastic method we set the
size of the $\eta$ bins to be $\Delta \eta_c=0.2$. The time scale
in Figs. \ref{tube1} and \ref{tube2} denotes the time in the local
frame of the central region. The solid and dotted curves depict the
results achieved in the lab and boosted frame respectively. The
thin solid lines show the ideal hydrodynamical limit calculated via 
a corresponding integral of the thermal phase space distribution
(\ref{bjorken}). Please note that we have taken the size of the
central region into account. Therefore the hydrodynamical results
(\ref{ntau}), (\ref{etau}), and (\ref{ttau}) are modified by 
\begin{eqnarray}
\label{nttau}
n(t) &=& a_n\, n(\tau=t)\,, \quad  a_n:=
\frac{\Delta \eta}{2\,\tanh(\Delta \eta/2)} \\
\epsilon(t) &=&  a_{\epsilon}\, \epsilon(\tau=t)\,, \quad a_{\epsilon}:=
\frac{1}{6\,\tanh(\Delta \eta/2)}\, 
\int_{-\Delta \eta/2}^{\Delta \eta/2} 
d\eta' \, \left (3+(\tanh \eta')^2 \right ) (\cosh \eta')^{4/3} \\
\label{tttau}
T(t) &=& a_T\, T(\tau=t)\,, \quad a_T:=\frac{a_{\epsilon}}{a_n}\,.
\end{eqnarray}
In the limit $\Delta \eta \to 0$ the additional factors go to $1$.

\begin{figure}[h]
\centerline{\epsfysize=10cm \epsfbox{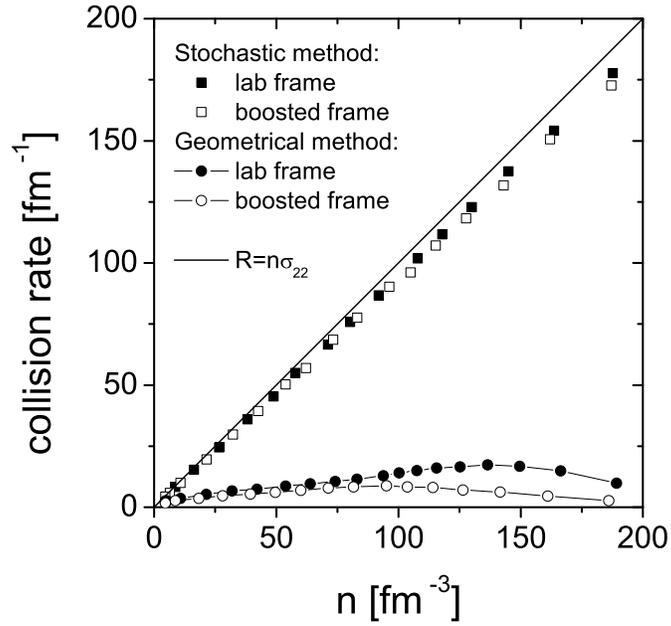}}
\caption{Collision rate in the central space-time rapidity region
for various particle
densities experienced during the expansion. The results are extracted
from the same simulations performed for the extractions of $n(t)$ and
$\epsilon (t)$ in Figs. \ref{tube1} and \ref{tube2}. 
The solid line shows the analytical expectation.
}
\label{tube3}
\end{figure}

\begin{figure}[h]
\centerline{\epsfysize=10cm \epsfbox{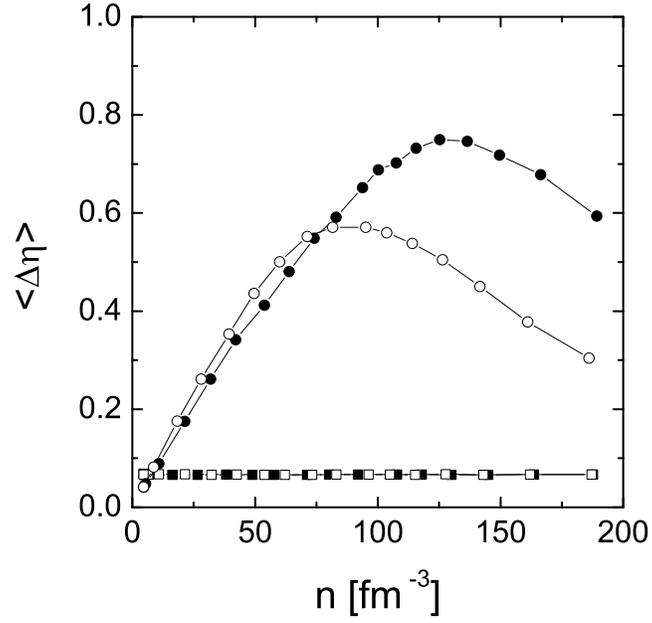}}
\caption{Averaged difference in space-time rapidity of colliding
particles extracted in the central space-time rapidity region for
various particle densities
experienced during the expansion. The results are extracted from the
same simulations performed for the extractions of $n(t)$ and
$\epsilon (t)$ in Figs. \ref{tube1} and \ref{tube2}. The labeling of
the symbols is identical to that of Fig. \ref{tube3}.
}
\label{tube4}
\end{figure}

From Figs. \ref{tube1} and \ref{tube2} we see that the frame
dependence of the considered quantities is quite noticeable in
the simulation when employing the geometrical method, while it
is rather weak in the simulation employing the stochastic method.
Moreover and astonishingly, the ``temperature'' in the simulation
with the geometrical method is increased at the beginning of the
expansion. This ``reheating'' \cite{CH02} is unphysical,
since the isotropic
initialization of the particle system does not give any reason
for an introversive pressure. The gradient of the pressure is
directed outward, so that in the further evolution the
longitudinal work done by the pressure should lead to a cooling
of the system. We also rule out any explanation based on a
possible viscous effect which might bring some effective net
energy flow into the local region, because there is no reheating
in the simulation with the stochastic method. From the
investigations within a static box we have realized that the
collision rate obtained in the simulation with the geometrical
method will be suppressed when the mean free path is in the same
order as (or even smaller than) the interaction length. This is
indeed the situation at the beginning of the expansion in the
tube. The suppression of collisions will obviously slow down the
cooling of the system, but this cannot lead to any reheating.
However, the fact that particles can interact with each other
over a larger distance than the mean free path makes it
reasonable that the pressure could be incorrectly built up.
The effect of the ``anti-pressure'' is thus a numerical artifact.
We extract the collision rate and the difference of space-time
rapidities of colliding particles per collision event
$<\Delta \eta>_{coll}$ in the central region from the simulations
carried out in the lab and boosted frame. The results are
depicted in Figs. \ref{tube3} and \ref{tube4}. The collision
rates are obtained by counting the collision events in the
central region within a time interval of $0.02$ fm/c. It is
clearly seen that the collision rates in the simulation with the
stochastic method agree well with the expectation. The slight
discrepancy can be understood as the consequence of the relative
large size of the $\eta$ bins ($\Delta \eta_c=0.2$). In contrast,
the collision rates in the simulation with the geometrical method
are strongly suppressed at high densities due to the relativistic
effect of the time spread of the two collision times, as explained
in Sec. \ref{sec:meth1}. The results of the $<\Delta \eta>_{coll}$
show that particles interact in fact over very large distance at
high densities in the simulation when employing the geometrical
method. The decrease of the $<\Delta \eta>_{coll}$ at the highest
densities corresponding to the very beginning of the expansion is
due to the fact that at the early times particles with large $\eta$
are still not formed. In the simulation employing the stochastic
method the interaction length is, however, controlled by the cell
structure. In summary, we suspect that the larger interaction
distance (compared with the mean free path) may be the reason for
the ``reheating''.

\begin{figure}[h]
\centerline{\epsfysize=10cm \epsfbox{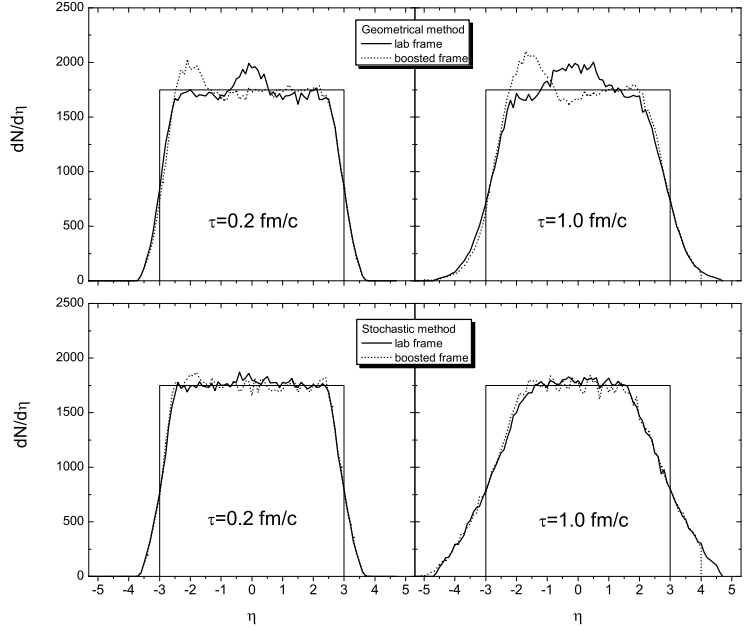}}
\caption{Particle distributions versus space-time rapidity at the proper time
$\tau=0.2$ and $1.0$ fm/c extracted from simulations employing the geometrical
and stochastic method in the lab and boosted frame. The initial condition
and collision cross section (and cell configuration) are the
same as in Fig. \ref{tube1} (and in Fig. \ref{tube2}). In order
to compare the distributions in the same physical regions directly, we
have shifted the distributions in the boosted frame by $-\eta_0=-2$.
Except that the distributions extracted from the simulations in the boosted
frame using the stochastic method are obtained by an average over ten
independent realizations, all other distributions are obtained from
$20$ independent realizations. The thin solid lines indicate the initial
distribution $dN/d\eta(\tau_0)=1748$. The cut at 
$\eta= 4$ in the distributions at $\tau=1.0$ fm/c for the expansion
in the boosted frame is due to the fact that the end time of the simulation
in the boosted frame is $t'=210$ fm/c and thus particles with $\eta$ 
being greater than $6$ (or $4$ after the shift) have smaller proper time
than $1$ fm/c.
}
\label{tube8}
\end{figure}

Figure \ref{tube8} shows the space-time rapidity distributions at
the proper time $\tau=0.2$ and $1.0$ fm/c extracted from the 
simulations in the lab and boosted frame with the geometrical
(upper panel) and the stochastic (lower panel) method respectively.
The solid (dotted) curves depict the distributions in the lab
(boosted) frame. The thin solid lines show the initial
distribution $dN/d\eta(\tau_0)=1748$ within $\eta \in [-3:3]$.
We see that the results obtained when employing the geometrical
method show a strong frame dependence.
A clear hump exists around the expansion center
$\eta=0$ in both frames and broadens gradually. (Note that the
expansion center in the boosted frame is at $\eta=-2$ after the
shift.) The humps indicate a net particle diffusion towards the
expansion center, which again can be explained as a consequence
of the ``antipressure'' effect: The introversive pressure drives
the particles back to the expansion center. In the distributions
obtained when using the stochastic method we see a relative tiny
hump at the expansion center which disappears at the later time.
The slight enhancement has been also noticed in Fig. \ref{tube20}.
We recognize that the size of the cell bins $\Delta \eta_c=0.2$
is not small enough to overcome the numerical artifact completely. 

\begin{figure}[h]
\centerline{\epsfysize=10cm \epsfbox{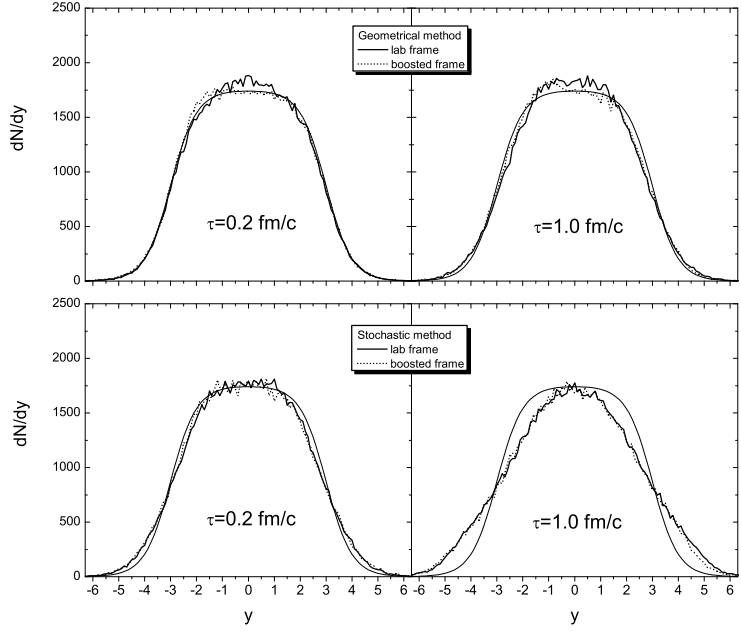}}
\caption{Particle distributions versus momentum rapidity at the
proper time $\tau=0.2$ and $1.0$ fm/c extracted from simulations
employing the geometrical and stochastic method in the lab and
boosted frame. The results are obtained from the same simulations
performed for the extractions of $dN/d\eta(\tau)$ in Fig. \ref{tube8}.
The thin solid curves indicate the initial distribution at
$\tau_0=0.1$ fm/c.
}
\label{tube9}
\end{figure}

In Fig. \ref{tube9} we depict the momentum rapidity distributions
at proper times. The thin solid curves show the initial rapidity
distribution 
\begin{equation}
\label{dndtau}
\frac{dN}{dy}(\tau_0) = \frac{R^2 T_0^3 \tau_0}{\pi} 
\frac{\sinh(2\eta_m)}{\cosh(2\eta_m)+\cosh(2y)} \,,
\end{equation}
where $\eta_m$ denotes the boundary of the initial system which
has been set to be $3$. In the upper panel of Fig. \ref{tube9} one
also recognizes the
particle diffusion towards the expansion center, though the effect
is not strong. The disributions obtained when using the stochastic
method show perfect ``no frame dependence'' and a collective flow
outwards to the higher rapidity at late times.
 
\begin{figure}[h]
\centerline{\epsfysize=10cm \epsfbox{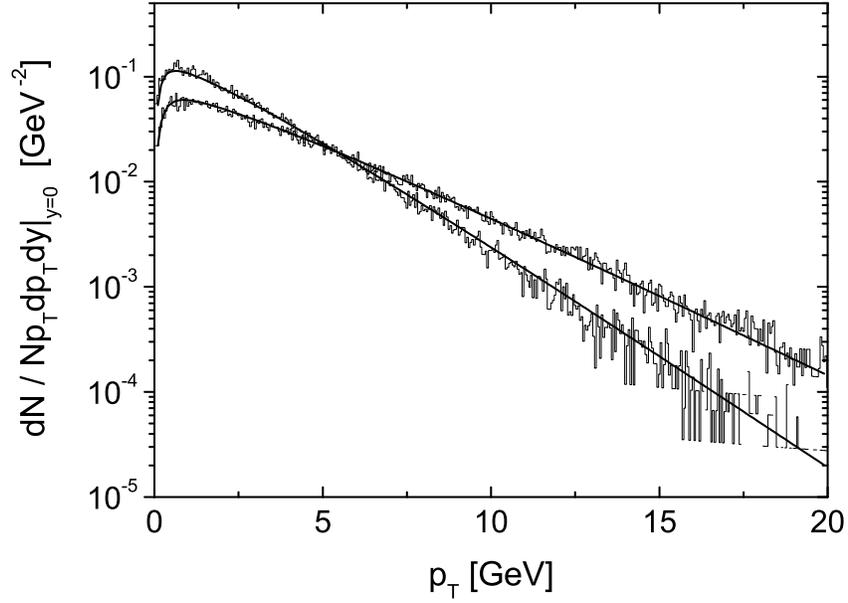}}
\caption{Distributions of the transverse momentum per unit rapidity 
at $y=0$ at $\tau=1.0$ and $4.0$ fm/c (from upper to lower histogram)
in a simulation employing the geometrical method in the lab frame.
The initial condition and collision cross section are the same 
as in Fig. \ref{tube1}. The results
are obtained by an average over $20$ independent realizations. The
solid lines show the analytical distributions (\ref{ptspect})
with the temperatures read off from Fig. \ref{tube1}.
}
\label{tube5}
\end{figure}

\begin{figure}[h]
\centerline{\epsfysize=10cm \epsfbox{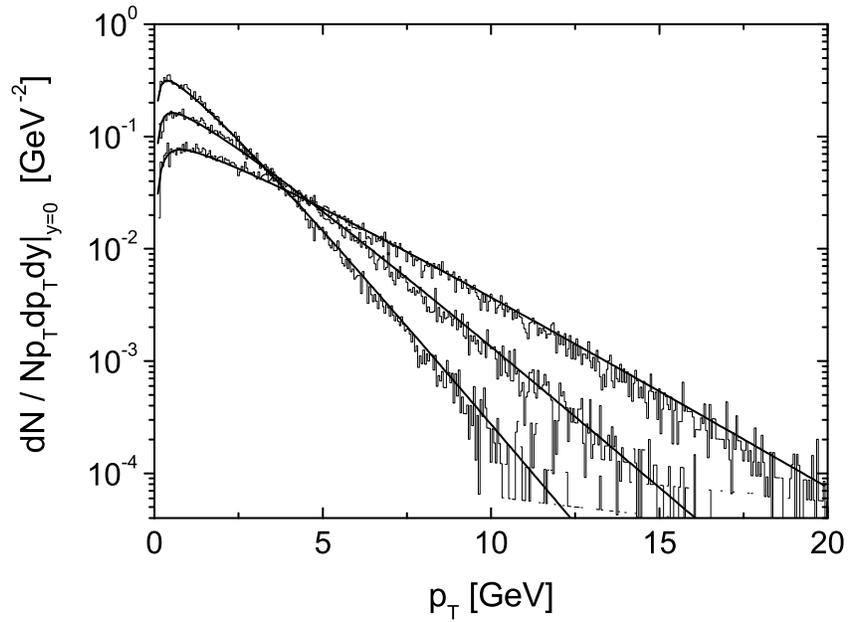}}
\caption{Distributions of the transverse momentum per unit rapidity 
at $y=0$ at $\tau=0.2$, $1.0$, and $4.0$ fm/c (from upper to lowest
histogram) in a simulation employing the stochastic method in the lab frame.
The initial condition, collision cross section, and cell
configuration are the same as in Fig. \ref{tube2}. The results
are obtained by an average over $20$ independent realizations. The
solid lines show the analytical distributions (\ref{ptspect})
with the temperatures read off from Fig. \ref{tube2}.
}
\label{tube6}
\end{figure}

For an initially thermal system it seems reasonable that the system
will be still locally in or close to kinetic equilibrium during
the further expansion. On the other hand, we have also realized
that numerical artifacts make strong effects at the beginning of
the expansion, especially in the simulations applying the
geometrical algorithm. Therefore it is essential to question whether
the encountered numerical problem does affect the maintenance of the
kinetic equilibrium in the cascade simulations of the
one dimensional expansion. For this we extract the transverse
momentum distributions at $y=0$ within an interval
$y\in [-0.5:0.5]$ at different proper times and compare them with
the analytical thermal distributions. In Figs. \ref{tube5} and
\ref{tube6} the $p_T$ distributions extracted from the
simulations in the lab frame are depicted. Figure \ref{tube5} shows
the results at $\tau=1.0$ and $4.0$ fm/c in the simulations with
the geometrical method and Fig. \ref{tube6} shows the results at
$\tau=0.2$, $1.0$ and $4.0$ fm/c in the simulations with the
stochastic method. The thermal distributions shown by the solid
lines are obtained as integral of the thermal particle
distribution function (\ref{bjorken}) by means of the
Cooper-Frye formula
\begin{equation}  
\label{ptspect}
\left .\frac{1}{N} \frac{dN}{p_T \,dp_T \, \Delta y} \right |_{y=0}(\tau)
=\frac{1}{N} \frac{2\, \pi^2 R^2}{(2\pi)^3} \frac{1}{\Delta y}
\int_{-3}^3 d\eta 
\int_{-\Delta y/2}^{\Delta y/2} dy \,p_T \, \tau \cosh (y-\eta) \,
e^{-\frac{p_T\,\cosh(y-\eta)}{T(\tau)}} \,,
\end{equation}
where $\Delta y=1$ and the temperature $T(\tau)$ is read off from
Fig. \ref{tube1} or Fig. \ref{tube2} at $t=\tau$. We see good
agreements between the numerical and the analytical distributions,
even for the case of the geometrical method. The analogous $p_T$
distributions, extracted from the simulations in the boosted frame
(at $y=\eta_0=2$), are also compared with the analytical spectra
(both not shown in figures). The agreements are perfect as those
presented in Figs. \ref{tube5} and \ref{tube6}. As a conclusion,
although the expansion does not proceed fully close to ideal
hydrodynamics, the expanding system is still kinetically
equilibrated at least until $\tau=4$ fm/c in the simulations with
the stochastic method as well as with the geometrical method,
although in the latter case the cooling of the system occurs much
slower.

\begin{figure}[h]
\centerline{\epsfysize=10cm \epsfbox{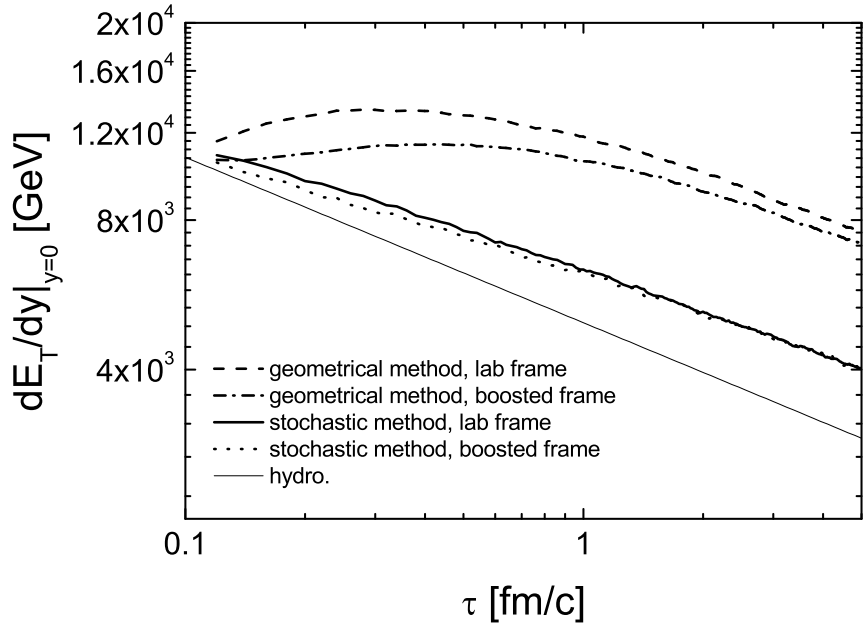}}
\caption{Proper time evolution of the transverse energy per unit
momentum rapidity at $y=0$ in the simulations employing the
geometrical and stochastic method in the 
lab and boosted frame. The initial condition
and collision cross section (and cell configuration) are the
same as in Fig. \ref{tube1} (and in Fig. \ref{tube2}).
The results are obtained by an average over $20$ 
independent realizations. The thin solid line shows the analytical
evolution in the hydrodynamical limit.
}
\label{tube7}
\end{figure}

As a last point, we show in Fig. \ref{tube7} the proper time
evolution of the transverse energy extracted at $y=0$ per unit
rapidity from both type of simulations in the lab and boosted frame
respectively. The thin solid line depicts the result in the
hydrodynamical limit
\begin{eqnarray}
\label{ettau}
\left . \frac{dE_T}{dy} \right |_{y=0}(\tau) &=& \frac{\pi R^2}{(2\pi)^3}
\int d\eta \, d^2p_T \, p_T^2\, \tau \cosh(y-\eta) \, 
e^{-\frac{p_T\,\cosh(y-\eta)}{T(\tau)}} \nonumber \\
&=& \frac{3}{4} R^2\,T^4\,\tau = \frac{3}{4} R^2 \, T_0^4\,\tau_0^{4/3}
\, \tau^{-1/3} \,.
\end{eqnarray}
The time evolutions of the transverse energy have similar
shapes like that of the temperature shown in Figs. \ref{tube1}
and \ref{tube2}. We also recognize the unphysical
``reheating'' occurring in the simulations with the standard
geometrical method.

Summarizing this section, we have studied the frame dependence
of a one dimensional expansion in a tube by employing the two
collision algorithms presented in this paper. The comparisons
show that quantities extracted in the simulations with the
geometrical method have a much pronounced and unphysical frame
dependence. Numerical artifacts are very significant in these
simulations, especially at the beginning of the expansion when
the system is very dense. In contrast, the results obtained
from the simulations when employing the stochastic method show
almost ``no frame dependence''.

\subsubsection{$2\leftrightarrow 2$ processes with test particles}
\label{22wTP}
The time evolutions of the particle density, energy density and
temperature depicted in Figs. \ref{tube1} and \ref{tube2} 
demonstrate that simulated dynamics does not undergo an ideal
hydrodynamical expansion. On the one hand, it is true that the
ideal hydrodynamics cannot be realized in simulations with finite
collision rate. One has to take the finite viscosity into account.
Thus it is interesting to make comparisons between the transport
results and those calculated from viscous hydrodynamics
\cite{DG85,GPZ97,M02}. This subject is, however, beyond the scope
of this paper. On the other hand, even the viscous expansion cannot
be solved exactly due to the limitation of the numerical
implementations. Especially, as observed in the simulations with
the geometrical method, the numerical artifacts make strong
unphysical effects. In this section we introduce the test particle
method into the dynamics to reduce this numerical uncertainty and
to study the convergency of the transport solutions.

From the experiences in the box calculations, one realizes that
the computing becomes more time consuming when more and more test
particles are added into the simulations. One way to reduce the
computing time in the present case is to consider a tube with a
smaller radius. The (real) particle density is however unchanged.
In simulations with the geometrical method we set the radius of
the tube to be $R'=R/\sqrt{N_{test}}$ with $R=5$ fm. However, in
simulations with the stochastic method we instead keep the radius
of $5$ fm, in order to be able to refine the cell configuration.

\begin{figure}[h]
\centerline{\epsfysize=15cm \epsfbox{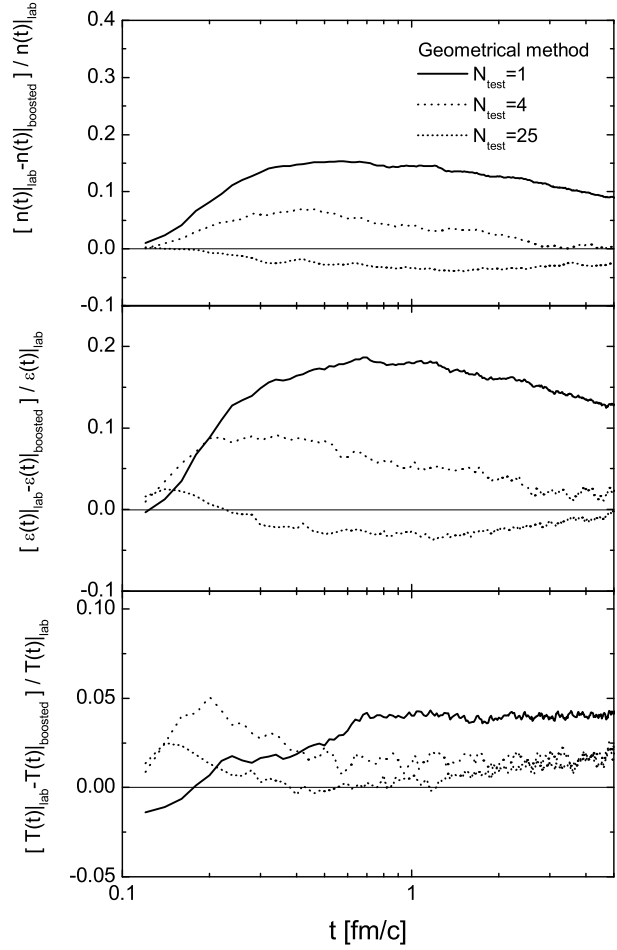}}
\caption{Relative frame dependence of the particle density, energy density,
and temperature in the simulations employing the geometrical method.
The initial condition and collision cross section are the
same as in Fig. \ref{tube1}. The results are
obtained by averaging $20$, $2$, and $20$ independent realizations for 
increasing test particles $N_{test}=1$, $4$, and $25$, respectively.
}
\label{tube11}
\end{figure}

\begin{figure}[h]
\centerline{\epsfysize=10cm \epsfbox{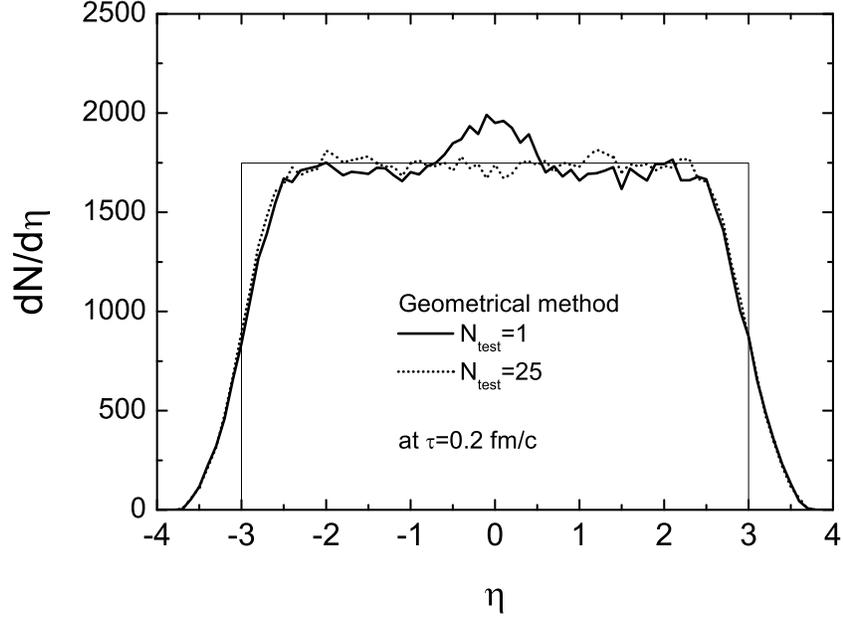}}
\caption{Comparison of the space-time rapidity distribution with
$N_{test}=25$ with the distribution without test particles at $\tau=0.2$ fm/c.
The distributions are extracted from the simulations employing the geometrical
method in the lab frame by averaging $20$ independent realizations.
The initial condition and the collision cross section are the
same as in Fig. \ref{tube1}. The thin solid line indicates the initial
distribution $dN/d\eta(\tau_0)=1748$ within $\eta \in [-3:3]$.
}
\label{tube13}
\end{figure}

Figure \ref{tube11} depicts the relative frame dependence of the
particle density, energy density, and temperature extracted in the
central region in the simulations with the geometrical method with
$N_{test}=1$, $4$, and $25$, respectively. The simulations are
performed in the lab frame. We obtain the results by averaging
$20$, $2$, and $20$ independent realizations, respectively. Note
that the simulation with $N_{test}=4$ is exceptionally carried out
with the default radius of $R=5$ fm. We see that the potential
frame dependence is more and more reduced when more and more test
particles are considered. The reduction of the frame dependence is
also clearly demonstrated in Fig. \ref{tube13}. Here the
distribution of the space-time rapidity obtained with
$N_{test}=25$ is compared with the distribution without test
particles (or $N_{test}=1$) at $\tau=0.2$ fm/c. The hump,
which exists in the
distribution without test particles due to the artificial back
diffusion, does not occur with $N_{test}=25$. For the case
employing the stochastic method it is not necessary to study the
reduction of the frame dependence with the test particle method,
since the frame dependence is actually very weak even without
test particles (see Fig. \ref{tube2}).

\begin{figure}[h]
\centerline{\epsfysize=12cm \epsfbox{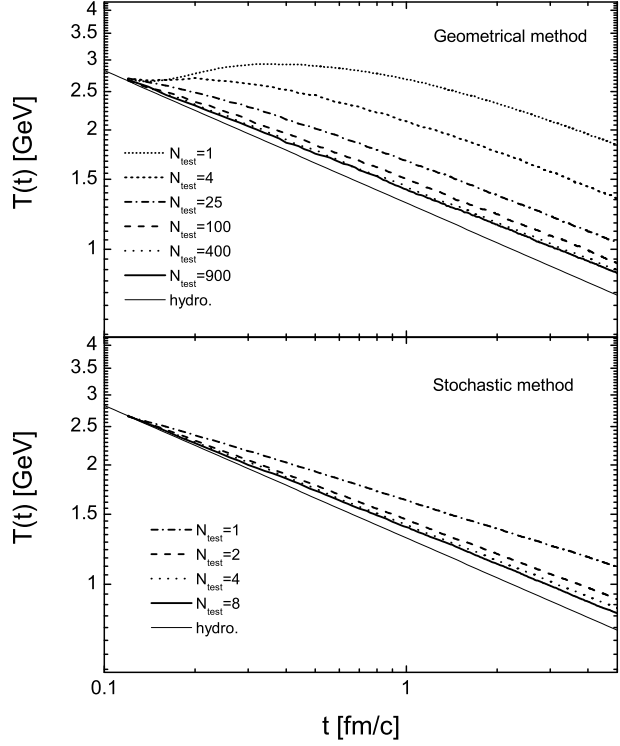}}
\caption{Convergency of temperature in the simulations in the 
lab frame with increasing test particles. The initial condition
and collision cross section (and cell configuration) are the
same as in Fig. \ref{tube1} (and in Fig. \ref{tube2}).
The results in the simulations employing the geometrical method are
obtained by averaging $20$, $2$, $20$, $5$, $5$, and $5$ independent
realizations for $N_{test}=1$, $4$, $25$, $100$, $400$, and $900$,
respectively. The results in the simulations employing the stochastic method 
are obtained by averaging $20$, $10$, $2$, and $1$ independent realizations 
for $N_{test}=1$, $2$, $4$, and $8$, respectively.
}
\label{tube10}
\end{figure}

We also employ the test particle method to study the convergency
of the transport solutions. Figure \ref{tube10} shows the time
evolution of the temperature extracted in the central region in
the simulations with increasing test particles in the lab frame.
The size of the $\eta$ bins constructed in the simulations with
the stochastic method is refined to $\Delta \eta_c = 0.2 / N_{test}$.
There are on average $11$ test particles in one cell. (We have
also performed simulations with doubled test particle number in
one cell to increase the statistics. The outcome shows almost no
changes.) From Fig. \ref{tube10} we see the clear tendency of
convergency. The time evolution of the temperatures extracted
from the simulations with the geometrical and stochastic method
converge towards almost the same curve. However, it is obvious
that the solution obtained with the stochastic method converges
more efficiently than the solution obtained with the geometrical
method. Therefore, we do favour the stochastic method to be
applied in transport simulations of system with high particle
density. Furthermore, we see that the effect of the artificial
reheating, appearing in the simulation with the geometrical
method with $N_{test}=1$, reduces and vanishes in the simulations
when employing higher test particles.

\begin{figure}[h]
\centerline{\epsfysize=12cm \epsfbox{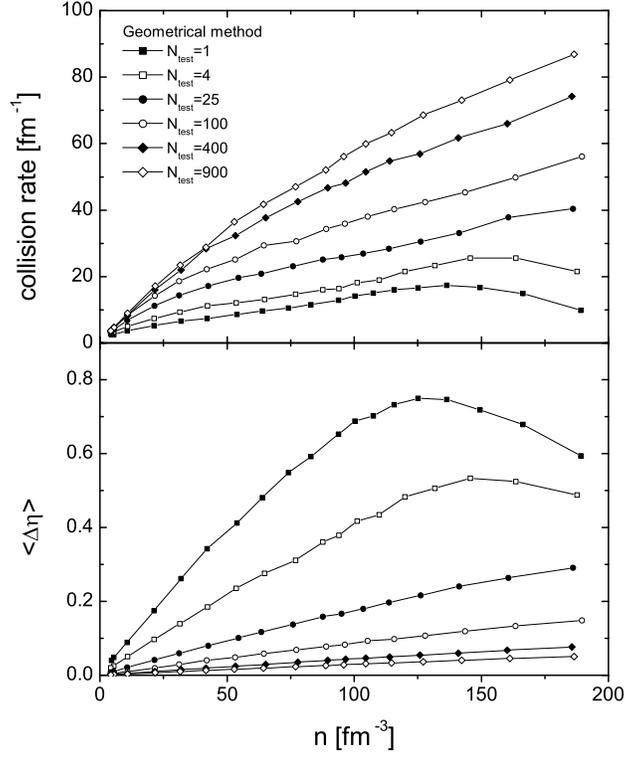}}
\caption{Collision rate and averaged difference in space-time rapidity 
of colliding particles. The results are extracted in the central
space-time rapidity region $\eta \in [-0.5:0.5]$ for various particle
densities experienced during the expansion. The simulations are the
same as performed in the upper panel of Fig. \ref{tube10} when
discussing the convergency of the temperature.
}
\label{tube12}
\end{figure}

In Fig. \ref{tube12} we depict the collision rate and the mean
difference of the space-time rapidities of colliding particles
per collision $<\Delta \eta>_{coll}$ in the simulations with the
geometrical method with increasing test particles. We see that
the collision rate increases when using more test particles.
However, even for $N_{test}=900$ the collision rates at high
densities are still suppressed. The reason is that the
interaction length decreases only with $1/\sqrt{N_{test}}$.
We also see that the $<\Delta \eta>_{coll}$ decreases when the
number of the test particles increases. Putting Fig. \ref{tube12}
in relation to Fig. \ref{tube10} confirms our suspicion in the
last subsection that unwanted collisions over large distances
may lead to the buildup of ``antipressure'' which then influences
the particle diffusion.
We mention that the same numerical artifact has been found in the
studies in Refs. \cite{MG00,CH02}.

\subsubsection{Including $2\leftrightarrow 3$ processes}
\label{23woTP}
We now include the inelastic $2\leftrightarrow 3$ processes into
the dynamics of the one dimensional expansion in the tube and
study the frame dependence for the new situation. The stochastic
method is applied to simulate the (in)elastic collisions whose
cross sections are set to be $\sigma_{22}=5$ mb and
$\sigma_{23}=2.5$ mb.
These parameters lead to the same rate of the elastic and
inelastic transitions. We consider isotropic collisions and set
the size of the $\eta$ bins to be $\Delta \eta_c=0.2/N_{test}$.

\begin{figure}[h]
\centerline{\epsfysize=13cm \epsfbox{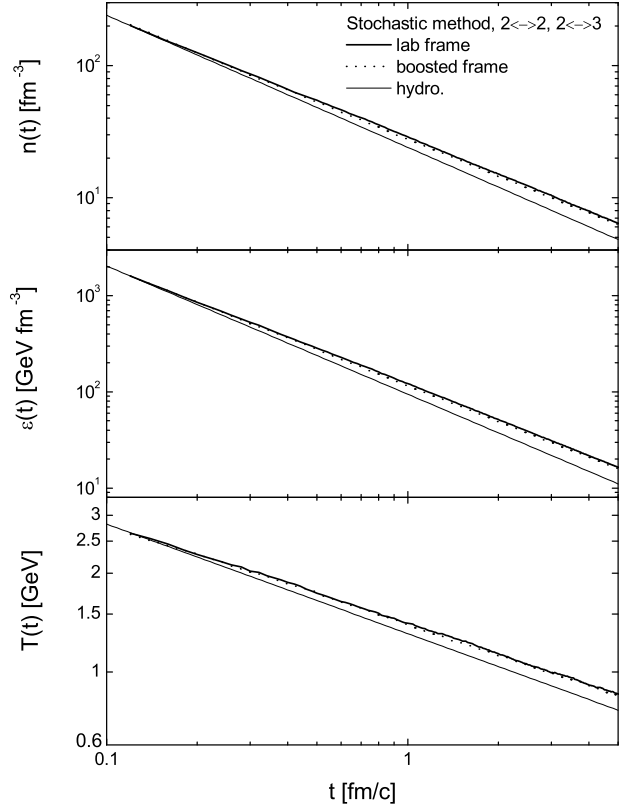}}
\caption{Time evolution of the particle density, energy density, and
temperature extracted in the central space-time rapidity region
$\eta \in [-0.5:0.5]$ from the simulations employing the stochastic method
in the lab and boosted frame. The initial condition and cell
configuration are the same as in Fig. \ref{tube2}. No test particles
($N_{test}=1$) are used. $2\leftrightarrow 2$ as well as 
$2\leftrightarrow 3$ processes are included in the simulations. We consider
isotropic collisions with constant cross sections of $\sigma_{22}=5$ mb
and $\sigma_{23}=2.5$ mb. The results are obtained by an average over
ten independent realizations. The thin solid lines indicate time evolutions
in the hydrodynamical limit.
}
\label{tube15}
\end{figure}

\begin{figure}[h]
\centerline{\epsfysize=10cm \epsfbox{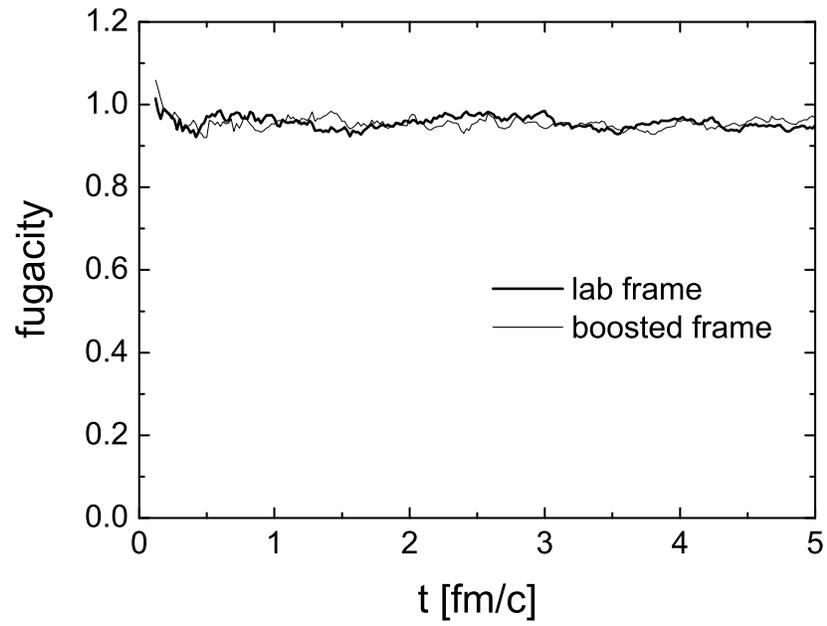}}
\caption{Time evolution of fugacity extracted from the 
same simulations performed for the extraction of $n(t)$ and $\epsilon(t)$
in Fig. \ref{tube15}.
}
\label{tube16}
\end{figure}

In Fig. \ref{tube15} we show the time evolutions of the particle
density, energy density, and temperature extracted in the central
space-time rapidity region from the simulations with $N_{test}=1$
carried out
in the lab and boosted frame. The results are absolutely frame
independent. Comparing to the results with only two-body
collisions shown in Fig. \ref{tube2}, we notice that the particle
density is slightly enhanced. This enhancement is not due to any
numerical artifacts, but the consequence of the chemical
equilibration: In the thermal equilibrium the particle density
is related with the temperature by $n=T^3/\pi^2$. Since during
the expansion the temperature is always higher compared to the
ideal hydrodynamical limit due to finite viscosity, therefore,
there have to be more particles being produced than annihilated
in order to account for the undersaturated system and to achieve
a new balance. To address the chemical equilibration we
concentrate on the time evolution of the fugacity which is
defined as $\lambda(t)=n(t)/n^{eq}(t)$, where
\begin{equation}
\label{fugaa}
n^{eq}(t)=a_n\, n^{eq}(\tau) = a_n\, \frac{T^3(\tau)}{\pi^2} =
\frac{a_n}{a_T^3}\, \frac{T^3(t)}{\pi^2} \,.
\end{equation} 
$a_n$ and $a_T$ are factors given in Eqs. (\ref{nttau}) and (\ref{tttau})
taking the size of the central region into account. The $T(t)$ in
Eq. (\ref{fugaa}) is just the extracted temperature from the simulation.
Figure \ref{tube16} depicts the time evolution of the fugacity. We see
that the chemical equilibrium is almost achieved and maintained
during the expansion in both frames. We have also extracted
the $p_T$ distributions and compared with the analytical spectra at
different times. The results show that the kinetic equilibrium is
also maintained during the expansion.

\begin{figure}[h]
\centerline{\epsfysize=10cm \epsfbox{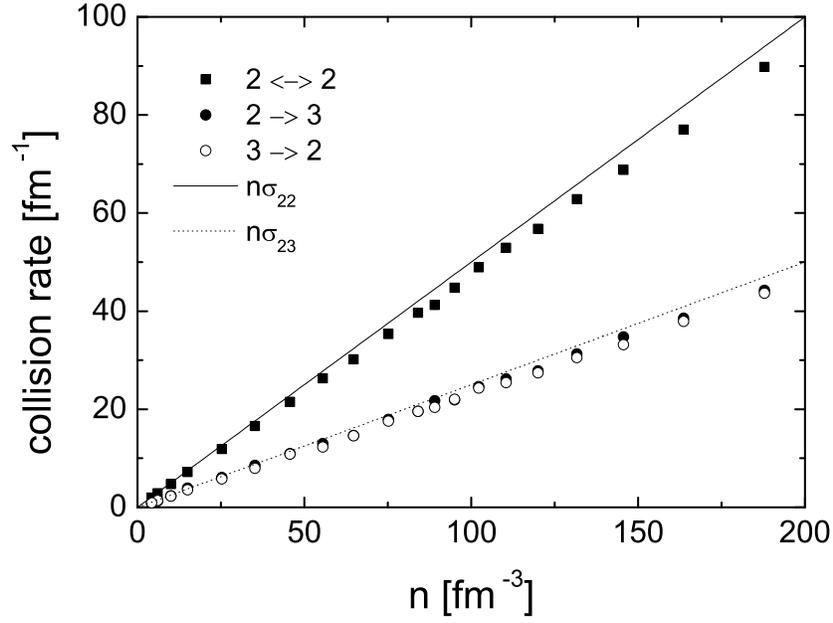}}
\caption{Collision rate in the central region for various particle
densities experienced during the expansion. The results are extracted
from the same simulations performed for the extractions of $n(t)$ and
$\epsilon (t)$ in the lab frame in Fig. \ref{tube15}.
The solid squares, solid circles and open circles depict, respectively,
the collision rates for $2\leftrightarrow 2$, $2\to 3$, and $3\to 2$
transitions. The solid and dotted line show the analytical expectations.
}
\label{tube17}
\end{figure}

The collision rates of $2\leftrightarrow 2$, $2\to 3$, and $3\to 2$
processes, extracted from the simulation in the lab frame, are
depicted in Fig. \ref{tube17}. We see perfect agreements of the
extracted collision rates with the expectations. Furthermore, the
collision rates of $2\to 3$ and $3\to 2$ processes are almost
identical, which demonstrates once more the maintenance of the
chemical equilibrium in the expanding system.

\begin{figure}[h]
\centerline{\epsfysize=10cm \epsfbox{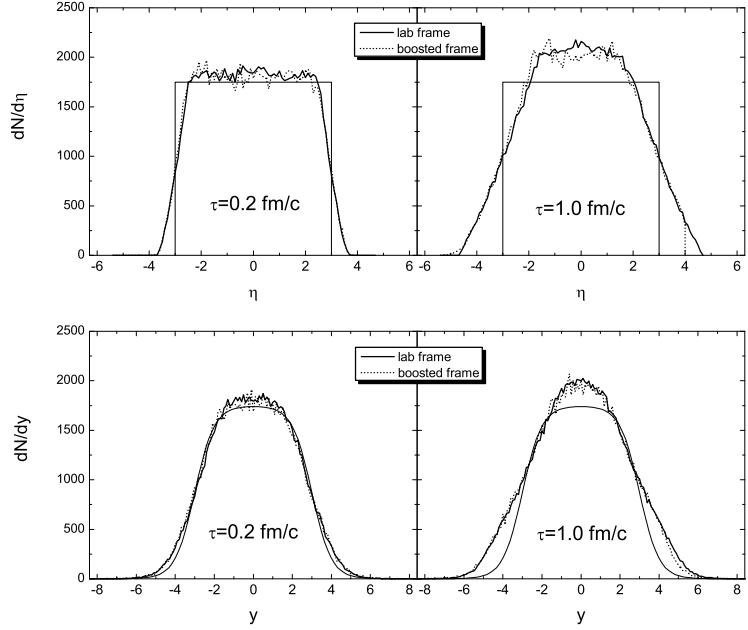}}
\caption{Particle distributions versus space-time rapidity and momentum
rapidity at the proper time $\tau=0.2$ and $1.0$ fm/c, extracted from the
simulations employing the stochastic method in the lab and boosted frame.
The initial condition, collision cross section, and cell 
configuration are the same as in Fig. \ref{tube15}. The distributions
extracted in the lab(boosted) frame are obtained by averaging $20$($6$)
independent realizations. The thin solid lines indicate the initial
distributions at $\tau_0=0.1$ fm/c.
}
\label{tube18}
\end{figure}

We show in Fig. \ref{tube18} the particle distributions 
versus the space-time rapidity $\eta$ and versus the momentum
rapidity $y$ at $\tau=0.2$ and $1.0$ fm/c, obtained from the
simulations with $N_{test}=1$ in the lab and boosted frame.
The frame dependence
is not noticeable and lies within the statistical errors. In
Fig. \ref{tube18} we also see the enhancement in the particle number
over a large range due to the slight particle production in the
ongoing chemical equilibration.

\begin{figure}[h]
\centerline{\epsfysize=12cm \epsfbox{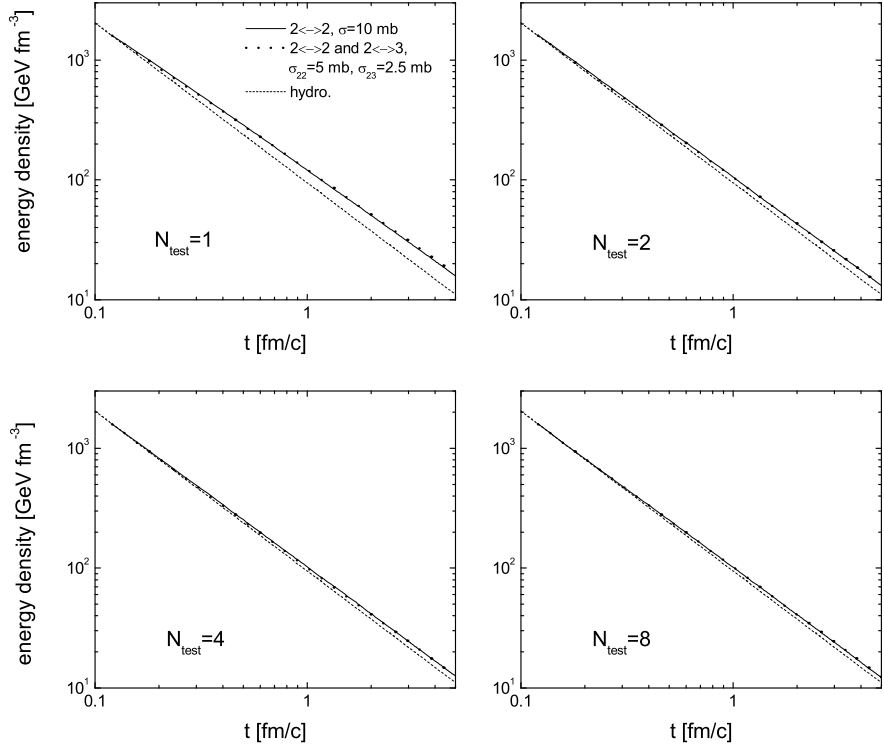}}
\caption{Convergence of energy density in the simulations in the 
lab frame with increasing test particles. The cascade simulations
are performed employing the stochastic algorithm.
The dotted lines depict the results with $\sigma_{22}=5$ mb and
$\sigma_{23}=2.5$ mb, while the thin solid lines depict the results
with purely elastic collisions and $\sigma_{22}=10$ mb.
The results are obtained by averaging $20$, $10$, $2$, and $1$
independent realizations for $N_{test}=1$, $2$, $4$, and $8$, respectively.
The thin dashed lines show the hydrodynamical limit.
}
\label{tube_m3}
\end{figure}

Finally, though the results from the simulations above are largely
frame independent, convergence to the correct Boltzmann solutions
requires using more test particles.
Due to the settings of $\sigma_{22}=5$ mb and $\sigma_{23}=2.5$ mb
one would expect that the total collision rate including elastic
and inelastic processes is the same as that in the simulation with
purely elastic collisions and $\sigma_{22}=10$ mb. Therefore,
the convergence with increasing test particles would be exactly
the same in both cases. Still, as realized from the above comparison
between Figs. \ref{tube15} and \ref{tube2}, the temperatures (and the
number densities as well) in the central space-time rapidity region
are slightly different due to the new balancing as explained above.
Therefore, the convergence of the temperature, for instance, will
not be the same as that shown in the lower panel of Fig. \ref{tube10}.
On the other hand, the new balancing should not affect the 
time evolution of the energy density. (This quantity is shown in
Fig. \ref{tube_m3} for the pure $2\leftrightarrow 2$ and the
$2\leftrightarrow 2$ + $2\leftrightarrow 3$ case with
increasing $N_{test}$.)
We see the exactly same convergence of the energy density in the
central region.

After this exhaustive discussion of testing the operation of the
cascade, we now proceed to describe real heavy ion collisions.

\section{Full $3+1$ dimensional operation of the parton cascade for 
central Au+Au collisions at RHIC: kinetic and chemical equilibration}
\label{sec:rhic}
In this section we take the step to simulate the space time evolution of
partons produced in a central Au+Au collision at maximal RHIC energy
$\sqrt{s}=200$ GeV by means of the well-tested stochastic collision
algorithm. The simulation is performed in the center-of-mass frame of
the colliding nuclei. For the present and first exploratory study we
include only the pQCD motivated gluonic interactions $gg\leftrightarrow gg$
and $gg\leftrightarrow ggg$ in the dynamical evolution. Simulations with
all parton degrees of freedom will be postponed to a sequent paper. 

\subsection{The initial conditions}
\label{rhic-init}
The initial conditions for the parton cascade are assumed to be generated
by minijet production in a central Au+Au collision modeled via multiple,
binary nucleon-nucleon collisions \cite{KLL87,EKL89}.
Of course, this is a strong
assumption. The picture of the very early stage of the collision, when
potentialy the partons are freed from the two nuclear wave functions and
do become on-shell particles, is crucial for all kinetic cascades which
can only describe the further evolution. Hence, one has to incorporate
a physical model for describing the very initial phase of liberated partons
serving as initial condition for cascades. One such physical picture
is based on the idea of a free superposition of minijets being produced
in the individual semihard nucleon-nucleon interactions. Another and
much celebrated picture is the so-called McLerran-Venugopalan model or
color glass condensate \cite{MV94}, which is based on the idea of gluon
saturation of the QCD structure function of the nuclei at sufficiently
low $x$. The so-called bottom up scenario \cite{BMSS01} of thermalization
relies on these initial conditions, where in the leading order of the
coupling constant $\alpha_s$ the various time scales of kinetic evolution
is parametrically estimated. We will leave this as an important task for
future investigation. At present, we choose the liberation of minijets
as the initial conditions. Minijets denote the on-shell
partons with transverse momentum being larger than a centain cutoff
value of $p_0 \sim 2$ GeV. Since no nuclear effects like
shadowing at small $x$ are considered in the present study, the
averaged number of produced partons is then just proportional to the
number of binary nucleon-nucleon collisions
\begin{equation}
\label{njet}
<N_{parton}>\,=\,2\,\sigma_{jet}\,T_{AA}(b=0) \, .
\end{equation}
$T_{AA}(b=0)$ denotes the nuclear overlap function for a central
nucleus-nucleus collision. The overlap function is given by
\begin{equation}
\label{overlap}
T_{AB}({\bf b}) = \int d^2x_{T1} dz_1\, d^2x_{T2} dz_2\,\, n_A({\bf r}_1)\,
n_B({\bf r}_2)\, \delta^2({\bf b}-({\bf x}_{T1}-{\bf x}_{T2})) \, .
\end{equation}
$n_{A/B}({\bf r})$ is the nuclear density. In physical terms,
$\sigma T_{AB}(b)$, where $\sigma$ denotes the total nucleon-nucleon
cross section, gives roughly the number of binary semihard nucleon-nucleon
collisions in a $A+B$ collision at impact parameter $b$ \cite{EKL89}.
The total jet cross section $\sigma_{jet}$ in a
nucleon-nucleon collision at $\sqrt{s}=200$ GeV is calculated by integrating
the differential jet cross section (\ref{csjet}). The factor $2$
in Eq. (\ref{njet}) indicates that minijets are produced in pair. Employing
the Woods-Saxon distribution for the nuclear density of a Lorentz
contracted nucleus
\begin{equation}
\label{woods}
n_A({\bf x}_{T1}, z_1)=\frac{\gamma \, n_0}
{1+Exp \left ( (\sqrt{x^2_{T1}+(\gamma z_1)^2}-R_A)/d \right )} \, ,
\end{equation}
where $d=0.54$ fm, $R_A=1.12A^{1/3}-0.86A^{-1/3}$ fm, and $n_0$ is
determined from the normalization $\int d^3r_1\, n_A =A$, one estimates
that with a cutoff $p_0=2$ GeV about $1200$ minijets will be produced
in a central Au+Au collision at maximal RHIC energy, in which $70 \%$
are gluons. We note that this number does crucially depend on
the cutoff $p_0$, which makes the minijet picture not so much promising.
On the other hand, one might improve this by choosing some
self-consistent relation for this crucial parameter \cite{EKRT00}.

The initializations of the individual produced partons in space-time 
and in momentum space are then realized statistically as follows:
The momenta are sampled according to the differential jet cross section
(\ref{csjet}). This sampling has already been performed in
Sec. \ref{sec:qgp} when the thermalization of a parton system was
studied within a fixed box. The space-time coordinates of the partons are
obtained by a simple geometrical picture when the two Lorentz contracted
nuclei do overlap. For convenience for the moment,
we set the zero point of the time scale to be the moment of the full overlap.
Then the longitudinal positions of the two nucleus centers are then at
$\pm \,v \,t$, respectively, where $v$ is the velocity of the nuclei.
One now identifies the intrinsic coordinates $z_1$ and $z_2$ in 
Eq. (\ref{overlap}) with the global space and time coordinate
\begin{equation}
z_1=z-v\,t \qquad \mbox{and} \qquad z_2=z+v\, t \,.
\end{equation}
Changing from $z_1$ and $z_2$ to $z$ and $t$ yields for Eq. (\ref{overlap})
for $b=0$ and $A=B$
\begin{eqnarray}
\label{overlap1}
T_{AA}({\bf b}=0) &=& \int d^2x_{T1} \, d^2x_{T2} \, 2\,v\, dt\,dz\,\, 
n_A({\bf x}_{T1},z-v\, t)\, n_A({\bf x}_{T2},z+v\, t)\,
\delta^2({\bf x}_{T1}-{\bf x}_{T2}) \nonumber \\
&=& \int d^2x_{T1} \, 2\,v\, dt\,dz\,\, 
n_A({\bf x}_{T1},z-v\, t)\, n_A({\bf x}_{T1},z+v\, t)\,.
\end{eqnarray}
One thus receives the statistical distribution for sampling the space-time
coordinates of the individual produced partons
\begin{equation}
\frac{d <N_{parton}>}{d^2x_{T1}\, dz\, dt} \sim 
n_A({\bf x}_{T1},z-v\, t)\, n_A({\bf x}_{T1},z+v\, t) \, .
\end{equation}
The probability for producing a parton at $({\bf r}, t)$ is thus
proportional to the convolution of the nuclear densities of the two
overlapping nuclei at the individual space-time point. Due to the
choice of the zero point in time, about half of the produced
partons are liberated at negative times. Therefore, with this
convention of the zero point in time the space-time
rapidity $\eta$ (\ref{rapidity}) is not a well-defined quantity.
In order to correct this, we shift all the times to be
larger than the absolute values of the corresponding longitudinal
positions, i.e., $t \to t+t_s > |z|$, with a uniquely chosen $t_s$.
This actually implies that $t_s \sim 0.5 R_A/\gamma$, i.e., half of
the overlapping time. Since we apply Woods-Saxon distribution
(\ref{woods}) for the nuclear density, we cannot exactly specify
when the first touch of the two colliding nuclei occurs.
$t_s$ is thus - strictly speaking - a parameter in our simulation.
For a larger $t_s$ particles pile up in the central space-time
rapidity region and for a smaller $t_s$ particles distribute within
a wider rapidity range during the very early evolution. On the other
hand, independently on the chosen $t_s$, most of the partons are in
fact produced in the central rapidity region due to the geometry of
the overlapping nuclei. In the simulations we determine $t_s$ with
the assumption that the initial partons are distributed within 
a space-time rapidity range of $\eta \in [-5:5]$.

In the above picture concerning the implementation of the
space-time production of minijets, it is assumed that
partons become immediately on-shell when the (semi)hard 
nucleon-nucleon collisions occur. Alternatively, one may
introduce an additional formation time for every minijet,
$\Delta t_f=\cosh y \, \Delta \tau_f \approx \cosh y \cdot 1/p_T$,
which models the off-shell propagation of the freed partons.
Within that time span, one assumes, for simplicity, that the
still virtual parton does not interact and moves with speed
of light. We realize, confirmed by numerical simulations, that
the introduction of such a formation time does not affect
our main findings too much. A further brief discussion will
follow in Sec. \ref{rhic-result}.
We note that the consideration of the initial conditions of
partons is a reasonable description of the minijet production
in space-time according to the overlapping of two heavy ions.
These are different from the Bjorken-type initial conditions as
considered in Refs. \cite{MG00,MG02}. Therefore, the initial
correlation between the space-time and momentum rapidity,
$\eta-y$ correlation, cannot be expected as the simple $\eta=y$.
Detailed analysis will be shown in Sec. \ref{rhic-result}.

\subsection{Cell configuration}
To be able to apply the stochastic method to simulate the full
collision sequences, we divide the space into appropriate cells.
The individual cell structure has to be considered selfconsistently
to be well suited to the details of the dynamical evolution of the
parton system. Since it is not clear whether the parton evolution
is invariant under the Bjorken boost, a configuration with constant
division in space-time rapidity, $\Delta \eta=$constant, as chosen
in Sec. \ref{sec:frame} when simulating one dimensional expansion
of a thermal system in a tube, is here not really reliable.
Cell structure should be refreshed every time step to adapt to
the dynamical parton evolution. In principle, one dimensional
expansion is still a good approximation for the whole parton
evolution for the first few fm/c after a nucleus-nucleus
collision. We thus still employ a static cell configuration
in the transverse plan: Cells are transversely set as squares
with a length of $0.5$ fm. Longitudinally, space is divided
into $\Delta z$ bins, where each bin contains about the same number
of test particles. This ensures the same statistics for each bin
and automatically adapts to the density profile of the evolving
parton system. This dynamical structuring begins at the center
of the fire-ball and then proceeds to the very outside. Test
particles from the far outside are not included into the cell
configuration, because there the density distribution is too
inhomogeneous. Instead, we then consider only elastic scatterings
among these partons treated via the geometrical method. To obtain
sufficient statistics, one has to tune the test particle number
in each bin to be large enough: It turns out that a number of $20$
test particles on average in each cell is sufficient during
first $4\sim 5$ fm/c. However, in the region with lower particle
density, especially in the transversal surface, there are not
enough test particles. If the test particle number in a cell is
less than a certain cutoff, which is set to be $4$ in the
simulations, we treat test particles in this cell again only by
means of elastic scatterings with the geometrical method. How fine
the longitudinal bins would be, depends on $N_{test}$, the number
of test particles per real particle. We set $N_{test}=60$ in the
simulations. In total, this leads to an equivalent division of roughly
equally sized bins in space-time rapidity with $\Delta \eta=0.1 \sim 0.2$.
Furthermore, in order to avoid that particles belong to the same cell
for too long time, as discussed in Sec. \ref{sec:frame}, we
randomly shift the cell structure by a small amount in the
longitudinal as well as transversal direction after every time step.

Besides this fine mesh of cells we also have to choose a sufficiently
small time step to prevent a too strong change of the spatial configuration 
in each local cell. In the simulations, this time step is time dependent and
is determined to be the one-fifth of the smallest occurring cell
length. For the case that a collision probability turns out to be
greater than $1$, all operations done within the current time step are
redone with an appropriately chosen smaller time step.

\subsection{Assumptions}
We calculate the dynamical screening mass $m_D^2$ in a similar way as 
done for the box calculations in Sec. \ref{sec:qgp} \cite{W96}
\begin{equation}
\label{md2a}
m_D^2=16\pi\alpha_s\int\frac{d^3p}{(2\pi)^3\,p}\, N_c\,f_g
\approx 16\pi\alpha_s N_c \frac{1}{V} \sum_i \frac{1}{p_i} \,.
\end{equation}
The evaluation is carried out (quasi)locally. $V$ denotes the volume of
a local region and the sum runs over all test particles in the region.
The presence of the cell structure makes it reasonable to calculate
the screening mass in each cell. However, the statistical uncertainty 
due to fluctuations is still large, since there are at maximum
$20 \sim 30$ test particles in one individual cell, and thus an
extraction of the particle phase space density $f$ is not precise.
If one assumes that the expansion in the first few fm/c is mainly
longitudinal, and further, that the transverse parton distribution is
homogeneous over a large transversal area, one can extend the sum
in Eq. (\ref{md2a}) over a more broader region compared to the individual
cell. In the simulations we consider a volume $V$ as a cylinder
with a radius of $6$ fm in the individual $\Delta z$ bin. Within each
bin $m_D^2/\alpha_s$ is assumed to be transversely constant.
This approximation will lead to an underestimate of $m_D^2/\alpha_s$
in the very central area and an overestimate in the outside area when
the transverse flow builds up, since within the same $\Delta z$ bin
the particles moving with larger transverse velocity have larger energy
and thus make a smaller contribution to the sum in Eq. (\ref{md2a}) than
the particles moving with smaller transverse velocity. The radius is
a parameter which we set to be roughly equal to the radius of a Au
nucleus. It turns out that the influence of this parameter on the
screening mass is still quite sensitive at least for late times.
A future improvement will be to simulate the parton evolution within
a parallel ensemble technique, which will give the possibility
to extract the particle phase space density locally more precise.

The coupling $\alpha_s$ is assumed to be
\begin{equation}
\label{as}
\alpha_s(Q^2)=\alpha_s(s)=
\frac{12\, \pi}{(33-2\,n_f)\, \ln (s/\Lambda_{QCD}^2)}
\end{equation}
for individual collisions, where $s$ denotes the invariant mass of
a particular colliding system of two or three particles. We set
the quark flavour $n_f$ to be $3$ and $\Lambda_{QCD}$ to be $200$ MeV.
In general, $Q^2$ in Eq. (\ref{as}) stands for the momentum transfer in
collision such as in deep inelastic scattering. For many-body
collisions, however, the scale $Q^2$ is not unambiguous.

The gluon collision rate $R_g$, which will be employed to determine
the effectively incorporated Landau-Pomeranchuk suppression in the
$gg\leftrightarrow ggg$ processes by means of a low-momentum cutoff,
is evaluated locally in cells
\begin{eqnarray}
R_g &=& R_{gg\to gg}+R_{gg\to ggg}+R_{ggg\to gg}\,,\nonumber \\
\mbox{where} \qquad && \nonumber \\
R_{gg\to f} &=& \frac{\sum_i P_i^{gg\to f}}{\frac{1}{2} N_g \Delta \tau}\,,
\quad f=gg, ggg, \quad \mbox{and} \\
R_{ggg\to gg} &=& \frac{\sum_i P_i^{ggg\to gg}}{\frac{1}{2} N_g \Delta \tau}
\,.
\end{eqnarray}
The $P_i$s denote the respective individual collision probabilities.
The sum over $P_i$ gives the mean number of collisions occurring
during a time step $\Delta t$ in a cell with $N_g$ gluons.
$\Delta \tau$ denotes the corresponding time interval in the comoving
frame $\Delta \tau = \Delta t/ \gamma$, where $\gamma=1/\sqrt{1-v^2/c^2}$
and $v$ is the collective velocity of the moving cell. For a cell
with about $20$ gluons there are totally $200$ individual possible
$gg \to gg$ and $gg \to ggg$ collisions each and $1200$ possible
$ggg \to gg$ collisions. The statistics is high enough to ensure
evaluations of the collision rates in local cells to be sufficiently
precise, in contrast to the calculation of the screening mass. Still
a problem remains, which is that the calculation of the collision
probability for a $ggg \to gg$ process by a two dimensional integral
is time consuming. To reduce the computing time we have to take the
following approximation, which has already been applied for the box
calculations in Sec. \ref{sec:qgp}: We randomly choose about $20$
gluon triplets instead of the total $1200$ combinations and compute
the amplified collision probabilities (\ref{prob_mod}). Therefore
the statistical fluctuation of the collision rate $R_{ggg \to gg}$
is stronger than that of the others. Also, when extracting the velocity
of an individual space element we encounter the same difficulty of
insufficient statistics as explained by calculating the screening mass.
We assume that all the cells in a $\Delta z$ bin have the same
longitudinal velocity
\begin{equation}
v=\frac{V\, \int \frac{d^3p}{(2\pi)^3}\,\frac{p_z}{E}\, f}
{V\, \int \frac{d^3p}{(2\pi)^3}\, f} \approx \frac{1}{N_g}\, 
\sum_i \frac{p_{iz}}{E_i} \,,
\end{equation}
where the sum runs over the test particles within a cylinder with
a radius of $6$ fm in the considered $\Delta z$ bin, and $N_g$
denotes the gluon number in the cylinder. The transverse component
of the velocity is set to be zero. In principle, this assumption
can be corrected when a parallel computing device is employed 
for achieving much higher statistics. Then one is able to look
for and calculate transverse flow of each individual cell more
accurately.

\subsection{Results}
\label{rhic-result}
\subsubsection{Rapidity distributions}
\begin{figure}[h]
\centerline{\epsfysize=10cm \epsfbox{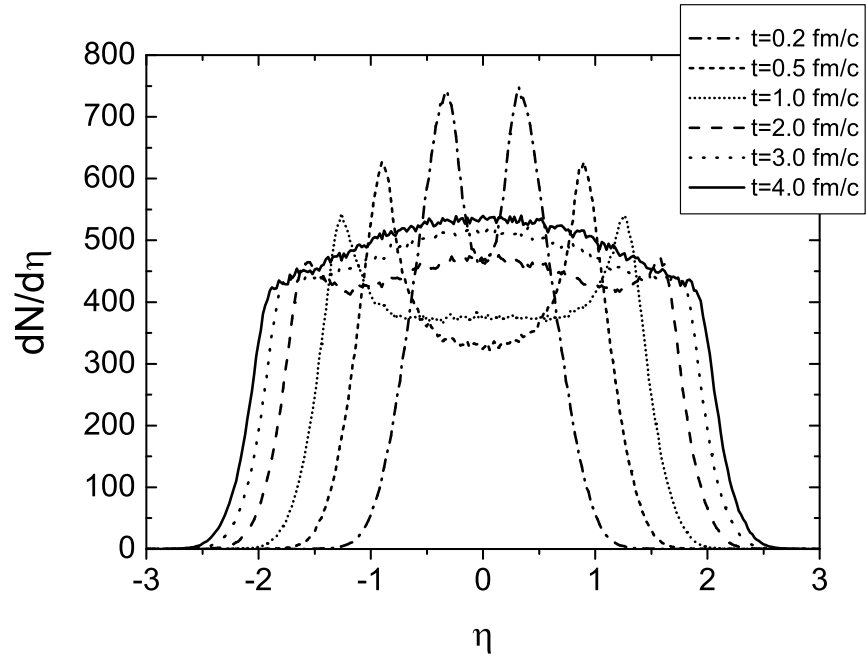}}
\caption{Gluon number distribution versus space-time rapidity at
the time $t=0.2$, $0.5$, $1.0$, $2.0$, $3.0$, and $4.0$ fm/c during the
expansion in a real, fully $3$D central Au+Au collision at the
maximal RHIC energy.
}
\label{rhic_dnde}
\end{figure}

\begin{figure}[h]
\centerline{\epsfysize=10cm \epsfbox{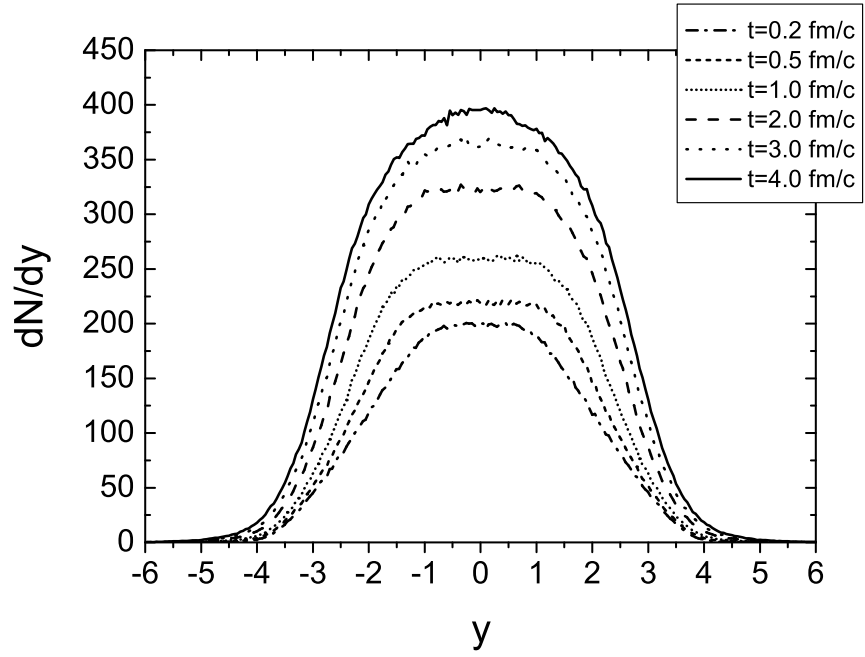}}
\caption{Gluon number distribution versus momentum rapidity at
the time $t=0.2$, $0.5$, $1.0$, $2.0$, $3.0$, and $4.0$ fm/c during the
expansion.
}
\label{rhic_dndy}
\end{figure}

We now present first numerical results obtained for the time evolution
of the gluons produced in a central Au+Au collision at RHIC energy
$\sqrt s=200$A GeV. In the simulations the number of the test
particles is set to be $N_{test}=60$. All results are obtained
by an average over $30$ independent realizations. Figures \ref{rhic_dnde}
and \ref{rhic_dndy} show the particle number distributions per
unit rapidity versus the space-time rapidity and the momentum rapidity
at the times $0.2$, $0.5$, $1.0$, $2.0$, $3.0$, and $4.0$ fm/c,
respectively. The time interval of the overlapping for the
two Au nuclei is about $0.17$ fm/c. Therefore,
the first extraction at $0.2$ fm/c is just after the end of the production
of the primary partons (or minijets). In Fig. \ref{rhic_dnde} one sees
a noticeable spreading of the $dN/d\eta$ distribution with progressing
time. The reason is that the initially produced partons are distributed
within a very small longitudinal region due to the Lorentz contraction of
the Au nuclei. Their momentum rapidities, however, have a wider
distribution, as can be seen in Fig. \ref{rhic_dndy}. The spreading of the
space-time rapidity distribution continues until its width reaches
a comparable magnitude with that of the momentum rapidity distribution.
For the special case of a simple noninteracting free streaming system,
the $dN/d\eta$ distribution will then have exactly the same shape as
the $dN/dy$ distribution at late times. In the present case we see that
at $4$ fm/c the width of space-time rapidity distribution is about $4.2$
and approaches nearly the width of the distribution
of the momentum rapidity being about $5$. It can be clearly seen that 
the spreading of the $dN/d\eta$ distribution indeed slows down at late times.
In the central space-time rapidity region the gluon density first decreases
due to this spreading, and then increases because of the ongoing gluon
production via the $gg \to ggg$ process. The gluon multiplication is most
clearly demonstrated by inspecting the momentum rapidity distributions in
Fig. \ref{rhic_dndy}, where for instance at $y=0$ the gluon number is
double amplified until $4$ fm/c. Moreover, at late times the net gluon
production slows down, which implies the completion of the ongoing chemical
equilibration. Of course, from the momentum rapidity distributions it is
difficult to recognize any evidences for kinetic equilibrium.
To investigate whether the system indeed does thermalize or not, one
needs more detailed analyses in sufficiently local regions. We
will present the results in next subsection. 

\begin{figure}[h]
\centerline{\epsfysize=12cm \epsfbox{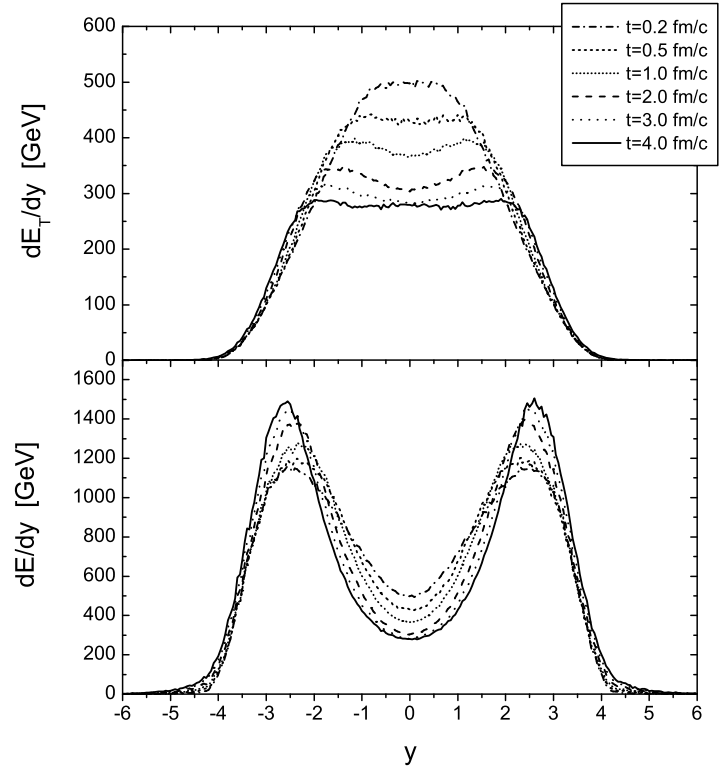}}
\caption{Momentum rapidity distributions of the transverse
energy (upper panel) and the total energy (lower panel) of gluons at
the time $t=0.2$, $0.5$, $1.0$, $2.0$, $3.0$, and $4.0$ fm/c during the
expansion.
}
\label{rhic_dedy}
\end{figure}

Figure \ref{rhic_dedy} shows the momentum rapidity distributions of the
transverse energy (upper panel) and the total energy (lower panel) at
the different times during the expansion. While the distributions
would not change during an evolution like free streaming, we see in
Fig. \ref{rhic_dedy} the decrease of the transverse energy and the
energy transport from the center towards the higher rapidity due to
the longitudinal work done by the pressure. This gives first
significant indications of collective behavior. In
addition, we note that when comparing Fig. \ref{rhic_dndy} with the
upper panel of Fig. \ref{rhic_dedy}, the shape of the latter
clearly looks more alike one dimensional Bjorken expansion than
that for the particle number distribution. Hence, one cannot really
conclude that the simple Bjorken expansion of
constant $dE_T/dy$ and $dN/dy$ manifests.

\subsubsection{Thermalization}
In the following we study possible gluon thermalization in the ``central
region'' being defined as a longitudinally expanding cylinder located
in the middle of the expanding system. The radius of the cylinder is
fixed to be $1.5$ fm and its length is $\Delta \eta=1.0$ from $-0.5$
to $0.5$. In view of the possible buildup of transverse flow, one
could consider a cylinder with varying radius which is comparable with
the longitudinal length. On the other hand, however, the statistics
within such cylinder would be very low at early times. Since the
analysis of transverse flow, which we want to address in a further
paper, shows that the transverse flow velocity is not large close to
the central region even at time of $4$ fm/c, the above choice with
fixed radius is a reasonable compromise.

\begin{figure}[h]
\centerline{\epsfysize=12cm \epsfbox{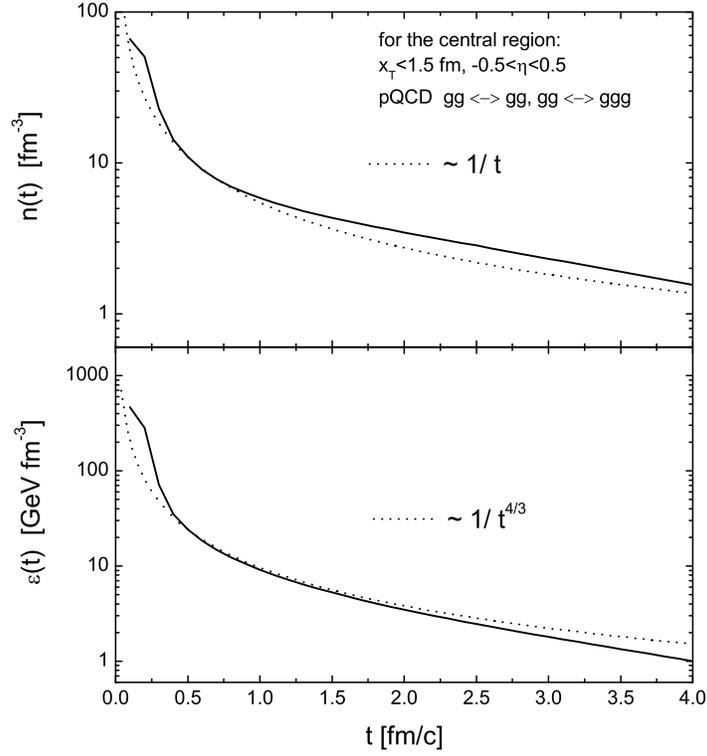}}
\caption{Time evolution of the gluon density and energy density in
the central region: radial transverse extension $x_T < 1.5$ fm and 
$\eta \in [-0.5:0.5]$ for a central Au+Au collision at the maximal
RHIC energy. The dotted curves denote the ideal hydrodynamical limit
with a fixed intercept at time $t=0.5$ fm/c.
}
\label{rhic_ne}
\end{figure}

In Fig. \ref{rhic_ne} we depict the time evolution of the gluon
density and energy density in the central region. The densities
are very high at early times. Alternatively, an
implementation of the formation time for gluons, as briefly
outlined at the end of Sec. \ref{rhic-init}, strongly reduces 
the densities at early times: $n \sim 20$ $\mbox{fm}^{-3}$ and
$\epsilon \sim 50$ $\mbox{GeV fm}^{-3}$ before $0.3$ fm/c. 
After that time the results for the particle and energy density
with and without the formation time are nearly identical throughout
the subsequent evolution. It shows that even if the densities are
very high at very early times with no formation time, the gluons
rather stream freely within the first $0.3-0.4$ fm/c.
At $4$ fm/c the energy density is still $1$ $\mbox{GeV fm}^{-3}$.
Thus the parton picture of particle interactions is valid for the
first $4$ fm/c in a central Au+Au collision at RHIC. After that
hadronization should occur and the system is then in a parton-hadron
``mixed phase''.

\begin{figure}[h]
\centerline{\epsfysize=10cm \epsfbox{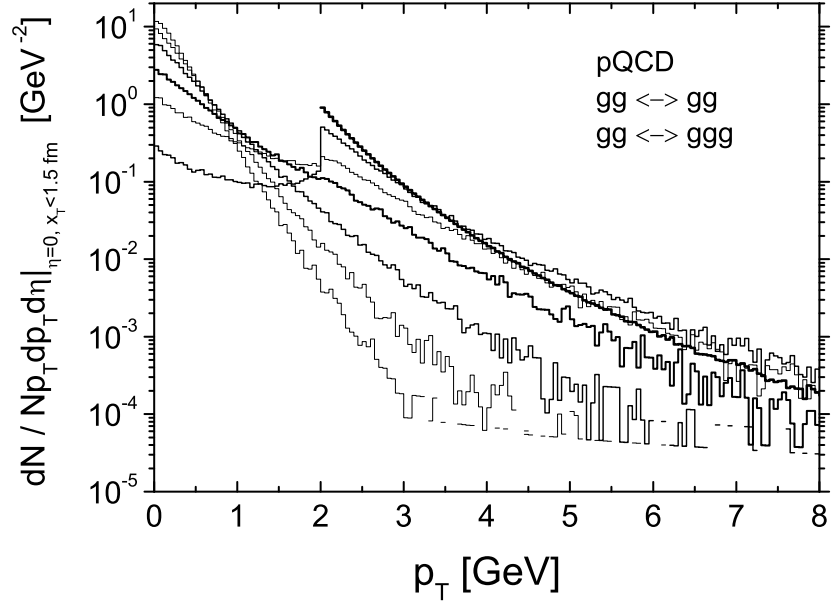}}
\caption{Transverse momentum spectrum in the central region at
different times ($t=0.2$, $0.5$, $1$, $2$, $3$, and $4$ fm/c from
second upper to lowest histogram) during the expansion.
The most-upper and bold-folded histogram with a lower cutoff
at $p_T=2$ GeV denotes the spectrum of the primary gluons (minijets).
}
\label{rhic_dndpt}
\end{figure}

Figure \ref{rhic_dndpt} shows the spectra of transverse momentum in 
the central region at different times during the expansion. The 
bold-folded histogram, which has a lower cutoff at $2$ GeV, depicts
the initial distribution of the primary gluons (minijets).
The spectrum possesses a typical power-law behavior. Already at $0.2$
to $0.5$ fm/c a tremendous population of the soft gluons below $2$ GeV
has taken place. However, still a remedy of the edge at $2$ GeV in spectra
is visible. The ``edge'' vanishes at about $1$ fm/c and the distributions
become nearly exponential and progressingly steepen at the later times
$2$, $3$, and $4$ fm/c. The ongoing steepening of the spectra in time
represents a further strong indication of a (quasi)hydrodynamical
expansion of an almost kinetically equilibrated system with
decreasing temperature.

To study kinetic equilibration in more detail, we first concentrate on
the momentum anisotropy $<p_T^2>/2<p_z^2>$. For an ideal,
one dimensional boost-invariant hydrodynamical expansion the value
of the anisotropy extracted within a region
$\eta \in [-\Delta \eta /2,\Delta \eta /2]$ is given by
\begin{eqnarray}
\frac{<p_T^2>}{2\, <p_z^2>} &=& 
\frac{\int_{-\tilde z}^{\tilde z} dz
\int d^2p_T dy \, E \, p_T^2\, e^{-\frac{p_{\perp}\cosh(y-\eta)}{T(\tau)}}}
{2\,\int_{-\tilde z}^{\tilde z} dz
\int d^2p_T dy \, E \, p_z^2\, e^{-\frac{p_{\perp}\cosh(y-\eta)}{T(\tau)}}}
\nonumber \\
\label{anisoth}
&=& \frac{\int_0^{\Delta \eta/2}d\eta \, (\cosh \eta)^{2/3}}
{6\int_0^{\Delta \eta/2}d\eta \, (\cosh \eta)^{8/3}
-5\int_0^{\Delta \eta/2}d\eta \, (\cosh \eta)^{2/3}} \,,
\end{eqnarray}
where $\tilde z=t\,\tanh (\Delta \eta/2)$. The expression (\ref{anisoth})
depends only on the longitudinal length of the local region where the 
momentum anisotropy is extracted, and goes to $1$  in the limit
$\Delta \eta \to 0$. In the central region with $\Delta \eta =1$,
the anisotropy is equal to $0.65$ for an ideal expansion.
In Fig. \ref{rhic_aniso} the time evolution of the momentum anisotropy
extracted from the present simulations is depicted by the solid curve.
Compared with the thermal value $(0.65)$, the curve in Fig. \ref{rhic_aniso}
shows first a significant increase during a short time $0.6$ fm/c and
then a smooth relaxation to that thermal value. 
The early increase of the momentum anisotropy is due to the initial
$p_T$ cutoff, $p_T > 2$ GeV, and the fact that the primary gluons with
large longitudinal momentum also have large rapidity and thus move
rapidly out of the central space-time rapidity region. The following
decrease of the anisotropy unambiguously implies the ongoing
persistance of kinetic equilibration. The reason why the anisotropy
is still larger than the thermal value is due to the fact that
particles with larger $p_T$ equilibrate later, as also seen from
the $p_T$ spectra in Fig. \ref{rhic_dndpt}. From that particular
analysis, quantitatively, the gluon system becomes approximately
fully equilibrated at $2.5$ fm/c. On the other hand, as just stated,
the clear bending over at a time of $0.75$ fm/c signals that the
strong thermalization has already started at that time, as one also
notices from the onset of the pronounced exponential behavior at
a similar time as seen in Fig. \ref{rhic_dndpt}. 

\begin{figure}[h]
\centerline{\epsfysize=10cm \epsfbox{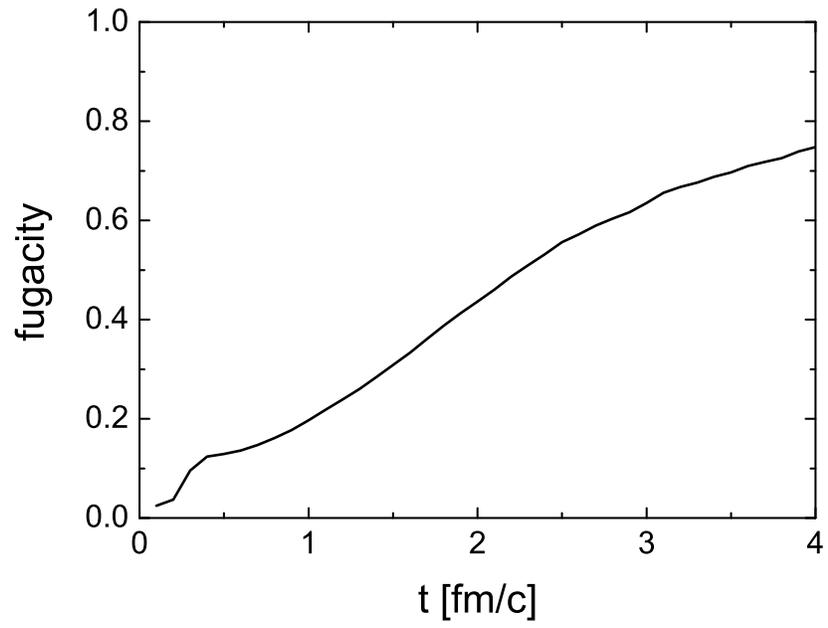}}
\caption{Time evolution of the gluon fugacity extracted
in the central region.
}
\label{rhic_fuga}
\end{figure}

The rapid streaming of the high-energy particles out of the central
region at the beginning of the expansion also explains the dramatic
decrease of the gluon density, energy density (both shown in
Fig. \ref{rhic_ne}) and the effective temperature $T=\epsilon/3n$ at
early times, which is shown in the upper panel of Fig. \ref{rhic_temp}
by the solid curve. The further decrease of the temperature until
$200$ MeV at $4$ fm/c is due to the fact that work is done by the
pressure and also due to the ongoing production of gluons. In case
of simple free streaming the effective temperature would be constant
over the whole time. To characterize the time dependence of the
temperature we assume that the temperature behaves like
$T \sim 1/t^{\alpha}$ with a time dependent exponent. $\alpha (t)$ is
shown in the lower panel of Fig. \ref{rhic_temp}. We see that the
exponent is almost constant, about $0.6$, at late times and is roughly
double the size of $1/3$, which one expects for an ideal,
one dimensional boost-invariant expansion. This might indicate the
buildup of transverse flow, but is mainly due to the further
production of gluons. For this we also extract the gluon fugacity from
the simulations and depict its time evolution in Fig. \ref{rhic_fuga}
in a way similar as in Sec. \ref{sec:qgp} [see Eq. (\ref{fuga})].
Until $4$ fm/c the chemical equilibration is still not fully achieved.
Inspecting again Fig. \ref{rhic_ne}, we have also plotted there for
comparison the standard Bjorken behavior $n\sim 1/t$ and
$\epsilon \sim 1/t^{4/3}$ with a fixed intercept at time $t=0.5$ fm/c.
One clearly recognizes that the particle number density decreases
more {\em slowly} (with an exponent of about $-0.7$) due to the
particle production. On the other hand, most interestingly, the
energy density more or less exactly follows the form which one
would expect from ideal Bjorken hydrodynamics. Indeed, the standard
relation $P=\epsilon /3$ is all what enters into ideal
hydrodynamical evolution for massless constituents, irrespective
whether the system is chemically saturated or not. This finding
gives another evidence that the system in the central region behaves
nearly as an ideal fluid.
We conclude that starting from the special, yet highly nonthermal
initial condition a gluon plasma, even not fully thermalized, may form
at $1$ fm/c in a central Au+Au collision at RHIC energy and its 
ongoing evolution in bulk behaves (quasi)hydrodynamically. 

We have to note here that, of course, this reasoning will depend
crucially on the initial conditions chosen. If we would only double
the number of initial gluons, thermalization should roughly occur
twice as fast. Indeed, our initial gluon number is lower compared
to other studies in the literature \cite{MG02,EKRT00}, where a factor
of $2-4$ more initial gluons is assumed. This will then clearly
imply that full gluon equilibration within a consistent pQCD
approach can have a full realization at RHIC. A detailed study,
addressing various initial conditions for the gluon number, 
i.e., different forms of minijet productions or color glass
condensate initial conditions, has to and will be done in
a further publication.

\begin{figure}[h]
\centerline{\epsfysize=10cm \epsfbox{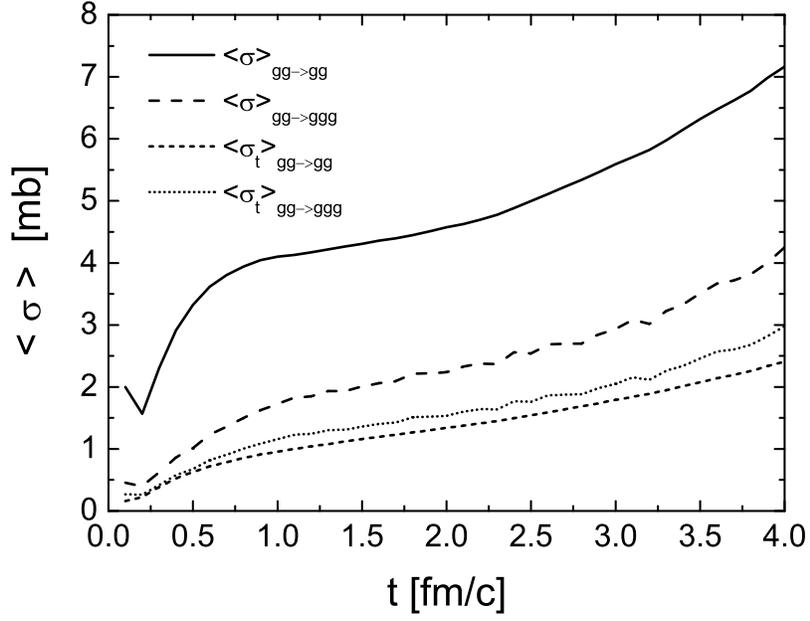}}
\caption{Time evolution of the averaged cross section
and the averaged transport cross section in the central
region. The solid and dashed (short-dashed and short-dotted)
curves depict the averaged cross sections (transport cross
sections) for the $gg \to gg$ and $gg \to ggg$ processes,
respectively.
}
\label{rhic_cs}
\end{figure}

Figure \ref{rhic_cs} shows the time evolution of the cross sections
which are first calculated as ensemble averages over all the
possible collisions in a cell and then averaged over all the cells
within the central region. As $<v_{rel}> \approx 1$ in the central
region, the collision rates of the $gg \leftrightarrow gg$ and
$gg \to ggg$ are obtained by $R=n<v_{rel}\sigma> \approx n <\sigma>$,
respectively. We have compared these collision rates with those counted
directly from the simulation and have seen nice agreements. The increase
in time of the two cross sections is due to the fact that the cross
sections are inversely proportional to the screening mass squared and
the latter is proportional to the temperature squared. One sees that
$\sigma_{gg\to gg}$ is always larger than $\sigma_{gg \to ggg}$. For
kinetic equilibration, however, not only a large total cross section
but also large scattering angle are essential for a possible fast
thermalization. In other word, the transport cross section
\begin{equation}
\label{transcs}
\sigma_t=\int d\sigma \, \sin^2 \theta_{c.m.} =\int d\theta_{c.m.}
\, \frac{d\sigma}{d\theta_{c.m.}} \,\sin^2 \theta_{c.m.} 
\end{equation}
is the key quantity controlling the ongoing of the equilibration
by given particle density $n$. $\theta_{c.m.}$ denotes
the scattering angle in the center of mass frame of the colliding
particles. For a $gg\to ggg$ process each outgoing particle has its
own scattering angle. In this case we modify Eq. (\ref{transcs}) by
$(\sin^2 \theta_1 + \sin^2 \theta_2 +\sin^2 \theta_3)/3$ instead of
$\sin^2 \theta_{c.m.}$. The averaged transport cross sections are
shown in Fig. \ref{rhic_cs} by the short dashed and short dotted
curves. Taking into account that at late times the collision rate
of the $ggg \to gg$ is comparable with the rate of the $gg \to ggg$
process, one realizes that the inelastic processes are actually the
dominant processes driving the system to kinetic equilibrium. The
ensemble averaged running coupling $<\alpha_s>$ is also
extracted within the central region during the gluon evolution. It
turns out that $<\alpha_s>$ increases almost logarithmically in
time from $0.2$ at $0.2$ fm/c to $0.5 \sim 0.7$ at $4$ fm/c. 
We note that when comparing the cross sections calculated in
thermal equilibrium (see Fig. \ref{ratevcs}), the cross sections
$\sigma_{gg\to gg}$ and $\sigma_{gg\to ggg}$ extracted from
the dynamical runs are $4\sim 5$ times larger at later times.
This is because  first $\alpha _s$ had been fixed to $0.3$ in
Fig. \ref{ratevcs} and is thus smaller. Second, the screening
mass is appreciably smaller in the dynamical calculation as the
gluons are not fully saturated in its occupation number. Both
effects add up to the difference.  

\begin{figure}[h]
\centerline{\epsfysize=10cm \epsfbox{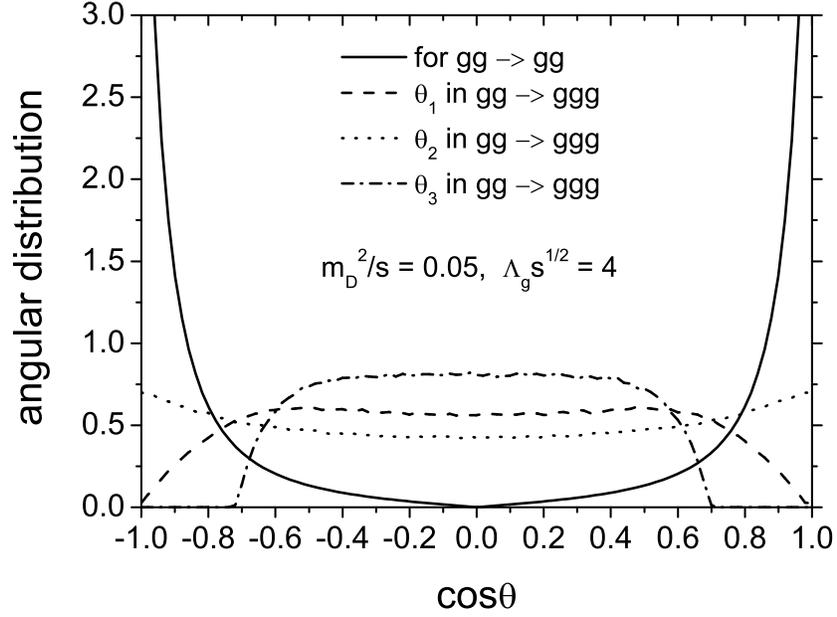}}
\caption{Angular distribution of the scattering processes $gg\to gg$
(solid curve) and $gg\to ggg$ for a representative situation during
the gluon evolution. $\theta_3$ denotes the scattering
angle of the radiated gluon and its radiation partner has
the angle $\theta_2$. The distributions are computed with the
parameters $m_D^2/s=0.05$ and $\lambda_g \sqrt s=4$ extracted
in the central region at an intermediate time during the evolution.
}
\label{rhic_angle}
\end{figure}

Following the expression of the differential cross section one knows
that the gluon elastic collisions favor small angle scatterings.
The transport cross sections in Fig. \ref{rhic_cs} indicate that the
angular distribution of the inelastic collisions is more moderate
than that of the elastic collisions. As can be realized from the
differential cross sections expressed in Appendices \ref{ppcs} and
\ref{ggcs}, the angular distribution of the elastic scatterings
depends on $m_D^2/s$, while it depends on $m_D^2/s$ and
$\lambda_g \sqrt s$ for the inelastic collisions. In
Fig. \ref{rhic_angle} we depict the angular distributions of the
$gg \to gg$ and $gg \to ggg$ scatterings for the parameters
$m_D^2/s=0.05$ and $\lambda_g \sqrt s=4$. The distributions are
calculated according to the differential cross sections. The two
parameters are chosen from an intermediate situation within the
simulation. We see that while the angular distribution of the elastic
collisions clearly shows forward scatterings as expected, the angular
distribution of the inelastic collisions is surprisingly almost
isotropic. The reason for this behavior is due to the effective
Landau-Pomeranchuk cutoff being implemented. For a larger
$\lambda_g \sqrt s$ the $gg \leftrightarrow ggg$ processes would
also favor the more the small angle scatterings. Notice that
$\theta_3$ denotes the angle of the radiated gluon and thus possesses
also a cutoff in its distribution due to the incorporation of the
Landau-Pomeranchuk suppression of low momentum gluon emissions.

\begin{figure}[h]
\centerline{\epsfysize=10cm \epsfbox{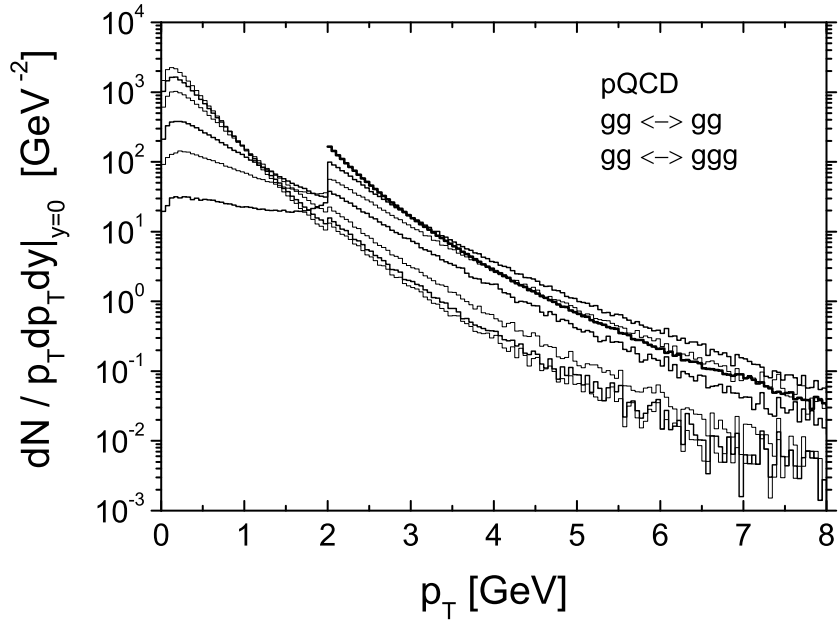}}
\caption{Transverse momentum spectrum in the central space-time rapidity
slice ($\eta \in [-0.5:0.5]$ and all gluons in the transverse plan are
counted for) at different times ($t=0.2$, $0.5$, $1.0$, 
$2.0$, $3.0$, and $4.0$ fm/c from second upper to lowest histogram).
The most-upper and bold-folded histogram with a lower cutoff at 
$p_T=2$ GeV denotes the spectrum of the primary gluons (minijets).
}
\label{rhic_dndpta}
\end{figure}

In Fig. \ref{rhic_dndpta} we present the $p_T$ spectra at different
times in the central momentum rapidity integrated now over the
whole transverse region. The spectra are arranged in the same way
as in Fig. \ref{rhic_dndpt}. Comparing the spectra in
Fig. \ref{rhic_dndpta} with those in Fig. \ref{rhic_dndpt}, we see
that there is no full global thermalization over whole transverse
region until $4$ fm/c. At least for the lower momenta we see
a nearly exponential population
and a clear steepening at the later stages.
Part of the minijet spectra, of course, survives as those gluons
might escape directly from the outer region without interactions.
In addition, Fig. \ref{rhic_dndpta} also demonstrates the potential
energy loss of gluons due to the Bremsstrahlung process. The new
developed parton cascade offers an alternative possibility to
investigate the phenomenon of the jet quenching in a more
quantitative way based on a full $3+1$ dimensional treatment of
the geometry. To be able to compare the numerical results with
the experimental data one has to model the mechanism of
the hadronization and include further hadronic interactions.
A detailed analysis is again one of possible future projects.

\subsubsection{Simulation only with elastic scatterings}
\begin{figure}[h]
\centerline{\epsfysize=10cm \epsfbox{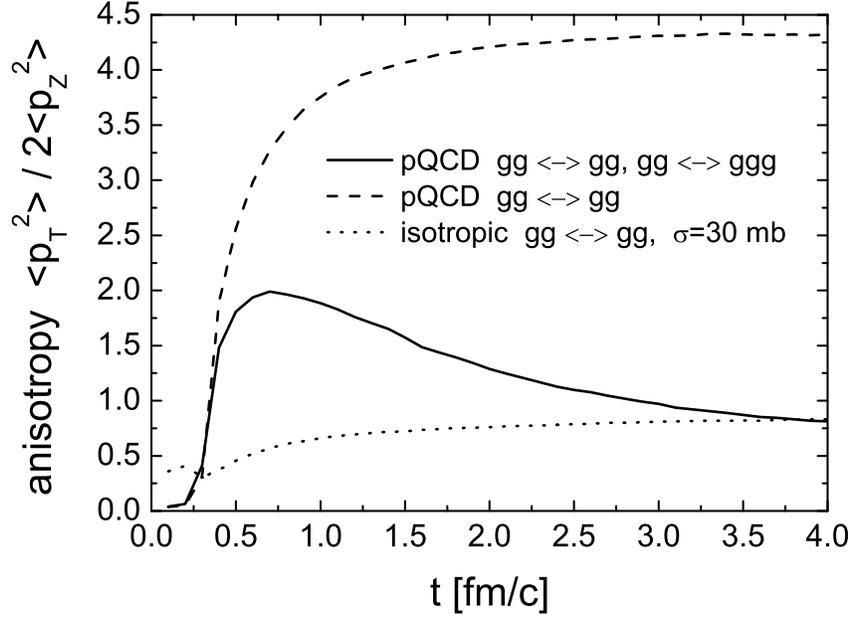}}
\caption{Time evolution of the momentum anisotropy extracted
in the central region. The solid curve shows the result from
the simulation with full dynamics, while the dashed curve shows the
result from the simulation with only elastic scatterings. The
dotted curve depicts the result from the simulation with 
isotropic elastic collisions and with (unrealistic) large cross
section of $\sigma=30$ mb.
}
\label{rhic_aniso}
\end{figure}

\begin{figure}[h]
\centerline{\epsfysize=12cm \epsfbox{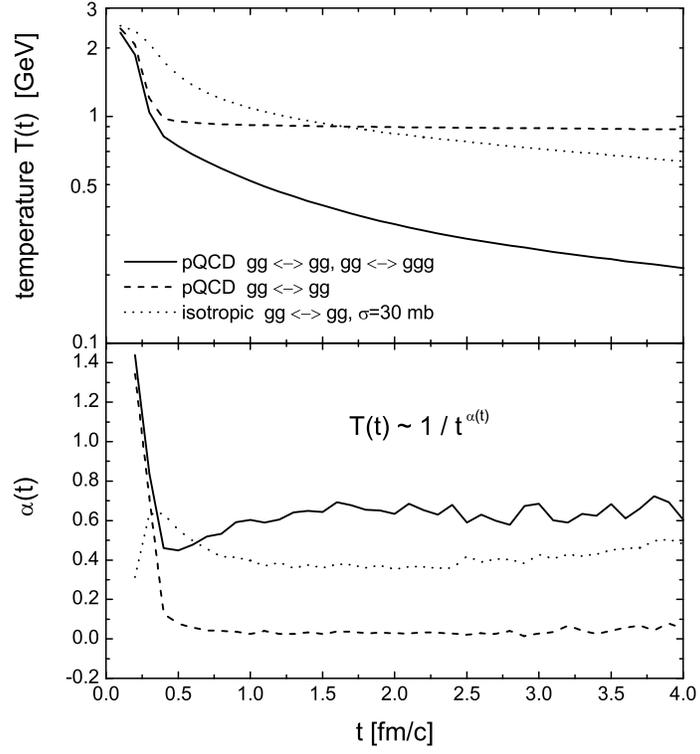}}
\caption{Time evolution of the effective temperature
(upper panel) and the exponent describing the cooling of
the system (lower panel) in the central region.
The curves are arranged in the same way as in Fig. \ref{rhic_aniso}.
}
\label{rhic_temp}
\end{figure}

\begin{figure}[h]
\centerline{\epsfysize=10cm \epsfbox{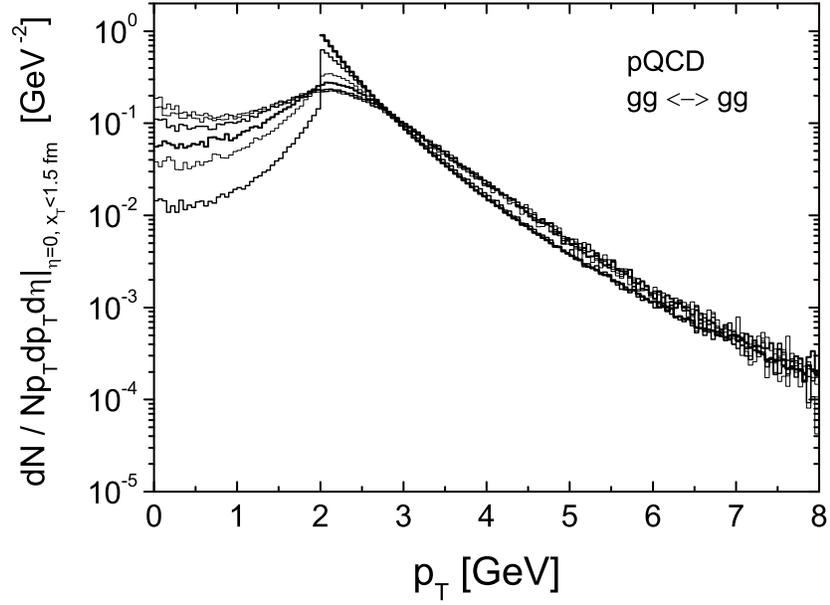}}
\caption{Transverse momentum spectrum in the central region
extracted from the simulation with only elastic collisions at different
times. The most-upper and bold-folded histogram with a lower cutoff
at $p_T=2$ GeV denotes the spectrum of the primary gluons
(minijets). According to the increase of the population of the soft
gluons below $2$ GeV, the other histrograms present the spectrum at
times $0.2$, $0.5$, $1.0$, $2.0$, $3.0$, and $4.0$ fm/c, respectively.
}
\label{rhic_dndpt22}
\end{figure}

In order to further focus on the importance of the inelastic
channels to the evolution, to the thermalization and to the potential
onset of nearly ideal hydrodynamical behavior of the partonic system,
we now perform simulations, for comparison, only with pure elastic
scatterings among the gluons.
Since in this case no gluons will be produced during
the evolution, more test particles are needed to build for a fine
cell structure. We set $N_{test}=240$. Fig. \ref{rhic_dndpt22}
depicts the spectra of the transverse momentum in the central
region at different times. The population of the soft gluons below
$2$ GeV is rather low and the distributions at large $p_T$ are
only slightly altered. Indeed gluons with highest momenta get more
populated. It is obvious that the gluon system is not thermalized
during the expansion. This can also be realized from the time
evolution of the momentum anisotropy presented in
Fig. \ref{rhic_aniso} by the dashed curve, where the anisotropy
saturates at much higher value than $1$ at late times. Furthermore,
the constant temperature shown in Fig. \ref{rhic_temp} indicates
that the evolution of the gluons is almost close to free streaming.
(Please note that the abrupt decrease before $0.5$ fm/c is also due
to the free streaming of the energetic gluons out of the central
space-time rapidity region.) We remark that in the full dynamics
with the inelastic channel, the contribution of the elastic
scatterings to kinetic equilibration is actually significantly
larger than that shown here, because in the full dynamics we have
more gluons due to the radiation and the transport cross section
also becomes larger compared to a nonequilibrated system.

\begin{figure}[h]
\centerline{\epsfysize=10cm \epsfbox{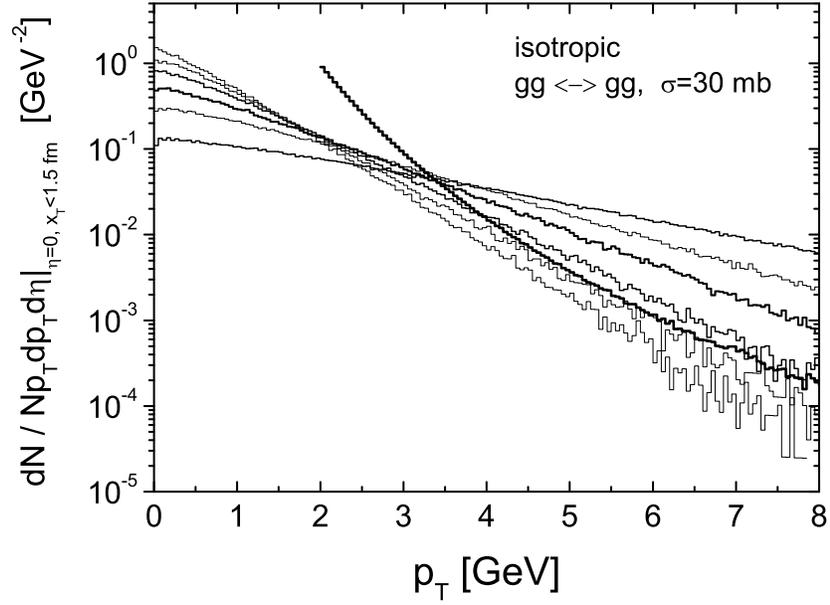}}
\caption{Transverse momentum spectrum in the central region 
extracted from the simulation with isotropic elastic scatterings and a large
cross section of $\sigma=30$ mb at different times ($t=0.2$, $0.5$, $1.0$, 
$2.0$, $3.0$, and $4.0$ fm/c from second upper to lowest histogram). 
The most-upper and bold-folded histogram with a lower cutoff at 
$p_T=2$ GeV denotes the spectrum of the primary gluons (minijets).
}
\label{dndpt_iso}
\end{figure}

In principle, kinetic equilibration can be achieved by elastic
scatterings alone, if ad hoc the transport cross section is chosen
sufficiently large. To demonstrate this we carry out simulations
with isotropic collisions and a large and constant total
cross section of $\sigma_{22}=30$ mb. The corresponding transport
cross section is thus $20$ mb.
Such extreme conditions of an assumed
large opacity in $2\leftrightarrow 2$ reactions have been used
in Ref. \cite{MG02} to study the
possible buildup of elliptic flow. Figure \ref{dndpt_iso} shows the
$p_T$ spectra in the central region at different times. Indeed we
observe fast equilibration. The spectrum at $0.5$ fm/c is almost
thermal. At the later times the distributions become more and more 
steeper, which indicates the cooling down of the system due to
(quasi)hydrodynamical expansion. The time evolution of the
momentum anisotropy, the dotted curve in Fig. \ref{rhic_aniso},
shows that from $1.0$ fm/c the anisotropy is almost constant and
slightly higher than the value of the ideal, one dimensional
boost-invariant expansion, $0.65$. Moreover, also the exponent
describing the cooling of the temperature (see 
Fig. \ref{rhic_temp}), $\alpha(t)$, is nearly constant from $1$
to $3$ fm/c and only slightly greater than the value of the ideal
expansion, $1/3$. All this demonstrates that for the given
extreme conditions the gluon system equilibrates indeed rapidly
and then expands nearly hydrodynamically according to the ideal
Bjorken scenario. However, of course, the constant and isotropic
cross section cannot be further motivated. In addition, following
that particular evolution, the system would stay for a rather
long time in a hot, but very dilute and undersaturated (in its
gluon number) deconfined state (see upper Fig. \ref{rhic_temp}).
Contrary, in the more realistic situation with inelastic collisions
included, the temperature drops much more dramatically 
and the system would stay only until $t\approx 4 $ fm/c in a pure
deconfined state, being then (nearly) fully saturated in
the gluonic degrees of freedom, and will then hadronize.

\begin{figure}[h]
\centerline{\epsfysize=10cm \epsfbox{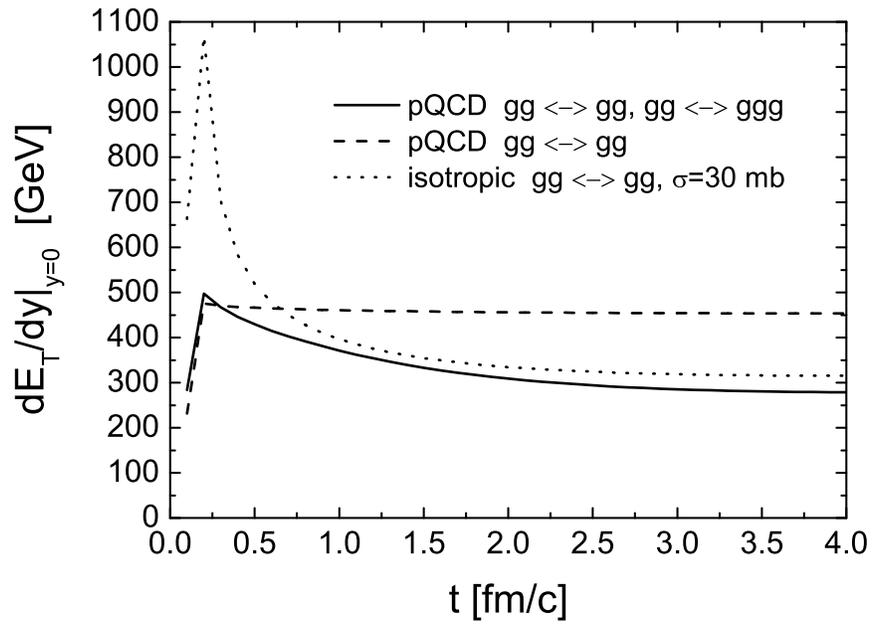}}
\caption{Time evolution of the transverse energy per unit
momentum rapidity at midrapidity.
The curves are arranged in the same way as in Fig. \ref{rhic_aniso}.
}
\label{rhic_et}
\end{figure}

Figure \ref{rhic_et} shows the time evolution of the transverse energy
per unit momentum rapidity at midrapidity for the three cases
compared also in Figs. \ref{rhic_aniso} and \ref{rhic_temp}.
We see that the transverse energy decreases in the simulation
including pQCD elastic and inelastic interactions and in the
simulation employing an isotropic, large cross section. In contrast
to the cooling of the temperature, to which the production of
gluons also contributes, the decrease of the transverse energy
within a unit rapidity is purely due to the longitudinal work done
by the pressure! In the simulation employing large and constant
cross section, energetic gluons are extremely stopped during their
formation periode, so that the transverse energy is very large at
very early times and pressure seems to be already built up during
the overlap of two nuclei. This leads to the following
(unrealistic) strong explosion with drastical cooling. The
unaltered behavior of the transverse energy in the simulation
including only pQCD elastic scatterings indicates again that in
this case the parton evolution resembles a free streaming.
One observes that from times $t\approx 0.5 $ fm/c both the 
full pQCD (including $gg\leftrightarrow ggg$) and the
``strongly coupled'' (with isotropic $\sigma_{22}=30$ mb) evolution
show almost the identical value and the same decrease in time for
the total transverse energy per rapidity. This again manifests
that both pathes resemble (quasi)hydrodynamical behavior by
performing a significant amount of (longitudinal) work.

\section{Summary and Outlook}
\label{sec:summary}
We have developed a new $3+1$ dimensional relativistic transport
model solving the kinetic on-shell Boltzmann equations. Besides
binary $2\leftrightarrow 2$ scatterings, inelastic
$2\leftrightarrow 3$ processes are also implemented in the cascade.
The numerical emphasis is put on the extension of the stochastic
collision algorithm for the back reaction $3\to 2$ which is treated
- for the first time - fully consistently within this scheme.
Although the development specifically aims at a simulation of the
parton evolution in an ultrarelativistic heavy ion collision, the
presented algorithm will certainly have more potential applications
beyond the scope of this paper. Also the standard geometrical collision
algorithm (based on the geometrical intepretation of cross section)
has been discussed in detail. In particular, we find out that for
the case that the mean free path of particles is in the same order
as or comes below the interaction length, which is always true in a
very energetic (and dense) high-energy heavy ion collision, the
results from the simulations employing the geometrical method have
shown several unphysical numerical artifacts. The convergency of the
numerical solution in this scheme for $N_{test} \to \infty$ turns out
to be not as efficient as it does in the simulations when employing
the stochastic method.

The operation of the newly developed cascade has been demonstated by
investigating gluon thermalization for a central Au+Au collision at
RHIC energy. The numerical results show that starting initially from
a nonthermal system made up of minijets (with cutoff $p_T > p_0=2$ GeV),
the gluons in the expanding center equilibrate kinetically on a scale
of $1$ fm/c and evolve further according to (quasi)hydrodynamics.
The system cools down due to the hydrodynamical expansion and ongoing
gluon multiplication. Full chemical equilibration follows on a longer
time scale of about $3$ fm/c. We have studied the contribution of the
elastic and inelastic collisions to kinetic equilibration. It turns
out that the inelastic scatterings are the main responsible processes
driving the system to equilibrium. Without any inelastic collision
channel, the collective behavior observed nowadays at RHIC cannot
be generated, unless one uses an unrealistic large cross section (or
equivalently a large gluon density) to mimic a strongly interactive
gluon system \cite{MG02}. We have also realized that the angular
distribution of the $gg \leftrightarrow ggg$ processes is almost
isotropic during the expansion. This leads to larger transport cross
section compared with the elastic scatterings.

\begin{figure}[h]
\centerline{\epsfysize=10cm \epsfbox{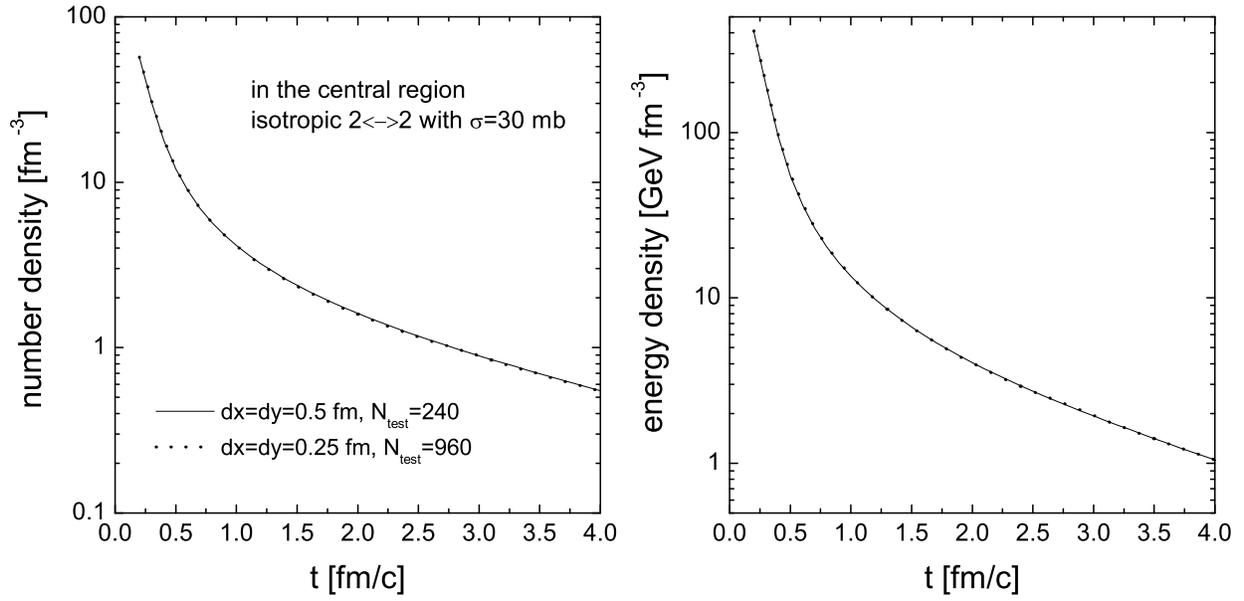}}
\caption{Time evolution of the number (left panel) and energy (right panel)
density extracted in the central region from the simulation with
$dx=dy=0.25$ fm and $N_{test}=960$ by the dotted lines, compared with the
results with the default settings $dx=dy=0.5$ fm and $N_{test}=240$, depicted
by the solid lines.
}
\label{rhic_m1}
\end{figure}

Even in the simulations applying the stochastic algorithm, particles
do collide at nonzero distance due to the nonzero spatial subvolume.
Therefore, one may worry about acausal effects due to larger signal
velocity than $c$ in the cells \cite{MG00,CH02}. In principle,
the spatial cell length should resolve the spatial inhomogeneities
in the dynamical system. For the situation when using a $30$ mb cross
section for mimicking a strongly interacting system, the mean free
path of the particles is initially smaller than the transverse
cell length. To explore whether any potential acausal effect makes some
numerical artifact, we now show a simulation employing half of the
default transverse cell length and four times enhanced number of
test particles (to keep the same statistics in cells).
In Fig. \ref{rhic_m1} we depict the time evolutions of the number and energy
density of gluons extracted in the central region from the simulation with
$dx=dy=0.25$ fm and $N_{test}=960$ by the dotted lines. Comparing the
results with those obtained with the default settings, depicted by the
solid lines, we do not recognize any visible difference.
This indicates that acausal effects seem to be not sensitive to
the cell length when the system is rather uniformly
distributed in space.

Moreover, we note once more that the two timescales for kinetic
and chemical equilibration depend crucially on the initial gluon
number. The one chosen here is rather low compared to other studies
in the literature \cite{MG02,EKRT00}, where a factor of $2-4$ more
initial gluons is assumed. This will clearly imply that the
timescales for gluon equilibration within the present pQCD approach
significantly shorten. Hence, In the future a lot of details have
to be explored: Thermalization (also of the light and heavy quarks
degrees of freedom) has to be investigated with various initial
conditions like minijets, with a detailed comparison to data,
or the color glass condensate, serving as input for the so-called
bottom up scenario of thermalization. How likely is the latter
picture for true coupling constants and not parametrically small ones?
Furthermore, the indication of the hydrodynamical behavior during the
expansion, which is one of the main findings from our first and
exploratory study concerning RHIC physics, gives strong motivation
for exploring transverse and elliptic flow using this new kinetic
parton cascade. Can the inelastic interactions generate the seen
elliptic flow $v_2$? Furthermore, one can also compare the present
calculations with some fixed and specified hydrodynamical initial
conditions directly with calculations based on viscous
relativistic hydrodynamics \cite{M02}, either assuming
Bjorken boost invariance within an expanding tube or for full
3+1 dimensions. Such a comparison can tell how viscous
the QGP really turns out to be. More practically, also
the phenomenon of the jet quenching or electromagnetic radiation
can be studied systematically within the new transport scheme.

Finally, the technique of the parallel programing is needed to
improve the practical operation of the cascade. With this
technique quantities like the screening mass can be calculated
and incorporated more precisely and quantum effects like the
Pauli-blocking and gluon enhancement can be then implemented
straightforwardly.

\acknowledgments
This work has been supported by BMBF and GSI Darmstadt.
The authors especially thank U.~Mosel for his constant interest and
motivation throughout the work. Z.X. and C.G. thank G.~Martens
and H.~St\"ocker for enlightening discussions and constructive
suggestions. C.G. also thanks W.~Cassing for helpful discussions.

\appendix

\section{Collision times in the geometrical method}
\label{app_colltime}
Within the algorithm implementing the geometrical picture
collisions occur if the considered particles approach each
other and their closest distance is less than the
interaction length $\sqrt{\sigma/\pi}$. This criterion will be
inspected in the center-of-mass frame of the colliding particles.
Suppose that $\hat r_i =(t_i, {\bf r}_i)$, $\hat p_i =(E_i, {\bf p}_i)$
and $\hat r'_i =(t'_i, {\bf r}'_i)$, $\hat p'_i =(E'_i, {\bf p}'_i)$,
$i=1,2$, are the space-time coordinates and four momenta of two particles
in the lab frame and in their c.m. frame, respectively. Defining
$H=(\hat r_2 -\hat r_1)\cdot (\hat p_1+\hat p_2)$, one has in the c.m.
frame: $t'_1 > t'_2$ if $H < 0$ and $t'_1 \le t'_2$ if $H \ge 0$.
For the case $t'_1 > t'_2$ (otherwise we change the indices of
the particles) the two particles will approach each other if
$\hat p_2^2\, [\hat p_1 \cdot (\hat r_2-\hat r_1)]
-(\hat p_1 \cdot \hat p_2)\, [\hat p_2 \cdot (\hat r_2-\hat r_1)] <0$.
The closest distance of the colliding particles in the c.m. frame is
\begin{equation}
\Delta r'_s=\sqrt{-f-\frac{a^2\,d+b^2\,c-2\,a\,b\,e}{e^2-c\,d}}\,,
\end{equation}
where
\begin{eqnarray}
a &=& (\hat r_2 - \hat r_1)\cdot \hat p_1\,, \quad 
b = (\hat r_2 - \hat r_1)\cdot \hat p_2 \,,\nonumber \\
c &=& \hat p_1^2 \,, \quad d = \hat p_2^2 \,,\quad 
e = \hat p_1 \cdot \hat p_2 \,, \nonumber \\
f &=& (\hat r_2 - \hat r_1)^2\,.
\end{eqnarray}
If $\Delta r'_s < \sqrt{\sigma/\pi}$,
the particles will collide at the same time $t'_{c1}=t'_{c2}$ at the
closest distance in the c.m. frame. Making Lorentz transformation back
to the lab frame gives
\begin{equation}
t_{c1}=t_1-E_1\, \frac{a\,d-b\,e}{e^2-c\,d}\,,\quad
t_{c2}=t_2+E_2\, \frac{b\,c-a\,e}{e^2-c\,d}\,.
\end{equation}
We call $t_{c1}$ and $t_{c2}$ the collision times. Due to the spatial
separation, the two collision times have, in general,
different values, $t_{c1} \ne t_{c2}$. This means that one of the particles
reacts later within the same collision. The transformed space coordinates
at the collision times are correspondingly denoted by ${\bf r}_{c1}$ and
${\bf r}_{c2}$. The new momenta of the particles 
are sampled in the c.m. frame according to the given differential cross
section and then transformed to the lab frame, which are denoted by
${\bf p}_{c1}$ and ${\bf p}_{c2}$. We thus label the particles with
$(t_{ci}, {\bf r}_{ci})$ and $(E_{ci}, {\bf p}_{ci})$, ($i=1,2$),
and keep the labels until their next respective collisions. For
example, $t_1$ denotes the time when the last collision of particle $1$
occurs. It is kinematically possible that the case
$t_1 < t_{c1} < t_2 < t_{c2}$ occurs. Such a collision sequence is not
causal, because at $t_{c1}$ when the particle $1$ experiences the
collision with the particle $2$, the particle $2$ is just on the way
to its last collision with some other particle.
To forbid those collisions we add an additional criterion: The collision
times $t_{c1}$ and $t_{c2}$ should be greater than $t_1$ as well as $t_2$.
Illustratively, the additional criterion means that during the time
interval $|t_1-t_2|$, the particle, which will change its trajectory later
(it is the particle $2$ in the example), is not considered for
dynamics for that particular interval.

\begin{figure}[h]
\centerline{\epsfysize=10cm \epsfbox{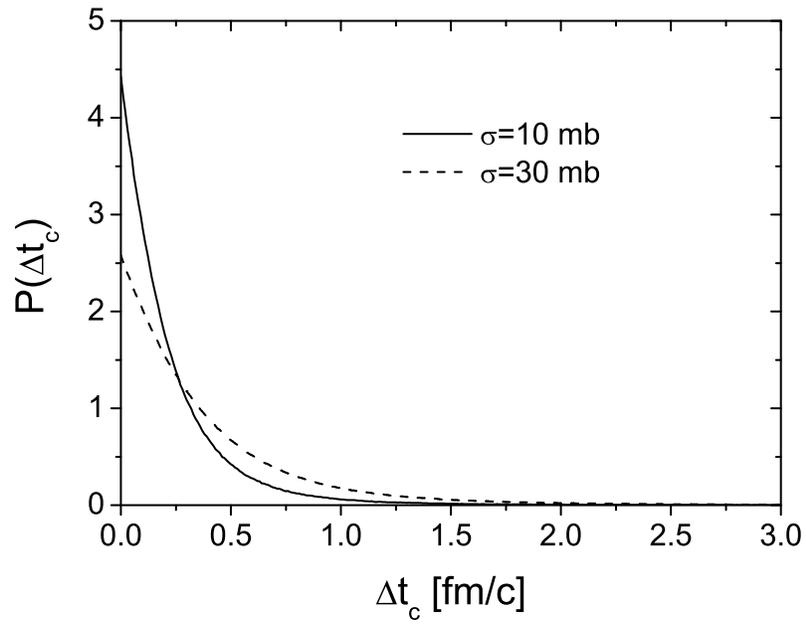}}
\caption{Probability distribution of difference in ``collision times''
within the geometrical collision algorithm. In the calculations
a thermal system is assumed and the cross section is set to be a constant. 
}
\label{app_pdtc}
\end{figure}

In the following we are interested in the probability distribution
of the difference of collision times, $\Delta t_c:=|t_{c1}-t_{c2}|$,
in a thermal system of massless particles. In this case we have $c=d=0$.
If $t_1 \ne t_2$ (e.g., $t_1 < t_2$), the particle with smaller time
($t_1$) can propagate freely to the larger time ($t_2$), which does not
give any effect on the whole evolution due to the additional criterion.
Thus we obtain
\begin{equation}
\label{dtc}
\Delta t_c= r_{12} \left | \frac{u_1+u_2}{1-\tilde u} \right | \,.
\end{equation}
$r_{12}$ denotes $|{\bf r}_2-{\bf r}_1|$ and $u_i=\cos \alpha_i$,
$\tilde u =\cos \theta$, where $\alpha_i$ is the angle between ${\bf p}_i$
and ${\bf r}_2-{\bf r}_1$ and $\theta$ is the angle between ${\bf p}_1$
and ${\bf p}_2$. Since $\tilde u$ relates $u_i$
according to $\tilde u=\sqrt{1-u_1^2}\,\sqrt{1-u_2^2}\,\cos(\phi_1-\phi_2)
+u_1\,u_2$, where $\phi_i$ is the polar angle of ${\bf p}_i$ around
${\bf r}_2-{\bf r}_1$, (\ref{dtc}) can be now expressed by
$\Delta t_c= r_{12}\,F(u_1,u_2,\phi)$ with $\phi := \phi_1-\phi_2$.
One obtains the probability distribution of $\Delta t_c$ by the integral
\begin{eqnarray}
\label{pdtc}
&&P(\Delta t_c) = \int_{-1}^1 du_1 \int_{-1}^1 du_2 \int_0^{2\pi} d\phi
\int_0^R dr_{12}\, P(r_{12},u_1,u_2,\phi)\, \delta(\Delta t_c-r_{12}F)\,
\Theta(\sqrt{\sigma/\pi}-\Delta r'_s) \nonumber \\
&& = \int_{-1}^1 du_1 \int_{-1}^1 du_2 \int_0^{2\pi} d\phi \,
P(r_{12},u_1,u_2,\phi)|_{r_{12}=\Delta t_c/F}\,
\frac{1}{F(u_1,u_2,\phi)}\,\Theta(\sqrt{\sigma/\pi}-\Delta r'_s)\,,
\end{eqnarray}
where $P(r_{12},u_1,u_2,\phi)$ is the multiple probability distribution.
Note that it is easy to realize that $\Delta r'_s$ can also be expressed
as a function of $r_{12}$, $u_1$, $u_2$
and $\phi$. Since $r_{12}$, $u_1$, $u_2$ and $\phi$ are independent
variables, $P(r_{12},u_1,u_2,\phi)$ can be factorized,
$P(r_{12},u_1,u_2,\phi)=P(r_{12})P(u_1)P(u_2)P(\phi)$. For a thermal
system we have $P(u_i)=1/2$, $P(\phi)=1/2\pi$ and $P(r_{12})=3r_{12}^2/R^3$,
where $R$ serves as a normalization factor and is set to be much larger
than the interaction length. We realize that the probability distribution 
(\ref{pdtc}) depends only on the size of the total cross section. For
a constant cross section we calculate the integral in (\ref{pdtc})
numerically. Figure \ref{app_pdtc} shows the results for $\sigma=10$ mb and
$\sigma=30$ mb. The distribution has a larger width for larger cross section.
We also calculate the mean value of $\Delta t_c$ and obtain
$<\Delta t_c>=0.24$ fm/c for $\sigma=10$ mb and $<\Delta t_c>=0.41$ fm/c
for $\sigma=30$ mb.

\section{Optimization of the computing time within the geometrical
method}
\label{app_optim}
Consider a system with $N$ particles in total. To get the next collision,
$N(N-1)/2$ operations have to be carried out to get all the ordering times
for each particle pair and $N(N-1)/2-1$ comparisons have to be made to
obtain the particle pair which collides next. Then these two particular
particles propagate freely until the two respective collision times when
the respective momenta will be sampled according to the differential
cross section. The same procedure will be repeated as long as needed.
Since the operation number in each step is proportional to $N^2$, the
computing time increases strongly with increasing particle number and
increasing collision number. However, a large amount of operations are
obviously futile, because after the update of two colliding particles
only the ordering times of particle pairs which involve one of the two
updated particles are indeed needed. Therefore only $2(N-2)$, but not
$N(N-1)/2$, operations are necessary if one, in principle, wants to save
all the ordering times from the last step. This, of course, reduce the
computing time enormously. On the other hand, however, a large storage for
those ordering times would be required. For an optimization we thus do
not store all the ordering times, but only do store for each particle
the informations of its possible next collision: the ordering time and
the collision partner. We need therefore $2N$, instead of $N(N-1)/2$,
memory places. The next collision will be obtained by comparing
the marked and stored ordering times. In a next step we compute only
the ordering times of the last colliding particles with the other
particles ($2(N-2)$ operations) and compare them with the other stored
times, respectively, to obtain the new informations of the next collision
for each particle. The ``worst'' case then occurs when the next
collision partner of a particle is one of the last colliding particles.
In this case the stored informations for this particle are out of use
and one has to compute the ordering times of the considered particle
with all the other particles (additional $N-3$ operations).
Fortunately those cases do not happen frequently in practice. We note that
our prescription is different from the optimization used by Zhang in
his parton cascade \cite{Zh98}, which follows the fact that particles
which are far away from each other most probably do not collide as next
pair. In this algorithm the space was divided into cells and only
particles from the same cell and the neighboring cells may collide next
within the geometrical method.

\section{Parton-Parton scattering cross sections}
\label{ppcs}
Differential pQCD parton-parton cross sections in leading order of $\alpha_s$
have been calculated in Ref. \cite{ORG78}. For elastic gluon scattering the
differential cross section reads
\begin{equation}
\label{gg}
\frac{d\sigma^{gg\to gg}}{dt} = \frac{9\pi\alpha_s^2}{2 s^2}
(3-\frac{t u}{s^2}-\frac{s u}{t^2}-\frac{s t}{u^2} ) \,,
\end{equation}
where $s$, $t$ and $u$ are the Mandelstam variables. $-t$ is equal to
the momentum transfer squared
\begin{equation}
-t=q^2 =\frac{s}{2}(1-\cos\theta) \,,
\end{equation}
where $\theta$ denotes the scattering angle in the
c.m. frame of colliding partons. For small angle scatterings the momentum
transfer is approximately equal to its transverse component 
$q_{\perp}$. Therefore we have $-t \approx q_{\perp}^2$.
Since the differential cross section (\ref{gg}) diverges at
small $t$ (and also at small $u$ due to the symmetry of
identical particles), Eq. (\ref{gg}) can be expressed approximately as
\begin{equation}
\label{gg1}
\frac{d\sigma^{gg\to gg}}{dq_{\perp}^2} \approx 
\frac{9\pi\alpha_s^2}{(q_{\perp}^2)^2}\,.
\end{equation}
We regularize the infrared sigularity in Eq. (\ref{gg1}) employing
the Debye mass and obtain
\begin{equation}
\label{gg2}
\frac{d\sigma^{gg\to gg}}{dq_{\perp}^2} =
\frac{9\pi\alpha_s^2}{(q_{\perp}^2+m_D^2)^2}\,.
\end{equation}
The other approximate differential cross sections are achieved in the same
way and read as follows:
\begin{eqnarray}
\frac{d\sigma^{gq\to gq}}{dq_{\perp}^2} &=&
\frac{2\pi\alpha_s^2}{(q_{\perp}^2+m_D^2)^2} \,,\\
\frac{d\sigma^{gg\to q\bar q}}{dq_{\perp}^2} &=&
\frac{\pi\alpha_s^2}{3s(q_{\perp}^2+m_q^2)} \,,\\
\frac{d\sigma^{qq\to qq}}{dq_{\perp}^2} &=&
\frac{16\pi\alpha_s^2}{9(q_{\perp}^2+m_D^2)^2} \,,\\
\frac{d\sigma^{qq'\to qq'}}{dq_{\perp}^2} &=& 
\frac{d\sigma^{q\bar q\to q\bar q}}{dq_{\perp}^2} =
\frac{8\pi\alpha_s^2}{9(q_{\perp}^2+m_D^2)^2} \,,\\
\frac{d\sigma^{q\bar q\to gg}}{dq_{\perp}^2} &=&
\frac{64\pi\alpha_s^2}{27s(q_{\perp}^2+m_q^2)} \,,\\
\label{qbq}
\frac{d\sigma^{q\bar q\to q'\bar q'}}{dt} &=&
\frac{4\pi\alpha_s^2}{9 s^2} \, \frac{t^2+u^2}{(s+4m_q^2)^2}\,,
\end{eqnarray}
where $m_D^2$ and $m_q^2$ denote, respectively, the Debye mass for
gluons and for quarks. In the last expression $-t$ is not replaced by
$q_{\perp}^2$, since $q\bar q\to q'\bar q'$ processes do not favor
small angle scatterings. Employing the fomulas (\ref{gg2})-(\ref{qbq})
the total cross sections can be obtained analytically by integration.
Equations (\ref{gg2})-(\ref{qbq}) also then dictate how to sample
new momenta for particles after an occurring collision.

\section{cross section for $gg \leftrightarrow ggg$ processes}
\label{ggcs}
For the multiplication process $gg\to ggg$, the Gunion-Bertsch
formula \cite{GB82} is used for the matrix element squared in leading
order of pQCD, and modified by implementing the Debye screening mass.
This is
\begin{equation}
\label{matrixgg}
| {\cal M}_{gg \to ggg} |^2 = \left ( \frac{9 g^4}{2} 
\frac{s^2}{({\bf q}_{\perp}^2+m_D^2)^2} \right ) 
\left ( \frac{12 g^2 {\bf q}_{\perp}^2}
{{\bf k}_{\perp}^2 [({\bf k}_{\perp}-{\bf q}_{\perp})^2+m_D^2]} \right ) \,,
\end{equation}
where $g^2=4\pi\alpha_s$, and ${\mathbf q}_{\perp}$ and 
${\mathbf k}_{\perp}$ are, respectively, the transverse component
of the momentum transfer and that of the momentum of radiated
gluon in the c.m. frame of the colliding gluons. In this
section we will give the derivations of the cross section $\sigma_{gg\to ggg}$
and $I_{ggg\to gg}$ defined in Sec. \ref{sec:multi} by an integral of
the scattering amplitude given in Eq. (\ref{matrixgg}) over momentum space.

Employing usual convention, the total cross section for a $gg \to ggg$
process is defined as
\begin{equation}
\label{csgg}
\sigma_{gg\to ggg} = \frac{1}{2s} \int 
\frac{d^3 p'_1}{(2\pi)^3 2 E'_1} \frac{d^3 p'_2}{(2\pi)^3 2 E'_2}
\frac{d^3 p'_3}{(2\pi)^3 2 E'_3} | {\cal M}_{12 \to 1'2'3'} |^2 (2\pi)^4
\delta^{(4)}(p_1+p_2-p'_1-p'_2-p'_3) \,,
\end{equation}
where $p_1$, $p_2$, $p'_1$, $p'_2$, and $p'_3$ are the four momenta and
all momenta are expressed in the c.m. frame of the two colliding gluons.
We assume that $3'$ denotes the radiated gluon.
Integrating over $d^3p'_2$ gives
\begin{eqnarray}
\label{csgg1}
\sigma_{gg\to ggg} &=& \frac{1}{256 \pi^5 s} \int 
\frac{d^3 p'_1}{E'_1} \frac{d^3 p'_3}{E'_3} \,
| {\cal M}_{12 \to 1'2'3'} |^2 \,
\delta((p_1+p_2-p'_1-p'_3)^2)  \nonumber \\
&=& \frac{1}{256 \pi^5 s} \int 
d^2 q_{\perp} dy'_1 d^2 k_{\perp} dy \,
| {\cal M}_{12 \to 1'2'3'} |^2 \,\delta(F) \,,
\end{eqnarray}
where $y'_1$ and $y$ denote the momentum rapidity of $1'$ and $3'$,
respectively, and
\begin{eqnarray}
\label{F}
F &=& (p_1+p_2-p'_1-p'_3)^2 \nonumber \\
&=& s-2\sqrt{s} q_{\perp} \cosh y'_1-2\sqrt{s} k_{\perp} \cosh y
+2 q_{\perp} k_{\perp} \cosh y'_1 \cosh y \nonumber \\
& & + 2{\mathbf q_{\perp}} \cdot
{\mathbf k_{\perp}}  -2 q_{\perp} k_{\perp} \sinh y'_1 \sinh y \,.
\end{eqnarray}
Further integration over $y'_1$ gives
\begin{equation}
\label{csgg2}
\sigma_{gg\to ggg} = \frac{1}{256 \pi^5 s} \int
d^2 q_{\perp} d^2 k_{\perp} dy \,
| {\cal M}_{12 \to 1'2'3'} |^2 \,\sum \frac{1}
{\left | \frac{\partial F}{\partial y'_1} \right |_{F=0}} \,,
\end{equation}
where all the solutions of $F=0$ contribute to the sum. The 
corresponding differential cross section has the form
\begin{equation}
\label{dcsgg2}
\frac{d\sigma_{gg\to ggg}}{d^2 q_{\perp} d^2 k_{\perp} dy} =
\frac{1}{256 \pi^5 s}\, | {\cal M}_{12 \to 1'2'3'} |^2 \,\sum \frac{1}
{\left | \frac{\partial F}{\partial y'_1} \right |_{F=0}} \,.
\end{equation}
This is different than in Ref. \cite{BDMTW93}, where the authors ignored
the factor $\sum 1/\left | \frac{\partial F}{\partial y'_1} \right |_{F=0}$.
However, to make the correct implementation of the detailed balance for
$gg \leftrightarrow ggg$ processes, one should take the exact formula of
the cross section. Expressing
$d^2 q_{\perp}$ and $d^2 k_{\perp}$ in polar coordinates and integrating
one of the two angles, one obtains
\begin{equation}
\label{in_polar}
\int d^2 q_{\perp} d^2 k_{\perp} dy \,\to \,
\pi \int dq_{\perp}^2 dk_{\perp}^2 dy \int_0^{\pi} d\phi \,,
\end{equation}
where $\phi$ denotes the angle between ${\mathbf q}_{\perp}$ and 
${\mathbf k}_{\perp}$.

We now turn to determine the integral boundaries for Eq. (\ref{in_polar}).
At first, the energies of the three particles in the final state cannot
be greater than $\sqrt{s}/2$ because of the energy conservation.
(Note that the total energy is equal to $\sqrt{s}$.) We then have
the upper boundaries for $q_{\perp}^2$ and $k_{\perp}^2$:
$q_{\perp}^2 < s/4$ and $k_{\perp}^2 < s/4$. Secondly, $k_{\perp}$
and $y$ will be further constrained by
$\Theta(k_{\perp} \Lambda_g-\cosh y)$ due to
the Laudau-Pomeranchuk suppression (compare Sec. \ref{sec:qgp}).
This leads to a lower cutoff for
$k_{\perp}$: $k_{\perp} > 1/\Lambda_g$. For given $q_{\perp}$ and
$k_{\perp}$, the constraints for $\cosh y$ are
\begin{equation}
\cosh y \le k_{\perp} \Lambda_g  \quad \mbox{and} \quad
\cosh y = \frac{E'_3}{k_{\perp}} \le \frac{\sqrt{s}}{2k_{\perp}} \,. 
\end{equation}
Thus the upper boundary of $y$, denoted by $y_m$, is the smaller one among
$Arcosh(k_{\perp} \Lambda_g)$ and $Arcosh(\sqrt{s}/2k_{\perp})$.
Finally we have
\begin{equation}
\label{csgg3}
\sigma_{gg\to ggg}  \sim \int_0^{s/4} dq_{\perp}^2 
\int_{1/\Lambda^2_g}^{s/4} dk_{\perp}^2 \int_{-y_m}^{y_m} dy 
\int_0^{\pi} d\phi \cdots \,.
\end{equation}
This integral actually scales with $s$, $\sigma_{gg\to ggg}=\bar \sigma/s$,
where
\begin{equation}
\bar \sigma \sim \int_0^{1/4} d\bar q_{\perp}^2 
\int_{1/\bar \Lambda^2_g}^{1/4} d\bar k_{\perp}^2 \int_{-y_m}^{y_m} dy 
\int_0^{\pi} d\phi \cdots 
\end{equation}
with $\bar q_{\perp}^2=q_{\perp}^2/s$, $\bar k_{\perp}^2=k_{\perp}^2/s$,
$\bar \Lambda_g=\Lambda_g \sqrt{s}$, and $\bar m_D^2=m_D^2/s$.
$\bar \sigma$ depends on two parameters: $\bar \Lambda_g$ and $\bar m_D^2$.
We evaluate the above integral numerically using the Monte Carlo integration
routine VEGAS \cite{PFTV}. For any sampled point 
$(\bar q_{\perp}^2, \bar k_{\perp}^2, y, \phi)$ one has to solve
$y'_1$ for $F=0$ in Eq. (\ref{F}). If there is no solution, then the
chosen point is out of the kinematic region and thus has no contribution
to the integral. Thus the equation $F=0$ serves as a further constraint
for the kinematic region of collisions.

For the annihilation process $ggg\to gg$, the analogous quantity as
$\sigma_{gg\to ggg}$, which sums all the possible final states, is
$I_{ggg\to gg}$ defined via
\begin{equation}
\label{ig}
I_{ggg\to gg}=\frac{1}{2} \int \frac{d^3 p'_1}{(2\pi)^3 2E'_1}
\frac{d^3 p'_2}{(2\pi)^3 2E'_2} | {\cal M}_{123\to 1'2'} |^2 (2\pi)^4
\delta^{(4)} (p_1+p_2+p_3-p'_1-p'_2) \,,
\end{equation}
where the factor $1/2$ takes the identical gluons $1'$ and $2'$ into
account and
\begin{equation}
| {\cal M}_{123\to 1'2'} |^2 = \frac{1}{\nu_g} | {\cal M}_{1'2'\to 123} |^2 \,,
\end{equation}
where $\nu_g=2\times 8$ is the gluon degeneracy factor.
Since $I_{ggg\to gg}$ is
invariant under Lorentz transformations, we evaluate the integral in the 
rest frame of the three incoming particles. Therefore it is
$p_1+p_2+p_3=(\sqrt{s}, {\mathbf 0})$. Integrating over $d^3p'_2$ in
Eq. (\ref{ig}) we find
\begin{eqnarray}
\label{i32}
I_{ggg\to gg} &=& \frac{1}{16 \pi^2} \int \frac{d^3 p'_1}{E'_1} \,
| {\cal M}_{123\to 1'2'} |^2 \, \delta ((p_1+p_2+p_3-p'_1)^2) \nonumber \\
&=& \frac{1}{16 \pi^2} \int d E'_1 \,d\cos \theta \,d\phi \,
E'_1 | {\cal M}_{123\to 1'2'} |^2 \, \delta (s-2\sqrt{s}\,E'_1)
\nonumber \\
&=& \frac{1}{64 \pi^2} \int_{-1}^{1} d\cos \theta \int_0^{2\pi} d\phi \,
| {\cal M}_{123\to 1'2'} |^2 \,,
\end{eqnarray}
where the solid angle $(\cos \theta, \phi)$ defines the orientation of 
${\mathbf p}'_1$ (${\mathbf p}'_2=-{\mathbf p}'_1$ because of the
energy-momentum conservation). We now express $q_{\perp}$, $k_{\perp}$ and 
${\mathbf q}_{\perp} \cdot {\mathbf k}_{\perp}$ in 
$| {\cal M}_{123\to 1'2'} |^2$ with the solid angle and momenta of the
incoming gluons. To do that one has to specify the fussion process $2\to 1$.
There are in total $6$ combinations. Each combination contributes to
$I_{ggg\to gg}$. One of them is
\[
123\to 1'2' \quad {\hat =} \quad \mbox{(a)}\, 23\to 2^*\, 
\mbox{and (b)} \, 12^* \to 1'2'
\]
and ${\bf p}_3$ corresponds to $({\mathbf k}_{\perp}, y)$.
In this particular case one can establish a coordinate system 
in momentum space whose $Z$ axis coincides with the orientation of 
${\mathbf p}_1$. We find out (after a direct but lengthy calculation)
\begin{eqnarray}
\label{app_qt}
&& q_{\perp} = E_1 \sin \theta \,,\\
&& k_{\perp} = E_3 \sqrt{1-(\sin \gamma \, \sin \theta \, \cos \phi
+ \cos \gamma \, \cos \theta )^2} \,,\\
&& {\mathbf q}_{\perp} \cdot {\mathbf k}_{\perp} = E_1 E_3 \,
(\sin \gamma \, \sin \theta \, \cos \theta \, \cos \phi
- \cos \gamma \, \sin^2 \theta ) \,,
\end{eqnarray}
where $\gamma$ denotes the angle between ${\mathbf p}_1$ and
${\mathbf p}_3$, and $(\cos \theta, \phi)$ is, as defined before,
the solid angle of ${\bf p}'_1$. Due to the Laudau-Pomeranchuk
suppression $\Theta (k_{\perp} \Lambda_g - \cosh y)$ and
$\cosh y = E_3/k_{\perp}$ we obtain the kinematic region for
the $ggg \to gg$ process
\begin{equation}
\label{app_kt}
k_{\perp} \ge \sqrt{\frac{E_3}{\Lambda_g}} \,.
\end{equation}
In analogy to $\sigma_{gg\to ggg}$, $I_{ggg\to gg}$ also scales with
$s$, $I_{ggg\to gg} \sim \bar I /s$, where $\bar I$ 
depends on five parameters, namely $E_1/\sqrt{s}$, $E_3/\sqrt{s}$,
$\gamma$, $\Lambda_g \sqrt{s}$, and $m_D^2/s$.

\section{Monte Carlo Sampling of momenta for outgoing particles}
\label{sampl}
Momenta of outgoing particles are sampled in the rest frame of the
incoming particles. Their momentum in the lab frame is obtained by
Lorentz transformations.

\subsection{$2\leftrightarrow 2$ processes}
In the rest frame the energy of each particle is $\sqrt{s}/2$. The only 
to be sampled quantity is the solid angle $(\cos \theta, \phi)$.
The scattering angle $\theta$ is samlped according to the differential cross
section and the polar angle $\phi$ is sampled uniformly within
$[0, 2\pi]$.

Since the pQCD differential cross sections (\ref{gg2})-(\ref{qbq})
can be integrated analytically, we can perform samplings for $q_{\perp}$
(or $\cos \theta$)
using the ``transformation method'' \cite{PFTV} from a uniform probability 
distribution. For isotropic collisions we sample the scattering 
angle $\theta$ according to the uniform distribution of $\cos \theta$.

\subsection{$gg\leftrightarrow ggg$ processes}
As shown in Appendix \ref{ggcs}, the differential cross section for
a $gg \to ggg$ process has the form
\begin{equation}
\label{dcsgg}
\frac{d\sigma_{gg\to ggg}}{dq_{\perp}^2 dk_{\perp}^2 dy d\phi} \sim
\frac{1}{(q_{\perp}^2+m_D^2)^2} \frac{q_{\perp}^2}
{k_{\perp}^2 [({\bf k}_{\perp}-{\bf q}_{\perp})^2+m_D^2]} \,\sum \frac{1}
{\left | \frac{\partial F}{\partial y'_1} \right |_{F=0}} \,,
\end{equation}
where $\phi$ denotes the angle between ${\mathbf q}_{\perp}$ and
${\mathbf k}_{\perp}$. We then first sample $q_{\perp}$, $k_{\perp}$,
$y$, and $\phi$ according to Eq. (\ref{dcsgg}). Since the differential
cross section cannot be integrated analytically, one cannot make
samplings by means of the transformation method, as done for
$2\to 2$ processes. Instead, we employ the ``rejection method'' \cite{PFTV}.

To make enough efficient samplings, we want to find out a special function
of $q_{\perp}$, $k_{\perp}$, $y$ and $\phi$, which should be always
greater than the right hand side of Eq. (\ref{dcsgg}) at every point set
($q_{\perp}$, $k_{\perp}$, $y$, $\phi$) in the kinematic region and, more
important, can be integrated out analytically over 
$q_{\perp}$, $k_{\perp}$, $y$, and $\phi$. Such a function is called as
a ``comparison function''. If one has the comparison function, one can
first use the transformation method to generate the random numbers
according to the comparison function. Then one needs a further
uniform sampling between zero and the value of the comparison function
at the particular sampled point. If this random number is less than
the value of the real distribution
[right hand side of Eq. (\ref{dcsgg})] at the sampled point, then we accept
this sampling, if not, we reject this sampling and start a next trial.
One possible choice of the comparison function is
\begin{equation}
\label{reject}
\frac{1}{q_{\perp}^2+m_D^2} \, \frac{1}{k_{\perp}^2} \,
\frac{1}{m_D^2} \, m \,,
\end{equation}
where $m$ denotes a constant with a sufficient large value, which
is greater than
$\sum 1/\left | \frac{\partial F}{\partial y'_1} \right |_{F=0}$
in Eq. (\ref{dcsgg}) at every point in the possible
kinematic region. Since, unfortunately, one cannot obtain the upper limit
for $\sum 1/\left | \frac{\partial F}{\partial y'_1} \right |_{F=0}$,
the value of $m$ is an empirical number.

We have to note that for an individual sampling one has to solve the
equation $F=0$, Eq. (\ref{F}). Therefore one also obtains $y'_1$, the momentum
rapidity of the particle $1'$, at the same time when $q_{\perp}$,
$k_{\perp}$, $y$, and $\phi$ are sampled. One sampling remains:
The direction of ${\bf q}_{\perp}$
is sampled uniformly in the transverse plan being perpendicular to the
scattering axis. Finally we obtain the momenta of the outgoing particles
\begin{eqnarray}
&&{\bf p}'_{1\perp}=-{\bf q}_{\perp} \,, \quad p'_{1z}=q_{\perp} \sinh y'_1\,,
\\
&&{\bf p}'_{3\perp}={\bf k}_{\perp} \,, \quad p'_{3z}=k_{\perp} \sinh y\,, \\
&&{\bf p}'_2=-({\bf p}'_1+{\bf p}'_3) \,.
\end{eqnarray}

For a $ggg\to gg$ process the solid angle ($\cos \theta, \phi$) is sampled
again by using the rejection method. We find
\begin{eqnarray}
\label{reject1}
\frac{dI_{ggg\to gg}}{d\cos \theta \, d\phi} &\sim&
\frac{1}{(q_{\perp}^2+m_D^2)^2} \frac{q_{\perp}^2}
{k_{\perp}^2 [({\bf k}_{\perp}-{\bf q}_{\perp})^2+m_D^2]} \nonumber \\
&<& \frac{1}{q_{\perp}^2+m_D^2} \, \frac{\Lambda_g}{E_3} \,
\frac{1}{m_D^2} = \frac{1}{E_1^2(1-\cos^2 \theta) +m_D^2} \, 
\frac{\Lambda_g}{E_3} \, \frac{1}{m_D^2} \,,
\end{eqnarray}
where we have employed the constraint (\ref{app_kt}) and the identity
(\ref{app_qt}) for the particular example presented in Appendix \ref{ggcs}.

\subsection{Isotropic $2\leftrightarrow 3$ processes}
A $2\to 3$ process, $12\to 1'2'3'$, is assumed to be composed of a two-body
scattering $12\to 1'2^*$ and a decay $2^*\to 2'3'$, where $2^*$ denotes
an intermediate state with an invariant mass of
$m^*=\sqrt{E_*^2-{\mathbf p}_*^2}$.
We employ the formula for the phase space integrations of \cite{BK73}
and obtain the differential cross section of an isotropic collision
\begin{equation}
\label{iso23}
\frac{d\sigma_{23}}{d\Omega_1 dm_*^2 d\Omega_2} \sim 
\lambda^{\frac{1}{2}}(s,0,m_*^2) \int dE'_2 \, \frac{E'_2}{E_*-E'_2} \,
\delta (f(E'_2)) \,,
\end{equation}
where $\Omega_1=(\cos \theta_1, \phi_1)$ denotes the solid angle of
${\mathbf p}'_1$ with respect to the collision axis, and
$\Omega_2=(\cos \theta_2, \phi_2)$ denotes the solid angle of
${\mathbf p}'_2$ with respect to ${\mathbf p}_*$, and
\begin{eqnarray}
&& f(E'_2)=E_*-E'_2-\sqrt{p_*^2+{E'}_2^2-2p_*E'_2 \cos \theta_2} \,,\\
&& \lambda(s,m_1^2,m_2^2)=s^2-2s(m_1^2+m_2^2)+(m_1^2-m_2^2)^2 \,.
\end{eqnarray}
The integral over $E'_2$ in Eq. (\ref{iso23}) gives
\begin{equation}
\label{iso231}
\frac{d\sigma_{23}}{d\Omega_1 dm_*^2 d\Omega_2} \sim 
\frac{(s-m_*^2)m_*^2}{(E_*-p_* \cos \theta_2)^2} \,.
\end{equation}
$\Omega_1$, $m_*^2$ and $\Omega_2$ are the to be sampled quantities. From
Eq. (\ref{iso231}) we realize that the differential cross section does not
depend on $\Omega_1$ and $\phi_2$. Thus they are sampled uniformly. Integral
over $\Omega_1$ and $\Omega_2$ gives the probability distribution for $m_*^2$.
It is simply proportional to $s-m_*^2$. We sample $m_*^2$ by employing the
transformation method. For given $\Omega_1$ and $m_*^2$ the momenta of
$1'$ and $2^*$ are fully determined due to the energy-momentum conservation.
Now Eq. (\ref{iso231}) just represents the probability distribution of
$\cos \theta_2$ for a given $\Omega_1$ and $m_*^2$. Its numerical sampling
is straightforward.

Sampling for an isotropic $3\to 2$ process is more trivial, since just one
solid angle is to be sampled and its probability distribution is uniform.



\newpage

\newpage

\newpage

\newpage

\newpage

\newpage

\newpage

\newpage

\newpage

\newpage

\newpage

\newpage

\newpage

\newpage

\newpage

\newpage

\newpage

\newpage

\newpage

\newpage

\newpage

\newpage

\newpage

\newpage

\newpage

\newpage

\newpage

\newpage

\newpage

\newpage

\newpage

\newpage

\newpage

\newpage

\newpage

\newpage

\newpage

\newpage

\newpage

\newpage

\newpage

\newpage

\newpage

\newpage

\newpage

\newpage

\newpage

\newpage

\newpage

\newpage

\newpage

\newpage

\newpage

\newpage

\newpage

\end{document}